\documentclass[final]{jfp-epi}

\usepackage{ifthen}

\newboolean{talk}
\setboolean{talk}{false}
\newboolean{paper}
\setboolean{paper}{false}

\newboolean{LMCSstyle}
\setboolean{LMCSstyle}{false}
\newboolean{IEEEstyle}
\setboolean{IEEEstyle}{false}
\newboolean{lipicsstyle}
\setboolean{lipicsstyle}{false}
\newboolean{eptcsstyle}
\setboolean{eptcsstyle}{false}
\newboolean{sigplanstyle}
\setboolean{sigplanstyle}{false}
\newboolean{PACMPL}
\setboolean{PACMPL}{false}
\newboolean{acmartstyle}
\setboolean{acmartstyle}{false}
\newboolean{JFPstyle}
\setboolean{JFPstyle}{false}

\newboolean{needstheorems}
\setboolean{needstheorems}{false}
\newboolean{withimages}
\setboolean{withimages}{false}
\newboolean{withproofs}
\setboolean{withproofs}{true}

\newboolean{french}
\setboolean{french}{false}

\setboolean{PACMPL}{true}

\usepackage{amsthm}

    \newtheorem{theorem}{Theorem}[section]
    \newtheorem{lemma}[theorem]{Lemma}
    \newtheorem{corollary}[theorem]{Corollary}
    \newtheorem{proposition}[theorem]{Proposition}
    \newtheorem{definition}[theorem]{Definition}
    \newtheorem{remark}[theorem]{Remark}
    \newtheorem{example}[theorem]{Example}

\ifthenelse{\boolean{talk}}{}{\usepackage[utf8]{inputenc}}
\ifthenelse{\boolean{PACMPL}}{}{
	\ifthenelse{\boolean{french}}{\usepackage[french]{babel}}{
	  \ifthenelse{\boolean{lipicsstyle}}{}{\usepackage[english]{babel}}}
	  }
\ifthenelse{\boolean{PACMPL}}{}{
	\ifthenelse{\boolean{acmartstyle}}{}{
		\ifthenelse{\boolean{JFPstyle}}{}{
			\usepackage{amssymb}
		}
	}
}
\usepackage{amsmath}
\usepackage{graphicx}
\usepackage{bussproofs}
\usepackage{fancybox}
\usepackage{cmll}
\usepackage{stmaryrd}
\usepackage{multirow}
\ifthenelse{\boolean{IEEEstyle}}{
	\usepackage[bookmarks={false}]{hyperref}
	}{
	\usepackage{hyperref}
	}
\usepackage{proof}
\usepackage{hhline}
\usepackage{xspace}
\usepackage{booktabs}
\usepackage{subcaption}
\usepackage{microtype}
\usepackage{wrapfig}
\usepackage{array}
\usepackage{tabularx}
\usepackage{arydshln}
\usepackage{commath}
\usepackage{thmtools}
\ifthenelse{\boolean{IEEEstyle}}{
\setboolean{needstheorems}{true}}{}

\ifthenelse{\boolean{eptcsstyle}}{
\setboolean{needstheorems}{true}}{}

\ifthenelse{\boolean{sigplanstyle}}{
\setboolean{needstheorems}{true}}{}

\ifthenelse{\boolean{needstheorems}}{
}{}

\newcommand{\myproof}[1]{
\ifthenelse{\boolean{withproofs}}{#1}{}}

\newcommand{\withproofs}[1]{
\ifthenelse{\boolean{withproofs}}{#1}{}}

\newcommand{\withoutproofs}[1]{
\ifthenelse{\boolean{withproofs}}{}{#1}}

\newcommand{\tm}{t}
\newcommand{\tmtwo}{u}
\newcommand{\tmthree}{r}

\newcommand{\var}{x}
\newcommand{\vartwo}{y}
\newcommand{\varthree}{z}

\newcommand{\Rew}[1]{\rightarrow_{#1}}

\renewcommand{\to}{\Rew{}}

\newcommand{\tob}{\Rew{\beta}}

\newcommand{\towh}{\Rew{wh}}

\newcommand{\symfont}[1]{\mathsf{#1}}

\newcommand{\ctxholep}[1]{[#1]}
\newcommand{\ctxhole}{\ctxholep{\cdot}}

\newcommand{\ctx}{C}

\newcommand{\ctxp}[1]{\ctx\ctxholep{#1}}

\newcommand{\nbvctxtwo}[1]{\nbvctxtwo{#1}}

\newcommand{\defeq}{:=}

\newcommand{\grameq}{::=}

\newcommand{\isub}[2]{\{#1/#2\}}
\newcommand{\esub}[2]{[#1/#2]}
\newcommand{\llbrace}{\{ \kern -0.27em \vert}
\newcommand{\rrbrace}{\vert \kern -0.27em \}}

\newcommand{\grammarpipe}{\mathrel{\big |}}

\renewcommand{\l}{\lambda}
\newcommand{\ie}{i.e.\xspace}
\ifthenelse{\boolean{LMCSstyle}}{}{
	\newcommand{\eg}{e.g.\xspace}
	}
\newcommand{\ih}{\textit{i.h.}\xspace}
\newcommand{\fv}[1]{\symfont{fv}(#1)}

\newcommand{\red}[1]{{\color{red} {#1}}}
\newcommand{\blue}[1]{{\color{blue} {#1}}}

\newcommand{\ignore}[1]{}

\newcommand{\myinput}[1]{\ifthenelse{\boolean{withimages}}{\input{#1}}{}}

\ifthenelse{\boolean{JFPstyle}}{}
{}

\newcommand{\lolli}{\multimap}

\newcommand{\size}[1]{|#1|}
\newcommand{\sizeparam}[2]{|#1|_{#2}}

\newcommand{\clos}{c}

\newcommand{\lenv}{e}
\newcommand{\lenvtwo}{\lenv'}

\newcommand{\stack}{S}

\ifthenelse{\boolean{PACMPL}}{\renewcommand{\state}{s}}{
	\ifthenelse{\boolean{acmartstyle}}{\renewcommand{\state}{s}}{
		\newcommand{\state}{s}
	}
}
\newcommand{\statetwo}{{\state'}}

\newcommand{\tokam}{\mathrm{\Rew{\KAM}}}
\newcommand{\tospkam}{\mathrm{\Rew{Sp\KAM}}}

\newcounter{numberone}

\newcounter{numbertwo}

\renewcommand{\ctxholep}[1]{\langle #1\rangle}

\newcommand{\dom}[1]{\mathsf{dom}(#1)}

\newcommand{\unfsym}{\rotatebox[origin=c]{-90}{$\rightarrow$}}
\newcommand{\unf}[1]{#1\unfsym}

\newcommand{\reflemma}[1]{Lemma~\ref{l:#1}}
\newcommand{\reflemmap}[2]{Lemma~\ref{l:#1}.\ref{p:#1-#2}}

\newcommand{\refprop}[1]{Prop.~\ref{prop:#1}}

\newcommand{\refthm}[1]{Theorem~\ref{thm:#1}}

\newcommand{\reffig}[1]{Fig.~\ref{fig:#1}}

\ifthenelse{\boolean{LMCSstyle}}{	
	
	}{
	
	}

\ifthenelse{\boolean{JFPstyle}}{}
	{}
\renewcommand{\esub}[2]{[#1{\shortleftarrow}#2]}
\renewcommand{\isub}[2]{\{#1{\shortleftarrow}#2\}}

\newcommand{\run}{\rho}
\newcommand{\runtwo}{\sigma}

\newcommand{\kstatetab}[3]{#1 & #2 & #3 }
\newcommand{\kstate}[3]{(#1,#2,#3)}

\newcommand{\cons}{{\cdot}}

\newcommand{\KAM}{KAM\xspace}
\newcommand{\NaKAM}{Naive KAM\xspace}

\newcommand{\SpKAM}{Space KAM\xspace}

\newcommand{\tomachhole}[1]{\rightarrow_{#1}}
\newcommand{\tomach}{\tomachhole{}}

\renewcommand{\tokam}{\rightarrow_{\textsc{KAM}}}

\newcommand{\stempty}{\epsilon}

\newcommand{\la}[1]{\lambda #1.}

\newcommand{\sem}[2]{\llbracket#1\rrbracket_{#2}}

\newcommand{\midd}{\; \; \mbox{\Large{$\mid$}}\;\;}

\newcommand{\rel}{\mathcal{R}}

\newcommand{\ccallbn}{Closed Call-by-Name\xspace}
\newcommand{\cbn}{CbN\xspace}
\newcommand{\ccbn}{Closed \cbn}

\ifthenelse{\boolean{JFPstyle}}{}
{

}

\newcommand{\Id}{\symfont{I}}

\newcommand{\mset}[1]{[#1]}
\newcommand{\emmset}{[\,]}

\newcommand{\initty}{\star}

\newcommand{\linty}{A}
\newcommand{\lintytwo}{\linty'}

\newcommand{\gty}{G}

\newcommand{\mty}{\mathcal{A}}
\newcommand{\mtytwo}{\mathcal{B}}
\renewcommand{\mty}{S}
\renewcommand{\mtytwo}{\mty'}
\newcommand{\arr}[2]{#1\rightarrow #2}

\newcommand{\tye}{\Gamma}
\newcommand{\tyetwo}{\Delta}

\newcommand{\tjudg}[3]{#1\vdash #2:#3}

\newcommand{\wtjudgone}[1]{\vdash^{\textcolor{violet}{#1}}}
\newcommand{\wtjudg}[4]{#1\stackrel{\textcolor{violet}{#2}}{\vdash}#3:#4}

\newcommand{\tjudgw}[4]{#1\vdash^{\textcolor{violet}{#2}} #3:#4}

\newcommand{\tyvar}{\textsc{T-var}}
\newcommand{\tylamstar}{\tylam_\star}
\newcommand{\tylam}{\textsc{T-}\lambda}
\newcommand{\tyapp}{\textsc{T-@}}
\newcommand{\tymany}{\textsc{T-many}}

\newcommand{\tymanysr}{\textsc{T-mr}}
\newcommand{\tymanysb}{\textsc{T-mb}}
\newcommand{\tynone}{\textsc{T-none}}

\newcommand{\tynonesr}{\textsc{T-nr}}
\newcommand{\tynonesb}{\textsc{T-nb}}

\newcommand{\tyenv}{\textsc{T-env}}
\newcommand{\tyclos}{\textsc{T-cl}}
\newcommand{\tystate}{\textsc{T-st}}

\newcommand{\tyd}{\pi}
\newcommand{\tydtwo}{\tyd'}
\newcommand{\tydthree}{\tyd''}
\newcommand{\pof}{\;\triangleright}

\newcommand\mydots{\hbox to .6em{.\hss.}}

\newcommand{\lm}[1]{\lambda \text{Meas}(#1)}
\renewcommand{\lm}[1]{\sizeparam{#1}{\mathsf{sp}}}

\newcommand{\weight}{w}
\newcommand{\weighttwo}{v}
\newcommand{\weightthree}{u}

\newcommand{\emmtype}{\mset{\,}}
\newcommand{\spkamstate}[3]{(#1,#2,#3)}

\newcommand{\sizeb}[1]{\sizeparam{#1}{\beta}}

\newcommand{\seasym}{\symfont{sea}}
\newcommand{\seavsym}{\symfont{sea}_{\symfont{v}}}
\newcommand{\seanvsym}{\symfont{sea}_{\neg \symfont{v}}}
\newcommand{\subsym}{\symfont{sub}}
\newcommand{\bwsym}{\beta_{\symfont{w}}}
\newcommand{\bnwsym}{\beta_{\neg\symfont{w}}}

\newcommand{\tokamsea}{\tomachhole{\seasym}}
\newcommand{\tokamb}{\tomachhole{\beta}}
\newcommand{\tokamsub}{\tomachhole{\subsym}}
\newcommand{\tokamseav}{\tomachhole{\seavsym}}
\newcommand{\tokamseanv}{\tomachhole{\seanvsym}}
\newcommand{\tokambw}{\tomachhole{\bwsym}}
\newcommand{\tokambnw}{\tomachhole{\bnwsym}}

\newcommand{\compil}[1]{\symfont{init}(#1)}

\renewcommand{\state}{q}

\newcommand{\PSPACE}{\mathsf{PSPACE}}

\renewcommand{\P}{\mathsf{P}}

\newcommand{\lmt}[1]{|#1|_\mathsf{ti}}

\newcommand{\relsym}{{\mathcal R}}
\newcommand{\Cctx}{\mathcal{\ctx}}
\newcommand{\Cctxp}[1]{\Cctx\ctxholep{#1}}
\newcommand{\lintys}{\mathcal{\linty}}

\newcommand{\cred}{\red{\mathsf{r}}}
\newcommand{\cblue}{\blue{\mathsf{b}}}

\newcommand{\cols}{\mathsf{C}}
\newcommand{\colf}[1]{\mathsf{col}(#1)}
\newcommand\doubleplus{\ensuremath{\mathbin{+\mkern-5mu+}}}
\newcommand{\pair}{p}

\newcommand{\wf}[1]{\mathsf{WF}(#1)}
\usepackage{xcolor}
\usepackage{microtype}
\usepackage{versions}
\includeversion{LONG}
\excludeversion{SHORT}

\renewcommand{\mty}{M}
\renewcommand{\mtytwo}{N}
\renewcommand{\stack}{\pi}
\renewcommand{\tyd}{\tau}
\renewcommand{\reffig}[1]{Figure~\ref{fig:#1}}

\begin{document}

\jfpJournal{JFP}
\jfpVolume{36}
\jfpArticle{5}
\jfpDOI{10.46298/jfp.17809}
\jfpYear{2026}
\received[Submitted]{July 2025}
\received[accepted]{May 2026}

\author{Beniamino Accattoli}
\orcid{0000-0003-4944-9944}           %% \orcid is optional
\affiliation{  
	\department{LIX}              %% \department is recommended
	\institution{Inria \& LIX, \'Ecole Polytechnique, UMR 7161}            %% 
	%%\institution is required  
	\country{France}             
	\authoremail{beniamino.accattoli@inria.fr}       %% \country is 
	%%recommended
}
          %% \email is recommended

%% Author with single affiliation.
\author{Ugo Dal Lago}
\orcid{0000-0001-9200-070X}  %% \orcid is optional
\affiliation{\institution{Universit\`a di Bologna}\country{Italy}}
\affiliation{\institution{ Inria}\country{France}
\authoremail{ugo.dallago@unibo.it}}

%% Author with single affiliation.
\author{Gabriele Vanoni}
\orcid{0000-0001-8762-8674}      %% \orcid is optional
\affiliation{\institution{CNRS, IRIF, Université Paris Cité}\country{France}
\authoremail{gabriele.vanoni@irif.fr}}

\title{Multi types and reasonable space}

%\received{20 March 1995; revised 30 September 1998}

\begin{abstract}
% !TeX spellcheck = en_US
% !TEX root = main.tex
%%%%%%%%%%%%%%%%%%%%%%%%%%%%%%%%%%%%%%%%%%%%%%%%%%%%%%%%%%%%%%%%%%%%%%%%%%%%%%%%
Accattoli, Dal Lago, and Vanoni have recently proved that the space used by the \textit{Space KAM}, a variant of the Krivine abstract machine, is a reasonable space cost model for the $\lambda$-calculus accounting for logarithmic space, solving a longstanding open problem. In this paper we provide a new system of \textit{multi types} (a variant of intersection types) and extract from multi type derivations the space used by the Space KAM, capturing into a type system the space complexity of the abstract machine. Additionally, we show how to capture also the \emph{time} of the Space KAM, which is a reasonable time cost model, and how to capture space consumption when pointers have different sizes, all via minor changes to the type system.
\end{abstract}

\maketitle

% !TeX spellcheck = en_US
% !TEX root = main.tex
%%%%%%%%%%%%%%%%%%%%%%%%%%%%%%%%%%%%%%%%%%%%%%%%%%%%%%%%%%%%%%%%%%%%%%%%%%%%%%%%
\section{Introduction}
Type systems are a key aspect of programming languages, ensuring good behavior during the execution of programs, such as absence of errors, termination, or properties such as productivity,  safety, and reachability properties. %Additionally, they ensure it in a \emph{compositional} way, that is, if programs are assembled according to the underlying type discipline, then the good behavior is ensured also for the composed program. 
For functional programming languages, types tend to have the structure of a logical system, to the point that, in the paradigmatic case of the simply typed $\l$-calculus, the type system is \emph{isomorphic} to the natural deduction presentation of intuitionistic logic. This is the celebrated \emph{Curry-Howard correspondence}. 

In the theory of the $\l$-calculus, there is another use of types that has been studied at length by experts, but it is less known to the wider research community on programming languages. Historically, it was introduced by Coppo and Dezani{-}Ciancaglini as the study of \emph{intersection types} \citep{DBLP:journals/aml/CoppoD78}. At first sight such types are simply a slight variation on simple types, in which one can assign a finite set of types to a term, thus providing a form of finite polymorphism. While some forms of intersection types can be used for programming purposes (see \citet{DBLP:conf/pldi/FreemanP91,DBLP:journals/mscs/Pierce97,DBLP:journals/jacm/FrischCB08,DBLP:conf/birthday/Dezani-Ciancaglini18} or the recent survey by \citet{DBLP:conf/lics/BonoD20}), they have mainly played a more theoretical role so far. 

Similarly to simple types, intersection types ensure termination. In contrast to most notions of type, however, they also \emph{characterize} termination, that is, they type \emph{all} terminating $\l$-terms. Intersection types have proved to be remarkably flexible, as by tuning details of the type system it is possible to characterize different forms of termination (weak/strong full normalization, head/weak/call-by-value evaluation) for various kinds of $\l$-calculi (with pattern matching~\citep{DBLP:conf/aplas/AlvesKR24}, state~\citep{DBLP:journals/tcs/deLiguoroT23,DBLP:conf/wollic/AlvesKR23,Alves_Kesner_Ramos_2025}, probabilities~\citep{DBLP:journals/pacmpl/LagoFR21}, algebraic effects~\citep{DBLP:conf/esop/GavazzoTV24,DBLP:journals/toplas/GalalGTV25} and more). Termination being only \emph{semi-decidable}, type inference cannot be decidable, which is why standard intersection types are somewhat incompatible with programming practice. As already pointed out, restricted forms of intersection types do find applications in programming languages, but the restrictions usually break the characterization of termination, namely their most distinctive trait.

\paragraph*{Abstracting computation.} The impractical aspect of intersection types can be understood as coming from the fact that inferring a type for $\tm$ is very similar to normalizing $\tm$. There is a variant of intersection types where the idempotency of intersections $A\cap A = A$ is dropped, which we refer to as \emph{multi types}, because non-idempotent intersections can be seen as multi-sets. \citet{DBLP:conf/icfp/NeergaardM04} show that, for multi types, type inference is \emph{equivalent} to normalization of the term to type. Careful here: this is \emph{not} a Curry-Howard correspondence, where normalization corresponds to cut (or detour) elimination from the type derivation, which goes through \emph{many} derivations. Here normalization corresponds to  inference of a \emph{single} type derivation, without any rewriting involved. 

The undecidability of type inference is balanced by the fact that, intuitively, multi-type derivations fully capture the computing mechanism itself in a \emph{compositional} way, despite looking very different from a computing device. The $\l$-calculus is already quite an abstract notion of computational model, if compared with Turing or random access machines, but it retains the idea of computation as a state transition system. Multi types push things a level further, as the entire dynamics of normalization is captured by a single \emph{static} type derivation. 

\paragraph*{Bounding computation.} The literature has not really focused on the relationship between the two processes of normalization and multi-type inference, because such a type inference is impractical. Instead, multi-type derivations have been used as \emph{exhaustive transcriptions} of the normalization process, from which to extract information about the normalization process itself. This information usually amounts to bounding: 
\begin{enumerate}
\item the number of steps to normal form, and 
\item  the size of the normal form itself,
\end{enumerate}
according to various notions of reduction (weak head/head/leftmost, for instance). The first such results are de Carvalho's bounds on the length of runs of the Krivine abstract machine~\citep{krivine_call-by-name_2007} (shortened to KAM) and of an extension of the KAM to (strong) leftmost evaluation \citep{deCarvalho07,deCarvalho18}. 
Such a \emph{de Carvalho correspondence} was shortly afterwards extended to strong 
normalization by \citet{Bernadet-Lengrand2013} and to linear 
logic proof nets by 
\citet{DBLP:journals/tcs/CarvalhoPF11}. The interest in such a correspondence is 
that it provides an abstract and compositional viewpoint on quantitative operational aspects. 
Moreover, the set of types a term $\tm$ can be typed with is a syntactic 
presentation of the relational semantics of $\tm$, a paradigmatic denotational 
model of the $\l$-calculus~\citep{DBLP:journals/apal/BucciarelliE01}.

%The compositionality of the type system plays a role in how tight are the bounds extracted from the type derivations. Terms can be assigned many different multi types, via different type derivations. The size of the multi type , the richer the interaction that the term can have with its environment, the less informative the bounds.

\paragraph*{Multi types and reasonable cost models.} When de Carvalho developed his 
correspondence in 2007, it was known that the number of steps in the weak 
$\l$-calculus is a reasonable time cost model (where \emph{reasonable} roughly 
means equivalent up to a \emph{polynomial} to the time cost model of Turing 
machines), as first proved by 
\citet{DBLP:conf/fpca/BlellochG95}. The reasonable time cost model of the strong 
(that is, unrestricted) $\l$-calculus was instead unclear (because the techniques 
for the weak case do not work in the strong case), thus it was also unclear to 
what extent the bounds extracted from multi-type derivations are related to the 
time complexity of $\l$-terms. In 2014 Accattoli and Dal Lago showed that the 
number of leftmost $\beta$ steps is a reasonable time cost model for strong 
evaluation \citep{accattoli_leftmost-outermost_2016}, clarifying the situation. In 
2018 Accattoli, Graham-Lengrand, and Kesner revisited the de Carvalho 
correspondence and refined it according to the new understanding of time for the 
$\l$-calculus \citep{accattoli_tight_2018}. Their work triggered a systematic 
extension of the de Carvalho correspondence to many $\l$-calculi and abstract 
machines \citep{DBLP:conf/aplas/AccattoliG18,DBLP:conf/esop/AccattoliGL19, 
DBLP:conf/fossacs/KesnerPV21,DBLP:conf/flops/BucciarelliKRV20, 
DBLP:conf/csl/KesnerV22,DBLP:conf/lics/KesnerV20,DBLP:conf/types/AlvesKV19,DBLP:journals/pacmpl/LagoFR21,POPL2021,DBLP:journals/tcs/LagoV26,DBLP:journals/pacmpl/FaggianPV24,DBLP:journals/pacmpl/AccattoliLMV25}.

In a recent work \citet{LICS2021} pushed  the 
correspondence into a new direction, by extracting the first \emph{space} bounds 
from multi-type derivations. Additionally, they use these bounds to disprove a 
conjecture about reasonable space for the $\l$-calculus. Namely, their bounds 
apply to Mackie's and Danos \& Regnier's \emph{interaction abstract 
machine}~\citep{mackie_geometry_1995,DBLP:conf/lics/DanosHR96,DR99,PPDP2020} (which 
is the computational machinery behind Girard's Geometry of 
Interaction~\citep{GIRARD1989221} and Abramsky, Jagadeesan, and Malacaria's game semantics \citep{DBLP:journals/iandc/AbramskyJM00}) and prove that its use of space---that was conjectured to be reasonable---is in fact \emph{unreasonable}.

\paragraph*{Reasonable space.} It is only very recently that results about reasonable space for the $\l$-calculus have started to appear, under the impulse of the advances about reasonable time. Forster, Kunze, and Roth showed that the maximum size of $\l$-terms during evaluation---the natural space cost model---is reasonable \citep{DBLP:journals/pacmpl/ForsterKR20} (that is, it is \emph{linearly} related to the space of Turing machines, as prescribed by Slot and van Emde Boas \citep{DBLP:journals/iandc/SlotB88,DBLP:books/el/leeuwen90/Boas90}). Such a notion of space, however, cannot account for logarithmic space, which is the smallest robust space class, playing the role of $\P$ for space. To study logarithmic space, indeed, one needs a separation between \emph{input} space and \emph{work} space, which cannot be modeled taking as space the size of $\l$-terms. 
Very recently, Accattoli, Dal Lago, and Vanoni have also shown that the space used 
by the \emph{\SpKAM}, a variant over the KAM, is a reasonable space cost model 
accounting for logarithmic space \citep{LICS2022}, solving the longstanding open 
problem. The \SpKAM tweaks the KAM in two orthogonal ways:
\begin{enumerate}
\item \emph{Space optimizations}: it performs two space optimizations, \emph{environment unchaining} and \emph{eager garbage collection}. Unchaining is a folklore optimization that prevents space leaks, that is, it keeps the environment compact. Eager garbage collection maximizes the reuse of space. Essentially, alive closures are compacted and dead closures are removed as soon as possible.

\item \emph{No environment sharing nor pointers}:  the sharing of environments 
used in the implementation of the KAM is \emph{disabled}, and additionally 
environments (as well as the argument stack) are \emph{not} represented as linked 
lists (via pointers). Environments and stacks are instead simply unstructured 
\emph{strings}. These changes are somewhat counter-intuitive but they are in fact 
mandatory for reasonable space, because sharing rests on pointers, which break the reasonability requirement, see \citep{LICS2022} for extensive discussions.
\end{enumerate}

Accattoli, Dal Lago, and Vanoni also show results about two \emph{time} cost 
models for the \SpKAM: if time is defined as the number of $\beta$ steps (which is 
the natural, high-level cost model), then the machine is unreasonable (more 
precisely, the overhead of the machine is not polynomial), while if time is 
considered as the time taken by the \SpKAM implementation, then it turns out to be 
a reasonable cost model. Therefore, the \SpKAM is reasonable for both time and 
space, even if for such a simultaneous reasonability one has to pick an unusual 
and low-level notion of time.

\paragraph*{Multi types and reasonable space.} In this paper we develop a de Carvalho correspondence for the \SpKAM from which we extract the \emph{space} cost of machine runs. It is the first such correspondence measuring a reasonable notion of space for the $\l$-calculus. And as it is typical of intersection and multi types, the type system characterizes---and thus provides the \emph{exact} space consumption---for \emph{all} terminating $\l$-terms (with respect to (weak) \emph{call by name} evaluation).

De Carvalho's seminal work is about the KAM, and relies on the fact that one can see the KAM transition rules as deductive rules of the multi-type system. Because the \SpKAM is an optimized machine, with modified transitions, we do modify the type system in order to obtain the same correspondence (Section~\ref{sec:type-system}). The key points of the type system are the following ones:
\begin{itemize} 
\item \emph{Unusual weakening}: the handling of garbage collection induces an unusual approach to \emph{weakening}. The reason is that the \SpKAM and the type system implement \emph{weak} evaluation, which does not inspect abstraction bodies, but the definition of garbage is \emph{strong}, as it depends also on occurrences inside abstractions. 

\item \emph{Indices}: multi types associated to arguments have an index, which is simply a natural number. These indices reflect the size of the \emph{closure} that shall be built for that argument by the \SpKAM. They seem arbitrary at first, but if a closed term is typable with a ground type then the indices are uniquely determined, and they capture \emph{exactly} the size of closures. Judgments carry a \emph{weight}, which is another natural number, this time corresponding to the maximum space used by the \SpKAM in the run represented by the derivation ending with that judgment.

\item \emph{Soundness proof}: soundness of the type system---that is, \emph{$\tm$ typable with weight $w$ implies that the \SpKAM run on $\tm$ ends and has space cost $w$} (Corollary~\ref{cor:soundness})---follows a slightly unusual pattern.  We merge together the statements of subject reduction and soundness, which are usually disentangled in the literature. This merging helps proving the space bound.

\item \emph{Abstraction of the space cost}: the type system actually counts 
\emph{the maximum number of closures} used by the \SpKAM. Roughly, the 
\emph{actual space} is obtained by multiplying such a number by the logarithm of 
the size of the initial term, which corresponds to the size of the \emph{code 
pointer} in every closure. More precisely, the simulation of Turing machines 
within a linear space overhead in \citep{LICS2022} rests on some technicalities for which 
not all code pointers have the same size. At first, we abstract away from those 
technicalities, and simply count the number of closures. Then, by fine tuning the 
type system, at the cost of losing clarity, we are able to recover all the 
information needed in the proof of space reasonability of the \SpKAM (Section~\ref{sec:split}).
\end{itemize}

A further contribution of the paper is that, after the study of space, we slightly modify the weights on judgments as to extract the exact \emph{time} cost of runs, instead of the space cost (Section~\ref{sec:time}). Therefore, we are able to measure both reasonable time and reasonable space using the same underlying type system decorated with different weights.

\paragraph*{Abstracting the \SpKAM} For proving that the two type systems capture the time and the space of the \SpKAM, we extend the typing rules to the states and data structures of the \SpKAM. The exact bounds for a term $\tm$, however, can be read out \emph{directly} from the typing of $\tm$, with no need to type states of the \SpKAM. Thus, the extension of the type system to machine states is only a technical tool for the proof of the correspondence. This fact shows that the type system allows us to manipulate the time and space complexities of a term abstracting away from the machine on which these measures are defined. This is the main contribution of the paper (Section~\ref{sec:correctness}).

\paragraph*{Journal vs. conference version.} This is the extended version of the paper appeared with the same title at ICFP 2022~\citep{ICFP2022}. Beyond providing full proofs (some of them located in Appendices~\ref{app:correctness} and \ref{app:time}), 
this extended journal version features three new sections. 
Just after the preliminaries (Section~\ref{sec:pre}) and an introduction to the \SpKAM (Section~\ref{sec:spkam}), Section~\ref{sec:intuitions} provides intuitions behind the design of the 
type system for measuring space. Section~\ref{sec:semantics} is devoted to 
denotational models: we show that the type system we have introduced does 
\emph{not} allow one to define a model, in contrast to what usually happens 
with intersection/multi types. This happens exactly because of our unusual 
use of weakening, which breaks the stability of typing by 
$\beta$-conversion. Finally, Section~\ref{sec:split} is devoted to 
fine-tuning the type system to account for the measure of different kinds of 
code pointers (with different sizes), as it is needed in the proof of space 
reasonability of~\citep{LICS2022}.

% !TeX spellcheck = en_US
% !TEX root = main.tex
%%%%%%%%%%%%%%%%%%%%%%%%%%%%%%%%%%%%%%%%%%%%%%%%%%%%%%%%%%%%%%%%%%%%%%%%%%%%%%%%
\section{Preliminaries}\label{sec:pre}
Here we gently introduce the technical tools that are going to be used and refined along the paper, namely the $\l$-calculus, Krivine's abstract machine, multi types, and the relational model.

\paragraph*{Reasonable cost models.} Before starting with the technical discussion, we recall Slot and van Emde Boas' invariance thesis \citep{DBLP:journals/iandc/SlotB88} (also called strong, extended, efficient, modern, or
complexity-theoretic Church(-Turing) thesis in the literature), that states: 
\emph{reasonable notions of computation\footnote{Slot and van Emde Boas speak of 
\emph{machines} rather than \emph{notions of computation} because they are mainly 
discussing Turing and random access machines. Since the $\l$-calculus is not 
formulated as a machine, we prefer to speak of \emph{notion of computation} (and avoid \emph{computational model} for the collision with \emph{cost model}).} 
simulate each other with polynomially bounded overhead in time and a linear 
overhead in space}. The rationale behind this thesis is that, under these 
constraints, complexity classes such as \textsf{LOGSPACE}, $\P$, and $\PSPACE$ 
become robust, \ie notion-independent. We invite the reader to consult 
\citep{DBLP:journals/entcs/Accattoli18,LICS2022} for more details.

\subsection{The Closed Call-by-Name $\l$-calculus}
Let $\mathcal{V}$ be a countable set 
of variables. 
Terms of the \emph{$\lambda$-calculus} $\Lambda$ are defined as follows:
\[
\begin{array}{rrcl}
\textsc{$\l$-terms}\quad\quad & \tm,\tmtwo,\tmthree & \grameq & 
x\in\mathcal{V}\midd \lambda x.\tm\midd 
\tm\tmtwo.
\end{array}
\]
\emph{Free} and \emph{bound variables} are defined as 
usual: $\la\var\tm$ binds $\var$ in $\tm$.
Terms are considered modulo $\alpha$-equivalence, and 
$\tm\isub\var\tmtwo$ denotes capture-avoiding (meta-level) substitution of 
all the free occurrences of $\var$ for $\tmtwo$ in $\tm$.
The operational semantics of the $\l$-calculus is given by just one rewriting 
rule, called the $\beta$-rule:
\[
(\la\var\tm)\tmtwo\mapsto_\beta\tm\isub\var\tmtwo
\]
Terms like $(\la\var\tm)\tmtwo$, called reducible expressions, or \emph{redexes}, can 
occur as sub-terms inside a bigger term. Restricting the 
$\beta$-rule to only some of the redexes, \ie defining a \emph{reduction 
	strategy}, gives rise to the different 
$\l$-calculi. In this paper we deal mainly with a very simple fragment, 
which we call \ccallbn (shortened to \ccbn). It is dubbed \emph{closed}, because we consider 
only 
closed terms, \ie terms which do not have any free variable, and 
\emph{call-by-name}, because 
arguments are substituted inside function bodies unevaluated. Moreover, \ccbn 
is a \emph{weak} strategy, sometimes called  \emph{weak head reduction}, because it 
does not 
evaluate under abstractions\footnote{Which is common practice in functional 
	programming languages.} (which are its all and only normal forms), and 
because it evaluates only the head redex. We can define \ccbn reduction via the following rule that can only be applied at top level:
\[
(\la\var \tm) \tmtwo \tmthree_1 \ldots 
\tmthree_h \ \ \towh \ \ \tm \isub\var \tmtwo 
\tmthree_1 \ldots \tmthree_h \ \ \ \\ \  h\geq 0.
\]
For instance, we have $(\la\var\la\vartwo\var\var) \tm \towh \la\vartwo\tm\tm$ and $(\la\var\la\vartwo\var\var) \tm \tmtwo \towh (\la\vartwo\tm\tm) \tmtwo$ but $\tmthree ((\la\var\la\vartwo\var\var) \tm \tmtwo) \not \towh \tmthree ((\la\vartwo\tm\tm) \tmtwo)$ and $\la\varthree((\la\var\la\vartwo\var\var) \tm \tmtwo) \not \towh \la\varthree((\la\vartwo\tm\tm) \tmtwo)$.

It is easily seen that weak head reduction is deterministic.

\begin{example}
	In the remainder of this paper, we are going to present different evaluation mechanisms for the $\l$-calculus. For each of them, we are going to develop a complete example. Each of these examples is based on the execution of the $\l$-term $(\la\var(\la\vartwo(\la\varthree\var)(\var\vartwo))\var)\Id$ where $\Id\defeq\la a a$ is the identity combinator. Here we show its \ccbn evaluation, that takes 3 $\beta$-steps.
	\[
	(\la\var(\la\vartwo(\la\varthree\var)(\var\vartwo))\var)\Id \towh (\la\vartwo(\la\varthree\Id)(\Id\vartwo))\Id \towh (\la\varthree\Id)(\Id\Id) \towh \Id
	\]
\end{example}

%%%
% !TeX spellcheck = en_US
% !TEX root = main.tex
%%%%%%%%%%%%%%
\begin{figure}
\begin{center}
	\fbox{$\begin{array}{c}
	\begin{array}{c@{\hspace{1.1cm}}c@{\hspace{1.1cm}}c@{\hspace{1.1cm}}c}
	\textsc{Closures} &
	\textsc{Environments} &
	\textsc{Stacks}&
	\textsc{States}\\
	\clos  \grameq  (\tm,\lenv) &  \lenv  \grameq  \stempty \midd \esub\var 
	\clos\cdot \lenv &
	\stack  
	\grameq  \stempty \midd \clos\cdot \stack&
	\state\grameq(\tm,\lenv,\stack)\\
	\end{array}
	\\[18pt]
	\hhline{=}\\
	\begin{array}{
			l@{\hspace{.4cm}}l@{\hspace{.4cm}}l | l |
			l@{\hspace{.4cm}}l@{\hspace{.4cm}}lll}
		\mathsf{Term} &  \mathsf{Env} & 
		\mathsf{Stack}
		&&	\mathsf{Term}  & \mathsf{Env} 
		& 
		\mathsf{Stack}
		\\
		\cline{1-8}
		&&&&\\[-8pt]
		\kstatetab{ \tm\tmtwo }{ \lenv }{ \stack } &
		\tokamsea &%\\[-2pt]
		\kstatetab{ \tm }{ \lenv }{ (\tmtwo,\lenv)\cdot \stack }
		\\[3pt]				
		\kstatetab{ \la\var\tm }{ \lenv }{ \clos\cdot \stack } &
		\tokamb &%\\[-2pt]
		\kstatetab{ \tm }{ \esub\var \clos\cdot \lenv }{ \stack }
		\\[3pt]  
		\kstatetab{ \var }{ \lenv }{ \stack } &
		\tokamsub &%\\[-2pt]
		\kstatetab{ \tmtwo}{ \lenvtwo }{\stack } & \mbox{if }\lenv(\var) = (\tmtwo,\lenvtwo)
		\\[3pt]
		%\cline{1-7}
		\end{array}
	\end{array}$}
\end{center}
	\caption{\KAM data structures and transitions.}
	\label{fig:KAM}
\end{figure}
%%%

\subsection{The Krivine Abstract Machine}
The KAM is a 
standard environment-based machine for \ccbn, often defined as in 
Figure~\ref{fig:KAM}. 
The machine evaluates closed $\l$-terms to weak head normal form via three 
transitions, the union of which is noted $\tokam$: 
\begin{itemize}
	\item $\tokamsea$ looks for redexes descending on the left of topmost 
	applications 
	of the active term, accumulating arguments on the stack; 
	\item $\tokamb$ fires a $\beta$ redex (given by an abstraction as active 
	term 
	having as argument the first entry of the stack) but delays the associated 
	meta-level substitutions, adding a corresponding explicit substitution to 
	the 
	environment;
	\item $\tokamsub$ is a form of micro-step substitution: when the active 
	term is $\var$, the machine looks up the environment $\lenv$ and retrieves the delayed replacement for $\var$, \ie $\lenv(\var)$.
\end{itemize}
The data structures used by the \KAM are \emph{local environments}, 
\emph{closures}, and a \emph{stack}. \emph{Local environments}, that we shall 
simply refer to as \emph{environments}, are defined by mutual induction with 
\emph{closures}.
The idea is that every (potentially open) term $\tm$ occurring in a state comes 
with an 
environment $\lenv$ that \emph{closes} it, thus forming a 
closure $\clos = (\tm,\lenv)$, and, in turn, environments are lists of entries 
$\esub\var\clos$ associating to each open variable $\var$ of $\tm$, a closure 
$\clos$ \ie, intuitively, a closed term.
The \emph{stack} simply collects the closures associated to the arguments met 
during the search for $\beta$-redexes.

A state $\state$ of the \KAM is the pair $(\clos,\stack)$ of a 
closure $\clos$ and a stack $\stack$, but we rather see it as a triple 
$(\tm,\lenv,\stack)$ by spelling out the two components of the closure $\clos = 
(\tm,\lenv)$.
Initial states of the \KAM
are defined as $\compil{\tm_{0}}\defeq\kstate{\tm_{0}}{\stempty}{\stempty}$ 
(where $\tm_{0}$ is a closed $\l$-term, and also the code). A state is \emph{final} if no transitions apply. A \emph{run} $\run: \state \tomach^*\statetwo$ is a possibly empty sequence of transitions, the length of which is noted 
$\size\run$. If $a$ is a transition label, $\sizeparam\run a$ 
is the number of $a$ transitions in $\run$. An \emph{initial run} is a run from an initial state $\compil\tm$, and it is also called \emph{a run from $\tm$}. A state $\state$ is \emph{reachable} if it 
is the target state of an initial run. A \emph{complete run} is an initial run ending on a final state.
In a state $\spkamstate\tm\lenv\stack$ that is reachable by an initial state $\compil{\tm_{0}}$, $\tm$ is always a sub-term of $\tm_0$. For this reason $\tm_{0}$ has a special role during the execution. Indeed, any sub-term $\tm$ of $\tm_{0}$ can be seen as a pointer to $\tm_{0}$ which we call the \emph{initial immutable code}. 

The key point in the proof of the correctness of the \KAM is that there is a bijection between $\towh$ steps and $\tokamb$ transitions, so that we can identify the two. In particular, \KAM states can be \emph{decoded} to $\lambda$-terms via the following decoding function $\unf\cdot$ for closures and states: 
\[\begin{array}{rrcl@{\hspace{2cm}}rcl}
\textsc{Closures}
&
\unf{(\tm, \stempty)} & \defeq & \tm
&
\unf{(\tm, \esub\var\clos\cons\lenv)} & \defeq & 
\unf{(\tm\isub\var{\unf\clos},\lenv)}

\\[3pt]
\textsc{States}
&
\unf{(\tm, \lenv, \stempty)} & \defeq & \unf{(\tm, \lenv)}
&
\unf{(\tm, \lenv, \clos\cons\stack)} & \defeq & \unf{(\tm, \lenv, 
	\stack)}\, \unf\clos
\end{array}\]
\begin{theorem}[KAM correctness~\citep{DBLP:conf/icfp/AccattoliBM14}]
	The \KAM implements \ccbn, that is, there is a complete $\towh$-sequence 
	$\tm \towh^{n} \tmtwo$ if and only if there is a complete run $\run: 
	\compil\tm \rightarrow_{\mathrm{\KAM}}^* \state$ such that $\unf\state = \tmtwo$ and 
	$\sizeb\run = n$.
\end{theorem}
\begin{example}
\label{ex:kam-run-bij}
	We provide the \KAM execution of our example term $(\la\var(\la\vartwo(\la\varthree\var)(\var\vartwo))\var)\Id$.
	\[
	\begin{array}{l|l|ll}
	\mathsf{Term}   & 
	\mathsf{Environment} & \mathsf{Stack}\\
	\cline{1-3}
	(\la\var(\la\vartwo(\la\varthree\var)(\var\vartwo))\var)\Id & \stempty & \stempty & \tokamsea\\
	\la\var(\la\vartwo(\la\varthree\var)(\var\vartwo))\var & \stempty & (\Id,\stempty) & \tokamb\\
	(\la\vartwo(\la\varthree\var)(\var\vartwo))\var & \esub\var{(\Id,\stempty)}=:\lenvtwo & \stempty & \tokamsea\\
	\la\vartwo(\la\varthree\var)(\var\vartwo) & \lenvtwo & (\var,\lenvtwo) & \tokamb \\
	(\la\varthree\var)(\var\vartwo) & \esub\vartwo{(\var,\lenvtwo)} \cdot \lenvtwo=:\lenv & \stempty & \tokamsea\\
	\la\varthree\var & \lenv & (\var\vartwo,\lenv) & \tokamb\\
	\var & \esub\varthree{(\var\vartwo,\lenv)}\cdot \esub\vartwo{(\var,\lenvtwo)} \cdot \esub\var{(\Id,\stempty)} & \stempty & \tokamsub\\
	\Id & \stempty & \stempty
	\end{array}
	\]
	We can observe that the evaluation takes 7 steps, of which 3 are $\beta$-steps, 3 are $\tokamsea$-steps, and 1 is a $\tokamsub$-step.
\end{example}

\subsection{\ccbn multi types}
In this section we give  an introduction to multi types, also known as \emph{non-idempotent 
intersection types}. First, we consider them from the 
\emph{qualitative} point 
of view, and then we refine our analysis \emph{quantitatively}. We will review 
them quite quickly, inviting the reader to consult \citep{BKV17} about the 
qualitative part, and \citep{DBLP:journals/jfp/AccattoliGK20} about the 
quantitative one. Multi types are 
deeply related to linear logic in how they keep track of the use of resources, and in 
particular their grammar is reminiscent of 
the (call-by-name, in our case) translation $(\cdot)^\dagger$ of intuitionistic 
logic into linear 
logic $(A\to B)^\dagger=\,!A^\dagger\lolli B^\dagger$. Semantically, they can 
be seen as a syntactical presentation of the relational 
model~\citep{DBLP:journals/apal/BucciarelliE01} of the $\l$-calculus, induced by the relational semantics of linear logic via the \cbn translation $(\cdot)^\dagger$. 

The grammar for types is based on \emph{two} layers of types, defined in a
mutually recursive way, linear types $\linty$ and finite multi-sets $\mty$ of 
linear types, which can also be considered a non-idempotent variant of 
\citet{DBLP:journals/tcs/Bakel92} strict types.
\[
\begin{array}{rrcll}
\textsc{Linear 
	types}&\linty,\lintytwo&\grameq&\initty\midd\arr{\mty}{\linty}

\\[3pt]
\textsc{Multi 
	types}&\mty,\mtytwo&\grameq&\mset{\linty_1,\ldots,\linty_n} \qquad 
n\geq 0
\end{array}
\]

Note that there is a ground 
type $\initty$, which can be thought as 
the type of normal forms, which in \ccbn are precisely abstractions. Note also 
that arrow (linear) types $\arr{\mty}{\linty}$ 
can have a multiset only on the left. The empty multiset is noted $\emmset$, 
and the union of two multisets $\mty$ and $\mtytwo$ is noted $\mty\uplus\mtytwo$. We write $\gty$ for either a linear type or a multi type.

Type environments, ranged over by $\tye,\tyetwo$ are partial finite maps
from variables to non-empty multi types, and we write $\tye = 
\var_1:\mty_1,\ldots,\var_n:\mty_n$ if 
$\dom\tye = \set{\var_1,\ldots,\var_n}$.
Given two type environments $\tye,\tyetwo$, the expression
$\tye\uplus\tyetwo$ stands for the type environment
assigning to every variable $\var$ the multiset
$\tye(\var)\uplus\tyetwo(\var)$.

Type 
judgments have the form $\tjudg{\tye}{\tm}{\linty}$ where $\tye$ is a type 
environment. The 
typing rules are in \reffig{mtypesystem}; type derivations are noted $\tyd$ 
and we write 
$\tyd\pof\tjudg{\tye}{\tm}{\linty}$ for a type derivation $\tyd$ with the 
judgment $\tjudg{\tye}{\tm}{\linty}$ as its conclusion.

% !TeX spellcheck = en_US
% !TEX root = main.tex
%%%%%%%%%%%%%%
\begin{figure}[t]
	\[
	\begin{array}{c@{\hspace{1cm}}c@{\hspace{1cm}}c}
	\infer[\tyvar]{\tjudg{\var:\mset{\linty}}{\var}{\linty}}{}
	&
	\infer[\tylam]{\tjudg{\tye}{\lambda\var.\tm}{\arr{\mty}{\linty}}} 
	{\tjudg{\tye,\var:\mty}{\tm}{\linty}}
	&
	\infer[\tylamstar]{\tjudg{}{\lambda\var.\tm}{\initty}}{}
	\\[8pt]
	\multicolumn{3}{c}{\infer[\tyapp]{\tjudg{\tye\uplus 
				\biguplus_{i\in\mset{1,\ldots,n}} \tyetwo_i  
			}{\tm\tmtwo}{\linty 
		}}{\tjudg{\tye}{\tm}{\arr{\mset{\lintytwo_1,\ldots,\lintytwo_n}}{\linty}}
			& 
			\mset{\tjudg{\tyetwo_i}{\tmtwo}{\lintytwo_i}}_{i\in\mset{1,\ldots,n}}}}
	\end{array}
	\]
	\caption{The multi-type system for \ccbn.}
	\label{fig:mtypesystem}
\end{figure}

Intuitively, a linear type $\linty$ corresponds to a single use of a term, while the cardinality of a multi type assigned to an argument $\tm$ stands for the number of times this term $\tm$ will be \emph{copied} during the reduction. More precisely, the cardinality corresponds to the number of times that $\tm$ is \emph{used} by the KAM, that is, the number of times that $\tm$ is substituted on the head variable during the KAM run.
Different uses of $\tm$ during a run are allowed to have different types: this is precisely the reason why we need the multiset operator. As an example, we show the type derivation for the term $(\la\var\var\var)(\la\vartwo\vartwo)$.

{\small \[
\infer[\tyapp]{\tjudg{}{(\la\var\var\var)(\la\vartwo\vartwo)}{\initty}}{
	\infer[\tylam]{\tjudg{}{\la\var{\var\var}}{\arr{\mset{\arr{\mset{\initty}}{\initty},\initty}}{\initty}}}{
		\infer[\tyapp]{\tjudg{\var:\mset{\arr{\mset{\initty}}{\initty},\initty}}{\var\var}{\initty}}{
			\infer[\tyvar]{\tjudg{\var:\mset{\arr{\mset{\initty}}{\initty}}}{\var}{\arr{\mset{\initty}}{\initty}}}{}
			& \infer[\tyvar]{\tjudg{\var:\mset\initty}{\var}{\initty}}{}}} &
	\hspace{-12pt}\infer[\tylam]{\tjudg{}{\la\vartwo\vartwo}{\arr{\mset{\initty}}{\initty}}}{
		\infer[\tyvar]{\tjudg{\vartwo:\initty}{\vartwo}{\initty}}{}} &
	\infer[\tylamstar]{\tjudg{}{\la\vartwo\vartwo}{\initty}}{}}
\]}
The reader can notice that the argument $\la\vartwo\vartwo$ is typed twice, and with different types. It is typed once with $\arr{\mset{\initty}}{\initty}$, when it is used as a function, and once with $\initty$, when it is used as a value. Moreover, the finite polymorphism of multi types allows us to type the term $\la\var\var\var$, which is not typable with simple types.

\paragraph*{Characterization of termination.} Multi types characterize Closed 
CbN termination, that is, they type all and only those $\l$-terms that terminate
with respect to Closed CbN. Differently from what happens with  idempotent intersection types, such a result has an easy proof with multi types. Soundness, that is, the fact that typable terms terminate, is usually the hard part, as it requires to exhibit a termination argument. For idempotent intersection types one needs to use the reducibility method, which is quite heavy machinery. For multi types, there is a strikingly simpler argument: a quantitative form of subject reduction holds, stating that the size of the type derivation $\size{\cdot}$, here defined as the number of $\tylam$ rules, decreases with each weak head $\beta$-step, which then provides the termination measure. A similar argument shall be used in Section \ref{sec:correctness}. As usual, we first need a quantitative substitution lemma, here restricted to close arguments, proved by induction on the structure of the type derivation, using an omitted quantitative splitting lemma.

\begin{lemma}[Quantitative substitution] Let $\tyd\pof\tjudg{\tye,\var:\mty}{\tm}{\gty}$ and $\tydtwo\pof\tjudg{}{\tmtwo}{\mty}$. Then, there exists $\tydthree\pof\tjudg{\tye}{\tm\isub\var\tmtwo}{\gty}$, where $\size\tydthree=\size\tyd+\size\tydtwo$.
\end{lemma}
Then, by induction on evaluation contexts, one can prove subject reduction.

\begin{proposition}[Quantitative subject reduction]\label{prop:sub-red} Let $\tm\towh \tmtwo$ and $\tyd\pof\tjudg{}{\tm}{\linty}$. Then there exist $\tydtwo\pof\tjudg{}{\tmtwo}{\linty}$ such that $\size\tydtwo =\size\tyd-1$.
\end{proposition}
Completeness of the type system, that is, the property that every terminating term is typable, is proved as for idempotent intersection types, by showing that all weak head normal forms are typable (immediate, via the rule $\tylamstar$) and via a subject \emph{expansion} property, again proved using a auxiliary anti-substitution and merging lemmas. In this case no quantitative reasoning is necessary.

\begin{proposition}[Subject expansion]\label{prop:sub-exp-calc}Let $\tm\towh \tmtwo$ and $\tyd\pof\tjudg{}{\tmtwo}{\linty}$. Then there exist $\tydtwo\pof\tjudg{}{\tm}{\linty}$.
\end{proposition}

The final quantitative characterization is then proved by induction on the size of the type derivation (soundness, \emph{if} direction) and on the length of the reduction (completeness, \emph{only if} direction).

\begin{theorem}[Characterization of weak head normalization]\label{thm:charact-wh}
	Let $\tm$ be a closed term. $\tm \towh^n\la\var\tmtwo$ if and only if there exists a multi-type  derivation
	$\tyd\pof\tjudg{}{\tm}{\initty}$ such that $\size{\tyd}=n$.
\end{theorem}

\paragraph*{Multi types and the \KAM time consumption.} Multi types have been 
successfully 
applied in quantitative 
analyses of normalization, starting with \citet{deCarvalho07,deCarvalho18}, who 
used them to give a bound to the length of \KAM runs. 
De Carvalho's technique can be re-phrased and distilled as a decoration of  
type derivations with \emph{weights}, that is, cost annotations, following the 
scheme of \reffig{weightskam}. This weight system just measures the size of the type derivation, since for each rule R, it adds 1 to the weight of the tree rooted in R. Please 
note that the 
weight assignment is blind to types (which is why terms and types are removed from \reffig{weightskam}), and thus relies only on the structure, an in particular the size, of the type derivation.  De Carvalho's result can be formulated as follows.
\begin{theorem}[\cite{deCarvalho07,deCarvalho18}]\label{thm:decarvalho}
	Let $\tm$ be a closed $\lambda$-term. Then there exists a complete \KAM run $\run$ from $\tm$ of length $w$ if and only if there exists a multi-type derivation $\tyd\pof\wtjudg{}{\weight}{\tm}{\initty}$.
\end{theorem}
The \KAM being deterministic, one has that all derivations 
$\tjudg{}{\tm}{\initty}$ induce the same weights. Moreover, there is a stronger 
correspondence between the rules of the type system, and the transitions of the 
\KAM, in the special case of closed terms typed with $\initty$. Every $\tyapp$ rule corresponds to a $\tokamsea$ transition, every 
$\tylam$ rule corresponds to a $\tokamb$ transition\footnote{This way, multi 
	type derivations can be used to count also the number of $\towh$-steps a 
	$\l$-term needs to normalize, which is the natural time cost model for \ccbn.}, and every $\tyvar$ rule 
corresponds to a $\tokamsub$ transition. The only $\tylamstar$ rule in a type 
derivation for $\tm$ corresponds 
to the final state of the \KAM run on $\tm$. The correspondence is deep, each 
state of the 
\KAM run for a $\l$-term $\tm$ corresponds to a type judgment occurrence in the type derivation of $\tm$, in such a way that the sub-term of the state is exactly the same sub-term typed by the judgment. In other words, the \KAM and multi types \emph{compute} in the \emph{same} way (in the special case of closed terms typed with $\initty$), they only differ in how such a computation is presented.

\begin{example}
\label{ex:tyd-bij}
	We provide the weighted multi-type derivation of our running example term $(\la\var(\la\vartwo(\la\varthree\var)(\var\vartwo))\var)\Id$.
	\[
	\infer[\tyapp]{\tjudgw{}{7} {(\la\var(\la\vartwo(\la\varthree\var)(\var\vartwo))\var)\Id}{\initty}}{
	\infer[\tylam]{\tjudgw{}{6}{\la\var(\la\vartwo(\la\varthree\var)(\var\vartwo))\var} {\arr{\mset{\initty}}\initty}}{
	\infer[\tyapp]{\tjudgw{\var:\mset\initty}{5}{(\la\vartwo(\la\varthree\var)(\var\vartwo))\var}{\initty}}{\infer[\tylam]{\tjudgw{\var:\mset\initty}{4}{\la\vartwo(\la\varthree\var)(\var\vartwo)} {\arr\emmset\initty}}{
	\infer[\tyapp]{\tjudgw{\var:\mset\initty}{3}{(\la\varthree\var)(\var\vartwo)}{\initty}}{
	\infer[\tylam]{\tjudgw{\var:\mset\initty}{2}{\la\varthree\var}{\arr\emmset\initty}}{
	\infer[\tyvar]{\tjudgw{\var:\mset\initty}{1}{\var}{\initty}}{}}}}}} & 
	\infer[\tylamstar]{\tjudgw{}{0}{\Id}{\initty}}{}}
	\]
	We note that there is a precise correspondence between this type derivation and the \KAM execution of the same term. Indeed, the final weight is 7, as the number of \KAM transitions. Moreover, each occurrence of rule $\tylam$ corresponds to a transition $\tokamb$, each occurrence of rule $\tyapp$ corresponds to a transition $\tokamsea$, and each occurrence of rule $\tyvar$ corresponds to a transition $\tokamsub$.
\end{example}

% !TeX spellcheck = en_US
% !TEX root = main.tex
%%%%%%%%%%%%%%
\begin{figure}
	\begin{center}
		\begin{tabular}{c}
			$\begin{array}{c@{\hspace{1.3cm}}c@{\hspace{1.3cm}}c@{\hspace{1.3cm}}c}
			\infer[\tyvar]{\wtjudgone{1}}{}
			&
			\infer[\tylam]{\wtjudgone{w+1}}{\wtjudgone{w}}
			&
			\infer[\tylamstar]{\wtjudgone{0}}{}
			&
			\infer[\tyapp]{\wtjudgone{w+\sum 
					v_i+1}}{\wtjudgone{w} & 
				\mset{\wtjudgone{v_i}}_{i\in\mset{1,\ldots,n}}}
			\end{array}$
		\end{tabular}
	\end{center}
	\caption{The weight assignment measuring the size of multi-type derivations/KAM execution time.}\label{fig:aaweightKAMtime}
	\label{fig:weightskam}  
\end{figure}

\renewcommand{\sem}[1]{\llbracket#1\rrbracket}

\paragraph*{The Relational model.} As we have already pointed out in the introduction, multi types allow for a syntactic presentation of the relational model of the untyped call by-name $\l$-calculus. In these preliminaries we avoid all details about denotational models (we shall say more when discussing whether our new type system induces a model, in Section~\ref{sec:semantics}) and only sketch the interpretation of a $\l$-term in the relational model, that can simply be given as follows:
\[
\sem{\tm} \defeq \{(\tye,\linty) \mid \exists\, \tyd \pof \tjudg{\tye}{\tm}{\linty}\}
\]
The just defined-relational interpretation is \emph{invariant} under $\beta$-reduction (a direct consequence of subject reduction (Proposition~\ref{prop:sub-red}) and expansion (Proposition~\ref{prop:sub-exp-calc}))\footnote{To be completely precise, one should prove subject reduction and expansion for all $\beta$-steps, not just weak head ones.} and \emph{adequate} for weak head reduction, in that $\sem{\tm}$ is empty if and only if $\tm$ diverges (Theorem~\ref{thm:charact-wh}).

\begin{example}
	We compute the interpretation of our running example term \\$(\la\var(\la\vartwo(\la\varthree\var)(\var\vartwo))\var)\Id$. Exploiting the invariance under $\beta$-reduction, we have:
	\[
	\sem{(\la\var(\la\vartwo(\la\varthree\var)(\var\vartwo))\var)\Id} = \sem{\Id}\defeq \{(\tye,\linty) \mid \exists\, \tyd \pof \tjudg{\tye}{\Id}{\linty}\}.
	\]
	We notice that the type derivation $\tyd \pof \tjudg{\tye}{\Id}{\linty}$ can only have the following shapes:
	\[
	\infer[\tylamstar]{\tjudg{}{\Id}{\initty}}{} \qquad \text{and} \qquad \infer[\tylam]{\tjudg{}{\Id\defeq \la s s}{\arr{\mset{\linty}}\linty}}{
		\infer[\tyvar]{\tjudg{s:\mset\linty}{s}{\linty}}{}} \qquad \text{for any }\linty.
	\]
	Then, $\sem{\Id}\defeq \{(\tye,\linty) \mid \exists\, \tyd \pof \tjudg{\tye}{\Id}{\linty}\} = \{(\emptyset,\initty)\} \cup \{(\emptyset,\arr{\mset{\linty}}\linty) \mid \linty \text{ is a linear type}\}$.
\end{example}

% !TeX spellcheck = en_US
% !TEX root = main.tex
%%%%%%%%%%%%%%%%%%%%%%%%%%%%%%%%%%%%%%%%%%%%%%%%%%%%%%%%%%%%%%%%%%%%%%%%%%%%%%%%
\section{The \SpKAM}\label{sec:spkam} Here, we present an optimization of the KAM, that we call \SpKAM, aiming at space \emph{reasonability}. More specifically, two modifications are implemented in order to achieve reasonability, namely \emph{unchaining} and \emph{eager garbage collection}.

\paragraph*{Unchaining.} It is a 
folklore 
optimization for abstract 
machines  bringing speed-ups with respect to both time and space, used \eg 
by \citet{DBLP:conf/birthday/SandsGM02}, 
\citet{DBLP:journals/lisp/Wand07}, 
\citet{DBLP:journals/lisp/FriedmanGSW07}, and 
\citet{DBLP:journals/jfp/Sestoft97}. Its first 
systematic study is by 
\citet{DBLP:journals/iandc/AccattoliC17}, with respect to time. This optimization prevents the creation of chains of \emph{renamings} in environments, that is, of delayed substitutions of variables for variables, of which the simplest shape in the \KAM is:
\begin{center}$\esub{\var_0}{(\var_1, \esub{\var_1}{(\var_2, 
			\esub{\var_2}{\ldots})})}$\end{center}
where the links of the chain are generated by $\beta$-redexes having a variable 
as argument. On some families of terms, these chains keep growing and growing, 
leading to the quadratic dependency of the number of transitions from $\sizeb\run$, and to the continuous increase of space consumption. An example of continuously growing chain of renamings may be obtained by computing finite prefixes of the KAM execution of the standard looping term $\Omega$.

\paragraph*{Eager garbage collection.} Beside the malicious chains connected to 
unchaining,  the \NaKAM is not parsimonious with space also because there is no 
garbage collection (shortened to ``GC''). In transition $\tokamsub$, the current 
environment is discarded, so something is collected, but this is not enough. 
It is thus natural to modify the machine as to maximize GC and space re-usage, that is, as to perform it \emph{eagerly}. 

% !TeX spellcheck = en_US
% !TEX root = main.tex
%%%%%%%%%%%%%%
\begin{figure}[t]
\fbox{$\begin{array}{c}
	\begin{array}{c@{\hspace{1cm}}c@{\hspace{1cm}}c@{\hspace{1cm}}c}
	\textsc{Closures} &
	\textsc{Environments} &
	\textsc{Stacks}&
	\textsc{States}\\
	\clos  \grameq  (\tm,\lenv) &  \lenv  \grameq  \stempty \midd \esub\var 
	\clos\cdot \lenv &
	\stack  
	\grameq  \stempty \midd \clos\cdot \stack&
	\state\grameq(\tm,\lenv,\stack)\\
	\end{array}
	\\[18pt]
	\hhline{=}\\
	\begin{array}{l@{\hspace{.4cm}} 
				l@{\hspace{.4cm}}l|l|l@{\hspace{.4cm}} 
				l@{\hspace{.4cm}}l@{\hspace{.4cm}}l}
			\mathsf{Term}   & 
			\mathsf{Env} & \mathsf{Stack}
			&&	\mathsf{Term} 
			& 
			\mathsf{Env}
			& \mathsf{Stack} 
			\\
			\cline{1-8}
			&&&&\\[-8pt]
			\tm\var  & \lenv  & \stack & 
			\tokamseav
			& \tm  & 
			\lenv|_\tm & \lenv(\var)\cdot\stack
			\\[3pt]
			 \tm\tmtwo  & \lenv  & \stack &
			\tokamseanv &%\\[-2pt]
			 \tm & \lenv|_\tm & (\tmtwo,\lenv|_\tmtwo)\cdot \stack
			 & \mbox{if }\tmtwo\not\in\mathcal{V}
			\\[3pt]	
			\la\var\tm  & \lenv & \clos\cdot \stack &
			\tokambw &%\\[-2pt]
			\tm    & \lenv  &\stack\qquad &\text{if }\var\not\in\fv{\tm}
			\\[3pt]  			
			 \la\var\tm  & \lenv & \clos\cdot \stack &
			\tokambnw &%\\[-2pt]
			 \tm    & \esub\var \clos\cdot \lenv  &\stack
			 &\text{if }\var\in\fv{\tm}
			\\[3pt]  
			 \var    &
			\lenv
			& \stack
			 &
			\tokamsub &%\\[-2pt]
			 \tmtwo  & \lenvtwo & \stack & \mbox{if }\lenv(\var) =(\tmtwo,\lenvtwo)
			\\[3pt]
			\cline{1-8}	
	\end{array}\\[45pt]
	\text{where $\lenv|_{\tm}$ denotes the restriction of $\lenv$ to the free 
		variables of $\tm$.}\end{array}$ % end of the space kam fbox
	}
	\caption{\SpKAM data structures and transitions.}
	\label{fig:spkam}
\end{figure}

\paragraph*{\SpKAM transitions and data structures.}
The \KAM optimized with both eager GC and unchaining 
(both optimizations are mandatory for space reasonability)
is here called \SpKAM and it is defined in \reffig{spkam}. The data structures, 
namely closures and (local) environments, are defined as for the \KAM, the changes 
concern the machine transitions only. Unchaining is realized by transition 
$\tokamseav$, while
eager garbage collection is realized mainly by transition $\tokambw$, which 
discards the argument if the variable of the $\beta$ redex does not occur in the body of the abstraction. 
Transitions $\tokamseanv$ and $\tokamseav$ also contribute to implementing the GC, by restricting the environment to the occurring variables, when the environment is propagated to sub-terms. As a consequence, we obtain the following invariant.

\begin{lemma}[Environment domain invariant]
	\label{l:spkam-invariants} % \reflemmap{spkam-invariants}{env}
	Let $\state$ be a \SpKAM reachable state. Then $\dom\lenv = \fv\tm$ for every 
	closure $(\tm,\lenv)$ in $\state$.
\end{lemma}
Because of this invariant, which concerns also the closure given by the active term 
and the local environment of the state, the variable transition $\tokamsub$ 
simplifies as follows:
\[\begin{array}{l@{\hspace{.3cm}} 
	l@{\hspace{.3cm}}l|l|l@{\hspace{.3cm}} 
	l@{\hspace{.3cm}}l}
\mathsf{Term}   & 
\mathsf{Env} & \mathsf{Stack}
&&	\mathsf{Term} 
& 
\mathsf{Env}
& \mathsf{Stack} 
\\
\cline{1-7}
\var    &
\esub\var{(\tmtwo,\lenvtwo)}
& \stack
&
\tokamsub &%\\[-2pt]
\tmtwo  & \lenvtwo & \stack
\end{array}\]

\paragraph*{Implementation and space consumption.} We consider the \SpKAM implemented in a rather unusual way, in order to obtain a space-reasonable machine. Indeed, we avoid any form of sharing of data structures. Transitions $\tokamseanv$ and $\tokamseav$ actually copy the environment itself, and not just a pointer to it. Moreover, all the data structures are implemented as unstructured strings, \ie not as linked lists or trees (just think of implementing the \SpKAM directly on a multi tape Turing machine, rather than on a RAM). In this way, the machine does not rest on pointers for structuring the stack and the environment data structures, and this is crucial, as such pointers would add an unreasonable space overhead. Pointers do not disappear altogether: pointers to the initial immutable code are used to represent the left component of every closure (that is, the term $\tmtwo$ in $(\tmtwo,\lenv)$), which, by the key \emph{sub-term invariant} of the KAM, is always a sub-term of the initial code. The use of these pointers is another crucial ingredient for space reasonability, as it allows to avoid the copy of sub-terms which would make the code grow during execution (that must be avoided to account for logarithmic space).

The avoidance of data structure pointers is also the reason why garbage 
collection can  be implemented eagerly, as it is assumed by the \SpKAM. If 
one instead adopts a more standard form of \emph{shared} environments (which are 
based on using pointers) then garbage collection becomes trickier, because of 
\emph{aliasing}, that is, the fact that shared environments can be referenced more 
than once. Our approach, instead, does not suffer from this difficulty, because 
\emph{nothing} is shared. Such a choice, of course, degrades the time performance, 
but here we are mainly interested in space behavior. All these implementation 
choices, necessary to obtain the space reasonability of the Space KAM, are 
discussed at length in~\citet{LMCS2024}.

%Once one has made explicit all these details, it is easy to define the size of a \SpKAM state. Intuitively, the size of a state is given by the number of pointers to the initial immutable code $\tmtwo_0$ multiplied the size of such pointers, that is $\log\size{\tmtwo_0}$. Counting the number of pointers is the same as considering the number of closures inside a state.

\paragraph*{Space and number of closures of the \SpKAM} Intuitively, the space occupied by a state $\state$ of the \SpKAM during a run of initial immutable code $\tm_0$ is the number $n$ of closures in $\state$ multiplied by the logarithm of the size of the initial immutable code $\log\size{\tm_0}$, accounting for the code in every closure (that is, the left component), which is represented as a pointer to the initial code (because of the sub-term invariant of the KAM mentioned above).

For the space reasonability result in \citep{LICS2022}, however, a trickier notion 
of space is used, as the initial code $\tm_0\defeq\tmtwo\tmthree$\footnote{The initial code can be considered an application without loss of generality, as terms are closed and, having $\lambda$-abstractions would mean to have already a final state.} is divided into \emph{two} address spaces, 
so that not every pointer takes space $\log\size{\tm_0}$ (in particular pointers to $\tmtwo$ use less space, \ie $\log\size{\tmtwo}$). For now, we 
abstract away from such a technical aspect, because it is needed to study the 
representation of Turing machines, but on $\l$-terms that do not belong to the 
image of the encoding of Turing machines it is irrelevant. We will come to this later in Section~\ref{sec:split}. Here, we use an 
abstract notion of space, namely \emph{the number of closures in a state}, which is roughly the space of the \SpKAM up to a factor $\log\size{\tm_0}$.

\begin{definition}[Abstract space]
	The abstract space $\size\state$ taken by a \SpKAM state $\state$ is defined on the structure of \SpKAM states as follows.\footnote{In the clause for states, we could have written $\size{(\tm,\lenv,\stack)} \defeq  \size{\lenv}+\size{\stack}+1$, counting also the space for the closure $(\tm,\lenv)$. Since the difference is just a constant, we can remove that ``$+1$'', without loss of generality. The removal of the $+1$ shall slightly simplify the extraction of the space usage from multi-type derivations in Section~\ref{sec:correctness}.}
\[\begin{array}{ccccccc}
	\textsc{Environments}
	&
	\textsc{Stacks}\\
	\begin{array}{rcl}
	\size{\stempty} & \defeq &0
	\\
	\size{\esub\var\clos\cdot\lenv} & \defeq & \size{\clos}+\size{\lenv}
	\end{array}
	&
	\begin{array}{rcl}
	\size{\stempty} & \defeq &0
	\\
	\size{\clos\cdot\stack} & \defeq &	\size{\clos}+\size{\stack}
	\end{array}
	\\\\
	
	\textsc{Closures} 
	&		
	\textsc{States}\\
	\begin{array}{rcl}
	\size{(\tm,\lenv)} & \defeq & 1+\size\lenv 
	\end{array}
	& 
	\begin{array}{rcl}			
	\size{(\tm,\lenv,\stack)} &\defeq & \size{\lenv}+\size{\stack}
	\end{array}
	\end{array}\]
\end{definition}

%\begin{definition}[Closure Measure of a State]
%	The number of closure of a \SpKAM state $\state$ is defined on the structure of $\state$ as follows.
%	\[\begin{array}{ccccccc}
%	\textsc{Environments}
%	&
%	\textsc{Stacks}\\
%	\begin{array}{rcl}
%	\size{\stempty} & \defeq &0
%	\\
%	\size{\esub\var\clos\cdot\lenv} & \defeq & \size{\clos}+\size{\lenv}
%	\end{array}
%	&
%	\begin{array}{rcl}
%	\size{\stempty} & \defeq &0
%	\\
%	\size{\clos\cdot\stack} & \defeq &	\size{\clos}+\size{\stack}
%	\end{array}
%	\\\\
%	
%	\textsc{Closures} 
%	&		
%	\textsc{States}\\
%	\begin{array}{rcl}
%	\size{(\tm,\lenv)} & \defeq & 1+\size\lenv 
%	\end{array}
%	& 
%	\begin{array}{rcl}			
%	\size{(\tm,\lenv,\stack)} &\defeq & \size{\lenv}+\size{\stack}
%	\end{array}
%	\end{array}\]
%\end{definition}

The size of a \SpKAM run is then obtained by considering the maximum size of the states reached along that run.

\begin{definition}[Run abstract space]
	Let $\run:\compil{\tm_0}\rightarrow_{\mathrm{Sp\KAM}}^*\state$ be a \SpKAM run. Then the (abstract) space consumption of the run $\run$ is defined as follows:
	\[\lm{\run}\defeq\max_{\statetwo\in\run}\size{\statetwo}\]
\end{definition}
%The size of the input pointers can be factored out. Thus, by defining $\lm{\run}\defeq\max_{\statetwo\in\run}\size{\statetwo}$, we have
%\[\rlm{\run}=\log\tm_0\cdot\max_{\statetwo\in\run}\size{\statetwo} = \log\tm_0\cdot\lm{\run}\]
%This consideration allows us to concentrate only on $\size\cdot$ and $\lm\cdot$. We just need to remember that in order to obtain the actual space consumption $\rsize\cdot$ and $\rlm\cdot$ we need to multiply the final result by $\log\tm_0$.

%After having specified all the cost measures, we are able to state the invariance theorem for the \SpKAM.
%\begin{theorem}[\citep{TRReasonableSpace}, The \SpKAM Is Reasonable for Space]
%	Closed CbN evaluation $\towh$ and the space of the Space KAM $\rlm{\cdot}$ provide a reasonable space cost model for the $\lambda$-calculus.
%\end{theorem}

\begin{example}\label{ex:spkam}
	We provide the \SpKAM execution $\run$ for the term $(\la\var(\la\vartwo(\la\varthree\var)(\var\vartwo))\var)\Id$.
	\[
	\begin{array}{l|l|ll}
	\mathsf{Term}   & 
	\mathsf{Environment} & \mathsf{Stack}\\
	\cline{1-3}
	(\la\var(\la\vartwo(\la\varthree\var)(\var\vartwo))\var)\Id & \stempty & \stempty & \tokamseanv\\
	\la\var(\la\vartwo(\la\varthree\var)(\var\vartwo))\var & \stempty & (\Id,\stempty) & \tokambnw\\
	(\la\vartwo(\la\varthree\var)(\var\vartwo))\var & \esub\var{(\Id,\stempty)} & \stempty & \tokamseav\\
	\la\vartwo(\la\varthree\var)(\var\vartwo) & \esub\var{(\Id,\stempty)} & (\Id,\stempty) & \tokambnw \\
	(\la\varthree\var)(\var\vartwo) & \esub\vartwo{(\Id,\stempty)} \cdot \esub\var{(\Id,\stempty)}=:\lenv & \stempty & \tokamseanv\\
	\la\varthree\var & \esub\var{(\Id,\stempty)} & (\var\vartwo,\lenv) & \tokambw\\
	\var & \esub\var{(\Id,\stempty)} & \stempty & \tokamsub\\
	\Id & \stempty & \stempty
	\end{array}
	\]
	Now the environment contains only closures bound to the free variables of the current term. Moreover, closures that will be never used, because the abstracted variable does not occur in the body of the abstraction, are simply deleted, such as $(\var\vartwo,\lenv)$ in the  $\tokambw$ transition. Finally, no chains that rename variables are present anymore. In the previous execution, on the \KAM, there was one such chain, namely $(\var,\esub\var{(\Id,\stempty)})$, which now is just $(\Id,\stempty)$. From the quantitative point of view, note that the maximum space consumed during the execution is 4 pointers: in the third to last state, the closure in the stack contains $\lenv$, which is itself an environment with two closures, as defined at the previous line. Thus, we have $\lm{\run}=4$.
\end{example}

% !TeX spellcheck = en_US
% !TEX root = main.tex
%%%%%%%%%%%%%%%%%%%%%%%%%%%%%%%%%%%%%%%%%%%%%%%%%%%%%%%%%%%%%%%%%%%%%%%%%%%%%%%%
\section{Intuitions about measuring space via types}\label{sec:intuitions}
In this section, we give an informal and intuitive explanation of the design of the type system for measuring space that shall be given in the next section.

\paragraph*{The key bijection and the length of stacks.} We have already observed (right after \refthm{decarvalho}) that there is a very strong correspondence between states of the KAM run $\run$ for $\tm$ and a type derivation $\tyd\pof\tjudgw{}{\weight}{\tm}{\initty}$. We pointed out a bijection between typing rules and machine transitions, but the correspondence goes further: there is a bijection between type judgments in $\tyd$ and states in $\run$ such that every judgment $\wtjudg{\tye}{w}{\tm}{\arr{\mty_1}{\arr{\cdots}{\arr{\mty_n}{\initty} } } }$ corresponds to a state $\spkamstate\tm\lenv{\clos_1\cdots\clos_n}$ for some $\lenv$. That is, the number of outermost arrows in the right-hand type corresponds exactly to the number of closures on the stack. Concisely: we can retrieve the length of the stack of states from the type derivation. This can be tested, for instance, by comparing Examples~\ref{ex:kam-run-bij} and \ref{ex:tyd-bij}.

\paragraph*{Length of environments.} Can we also retrieve the length of the environment from the type derivation? There is a certain relationship between the type environment $\tye$ of a judgment and the local environment $\lenv$ of the associated state. But the answer is no, unfortunately there is a mismatch between the two. If the KAM run on $\tm$ has final state $\spkamstate{\la\var\tmtwo}\lenv\stempty$ and $\tmtwo$ has some free variable $\vartwo\neq\var$ then $\lenv$ contains an entry $\esub\vartwo\clos$ while the corresponding judgment $\tjudg{}{\lambda\var.\tm}{\initty}$ for $\la\var\tmtwo$ is obtained via rule $\tylamstar$ and has empty type environment.

Intuitively, the machine keeps track of substitutions that might happen if evaluation would go under abstraction, whereas the type system forgets them.

\paragraph*{Tweaking the type system.} The first step towards our result is to modify the rules of the type system so that the type environment and the local environment of the corresponding state have exactly the same length. This shall be obtained via a modification of rule $\tylamstar$ (plus some other detail). In this way we shall have a refined bijection between judgments and states summed up by the following schema:
\begin{center}
$\begin{array}{c}
\wtjudg{\var_1:\mtytwo_1,\ldots,\var_n:\mtytwo_n}{w}{\tm}{\arr{\mty_1}{\arr{\cdots}{\arr{\mty_m}{\initty} } } }
\\
\Updownarrow
\\
\spkamstate\tm{\esub{\var_1}{\clos'_1}\cdots\esub{\var_n}{\clos'_n}}{\clos_1\cdots\clos_m}
\end{array}$
\end{center}
Now, we can read all the closures managed by the KAM out of the type derivation. Note that closures correspond to multisets.

The next step is to decorate each multiset $\mty$ with an index $k$ meant to represent the size of the associated closure $\clos$. In this way we obtain a refined bijection schema:
\begin{center}
$\begin{array}{c}
\wtjudg{\var_1:\mtytwo_1^{\size{\clos'_1}},\ldots,\var_n:\mtytwo_n^{\size{\clos'_n}}}{w}{\tm}{\arr{\mty_1^{\size{\clos_1}}}{\arr{\cdots}{\arr{\mty_m^{\size{\clos_m}}}{\initty} } } }
\\
\Updownarrow
\\
\spkamstate\tm{\esub{\var_1}{\clos'_1}\cdots\esub{\var_n}{\clos'_n}}{\clos_1\cdots\clos_m}
\end{array}$
\end{center}
At this point the size of the state $\state$ associated to a judgment $\wtjudg{\tye}{w}{\tm}{\linty }$ is simply the sum of the indices appearing in $\tye$ and $\linty$.

Lastly, one redefines the weight $w$ of the judgment in a way that tracks the maximum of the sizes of states above the current judgment. In this way, the weight of the final judgment is the needed space measure.

% !TeX spellcheck = en_US
% !TEX root = main.tex
%%%%%%%%%%%%%%%%%%%%%%%%%%%%%%%%%%%%%%%%%%%%%%%%%%%%%%%%%%%%%%%%%%%%%%%%%%%%%%%%
\section{The closure-type system}\label{sec:type-system}
Here we define our variant of multi types, dubbed \emph{closure types}, that we are going to use to measure the space consumption of (typable) terms. The definition of types is standard, but for the fact that multi-sets come labeled with an index $k$. The idea is that multi-sets of types are associated to arguments (according to the call-by-name translation of the $\lambda$-calculus into linear logic), and arguments give rise to closures (hence the name closure types): the index represents the size $\size\clos$ of the closure $\clos$ (\ie the number of pointers to implement it) that shall be associated to that argument/multi-set.
\[
	\begin{array}{rrcll}
		\textsc{Linear 
		Types}&\lintys\ni\linty,\lintytwo&\grameq&\initty\grammarpipe\arr{\mty^k}{\linty} 
		\\[3pt]
		\textsc{Closure 
		Types}&\mty^k,\mtytwo^k&\grameq&\mset{\linty_1,\ldots,\linty_n}^k \qquad 
		k>0,n\geq 0
	\end{array}
\]

Type environments are now partial finite maps
from variables to closure types. Type judgments and type derivations are defined exactly as in the traditional multi type case. Please note that some subtleties become apparent, regarding how labeled multi-sets are handled. The empty multi-set is noted $\emmset^{k}$ and it also comes labeled with $k$. Note that $k$ has to be strictly positive. The sum $\uplus$ of multi-sets requires the two multi-sets to have the same index, that is, we have that $\mty^{k}\uplus \mtytwo^{k} \defeq (\mty\uplus\mtytwo)^{k}$, where on the right hand side we treat $\mty$ and $\mtytwo$ as ordinary multi-sets, while $\mty^{k}\uplus \mtytwo^{h}$ is undefined for $h\neq k$.

\paragraph*{Space and weakenings.} The study of space requires an unusual approach 
to weakening. In a weak evaluation setting a judgment 
$\tjudg{\tye}{\tm}{\linty}$ usually implies only that $\dom\tye \subseteq \fv\tm$ 
and not necessarily that $\dom\tye = \fv\tm$, because there can be untyped free 
variables, that is, free variables occurring under abstraction that are not 
touched by weak evaluation and thus not typed. Indeed, given an abstraction 
$\la\var\tm$ with $\fv{\la\var\tm} \neq \emptyset$, usually one can type it with 
$\initty$, deriving a judgment $\tjudg{}{\la\var\tm}{\initty}$ with an 
\emph{empty} type context. Here this shall not be possible, because in the \SpKAM 
the variables in $\fv{\la\var\tm}$ play a role in the space usage, as they forbid 
to garbage collect some closures. Therefore, we modify the type system as to 
enforce the property $\dom\tye = \fv\tm$, even if evaluation is weak, and this is 
done via an unusual use of weakening. In particular, we distinguish between a 
variable $\var\notin\dom\tye$ and a variable $\var$  that gets assigned an empty multi-set, \ie such that $\tye(\var)=\emmset^{k_{\var}}$. The intuition is that $\var\notin\dom\tye$ means 
that $\var$ does not appear in the term at all, and it would correspond to $\var$ 
being typed with $\emmset^{0}$, which is however not a valid type because the 
index $k_{x}$ must be $>0$. Instead, $\tye(\var)=\emmset^{k_{\var}}$ means that 
$\var$ appears but it is not going to be used, and it shall nonetheless have an 
associated closure of size $k_{\var}$. We point out that also 
\citet{DBLP:journals/iandc/LengrandLDDB04} have studied intersection types in a 
calculus with explicit substitutions and a garbage collection rule.

We shall need a notion of type context assigning the empty multi-set (with a positive index) to all the variables in its domain.
\begin{definition}[Dry type contexts]
	A type context $\tye$ is \emph{dry} if for every $\var\in\dom\tye$, there 
	exists $k_{\var}$ such that $\tye(\var)=\emmset^{k_{\var}}$.
\end{definition}

\paragraph*{Typing Rules.}
%Type judgment are indexed with a weight $\weight\in\nat$, and noted $\wtjudg{\tye}{\weight}{\tm}{\linty}$.
In order to define the typing rules we need two further notions.
First, we need a notion of size of types and type contexts.
\[
\begin{array}{c@{\hspace{2cm}}c@{\hspace{2cm}}c}
	\size{\star}\defeq 0 &
	\size{\arr{\mty^k}\linty}\defeq k+\size{\linty}&
	\size{\var:\mty^k,\tye}\defeq k+\size{\tye}
\end{array}
\]
Note that for a type context one sums only the indices over the multi-sets, ignoring the size of linear types inside the multi-sets themselves.

To define the type system, we need a predicate over type environments to ensure that they are summable, because multi-sets sum is restricted to multi-sets having the same index.

\begin{definition}[Summability]
	 Two type environments $\tye$ and $\tyetwo$ are \emph{summable}, noted $\tye\# \tyetwo$, if, when $\tye(\var) = \mty^{k}$ and 
$\tyetwo(\var) = \mtytwo^{h}$, then $k =h$, for all 
$\var\in\dom\tye\cap\dom\tyetwo$. The notion can be naturally generalized to an 
arbitrary number of type environments $\tye_i$ as $\#_i\tye_i$.
\end{definition}

% !TeX spellcheck = en_US
% !TEX root = main.tex
%%%%%%%%%%%%%%%%%%%%%%%%%%%%%%%%%%%%%%%%%%%%%%%%%%%%%%%%%%%%%%%%%%%%%%%%%%%%%%%%
\begin{figure}[t]
	{\small\[
	\begin{array}{c@{\hspace{0.4cm}}c}
	\infer[\tyvar]{\wtjudg{\var:\mset{\linty}^k}{k+\size{\linty}}{\var}{\linty}}{}
	&
	
	\infer[\tylamstar]{\wtjudg{\tye}{\size{\tye}}{\lambda\var.\tm}{\initty}}{
		\dom\tye = \fv{\la\var\tm} & \tye \mbox{ is dry}}
	\\[10pt]
	\infer[\tylam_{1}]{\wtjudg{\tye}{w}{\lambda\var.\tm}{\arr{\mty^{k}}{\linty}}} 
	{\wtjudg{\tye,\var:\mty^{k}}{w}{\tm}{\linty}}
	&
	\infer[\tylam_{2}]{\wtjudg{\tye}{\max\{w,\size\tye+\size\linty+k\}}{\lambda\var.\tm}{\arr{\emmtype^{k}}{\linty}}}
	{\wtjudg{\tye}{w}{\tm}{\linty} & \var\notin\dom\tye}
	\\[10pt]
	\infer[\tymany]{\wtjudg{\uplus_{i=1}^n\tye_i}{\max_{i}\set{v_{i}}}{\tm}{\mset{\linty_1\mydots\linty_n}^{1+\size{\uplus_{i=1}^n\tye_i}}}}
	{\wtjudg{\tye_i}{v_{i}}{\tm}{\linty_i} & 1\leq i\leq n & \#_i\tye_{i}}
	& \infer[\tynone]{\wtjudg{\tye}{0}{\tm}{\emmset^{1+\size\tye}}}{\dom\tye = \fv{\tm} & \tye  \mbox{ is dry}}
	\\[10pt]
	
	\infer[\tyapp_{1}]{\wtjudg{\tye\uplus 
			\tyetwo}{\max\{w,v\}}{\tm\tmtwo}{\linty 
	}}{\wtjudg{\tye}{w}{\tm}{\arr{\mty^{k}}{\linty}}
		& 
		\wtjudg{\tyetwo}{v}{\tmtwo}{\mty^{k}}& \tye\# \tyetwo & \tmtwo\not\in\mathcal{V}} 
	&
	
	%	\multicolumn{3}{c}{\infer[\tyapp]{\wtjudg{\tye\uplus 
	%	\sum_{i=1}^n \tyetwo_i  }{\max\{w,v_i\}}{\tm\tmtwo}{\linty 
	%		}}{\wtjudg{\tye}{w}{\tm}{\arr{\mset{\lintytwo_1,\ldots,\lintytwo_n}^{\size{
	%		  \tyetwo}+1}}{\linty}}
	%			& 
	%			\mset{\wtjudg{\tyetwo_i}{v_i}{\tmtwo}{\lintytwo_i}}_{i\in\mset{1,\ldots,n}}}}
	%		\\[10pt]
	%
	%		\multicolumn{3}{c}{\infer[\tyapp]{\wtjudg{\tye 
	%		}{w}{\tm\tmtwo}{\linty 
	%				}}{\wtjudg{\tye_{|\tm}}{w}{\tm}{\arr{\mset{\cdot}^{\size{
	%							\tye_{|\tmtwo}}+1}}{\linty}}}}
	%	\\[10pt]
	\infer[\tyapp_{2}]{\wtjudg{\tye\uplus
			\var:\mty^k  
		}{w}{\tm\var}{\linty 
	}}{\wtjudg{\tye}{w}{\tm}{\arr{\mty^k}{\linty}}
		&&\tye\#\var:\mty^k}
	\end{array}
	\]}
	
	\caption{The closure-type system.}
	\label{fig:typing-rules}
\end{figure}

The typing rules for closure types are in \reffig{typing-rules}. For now, just ignore the weights. As in the case of the multi-type system and of the \KAM, we have crafted this system in such a way that, given a type 
derivation $\tyd\pof\tjudg{}{\tm}{\initty}$, there is a one-to-one correspondence 
between the transitions used by the \SpKAM run on $\tm$ and the occurrences of 
the typing rules (excluded $\tymany$ and $\tynone$) of $\tyd$. Moreover, we ensure that $\tylamstar$ is used to type final states. Namely, $\tokamseav$ corresponds to $\tyapp_{2}$, $\tokamseanv$ to $\tyapp_{1}$, $\tokambnw$  to $\tylam_{1}$, $\tokambw$  to $\tylam_{2}$, and $\tokamsub$ to $\tyvar$. The intuition behind the closure types is that there is a correspondence between \SpKAM data 
structures 
and the type-theoretic side. The idea is that every closure type $\mty^k$ 
corresponds to a closure $\clos$ such that $\size\clos=k$. In particular, given 
a state $\state = \spkamstate{\tm}{\lenv}{\stack}$ and 
a judgment $\tjudg{\tye}{\tm}{\linty}$, the 
type environment $\tye\defeq\var_1:\mty^{k_1}\ldots\var_n:\mty^{k_n}$ 
corresponds to the environment 
$\lenv\defeq\esub{\var_1}{\clos_1}\ldots\esub{\var_n}{\clos_n}$ 
and moreover $\size{\clos_i}=k_i$ for each $1\leq i\leq n$. Similarly, 
there is a correspondence between the stack and the type $\linty$, namely  
$\linty\defeq\arr{\mty_1^{k_{1}}}{\arr{\cdots}{\arr{\mty_m^{k_{m}}}{\initty}}}$ 
corresponds to the stack $\stack\defeq\stack_1\ldots\stack_m$ and moreover 
$\size{\stack_i}=k_i$ for each $1\leq i\leq m$. Essentially, this means that we are able to read from the type derivation of a term $\tm$ the sizes of all the states reached by the \SpKAM evaluating $\tm$. 

Multi-sets are introduced on the right with rules $\tymany$ and $\tynone$. To explain the index on the introduced multi-set, let us first consider $\tynone$. The idea is that $\tm$ shall be paired by the machine with an environment $\lenv$ to form a closure $\clos = (\tm,\lenv)$. By the invariant of the machine (\reflemma{spkam-invariants}), $\dom\lenv$ is exactly $\fv\tm$, that is, $\lenv$ contains a closure for each variable $\var$ in $\fv\tm$. Now, each such closure shall have size $k_{\var}$, where $k_{\var}$ is the index given to $\emmset$ by the dry context $\tye$. Therefore, $1+\size\tye$ shall correspond to $1+\size\lenv$, which is exactly the size of $\clos$.

For $\tymany$, the reasoning is analogous. Let us explain a point about the 
quantity $\size{\uplus_{i=1}^n\tye_i}$ in its conclusion. An invariant 
(forthcoming \reflemma{tye-invariant}) shall guarantee that $\dom\tye = \fv\tm$, 
whenever $\tjudg{\tye}{\tm}{\linty}$. Then in $\tymany$, the type contexts
$\tye_{i}$ have all the same domain, and by the summable hypothesis, they all give 
the same index $k$ to the (potentially different) multi-sets $\tye_{i}(\var)$ for 
a same variable $\var$. Thus the various $\size{\tye_j}$ all coincide (for 
$j\in\set{1,\ldots,n}$) and also coincide with the quantity 
$\size{\uplus_{i=1}^n\tye_i}$ (so one could have as well decorated the conclusion 
of $\tymany$ with the index $\size{\tye_1}$ rather than 
$\size{\uplus_{i=1}^n\tye_i}$, as they coincide).

Some of the rules of the type system seem to introduce some \emph{arbitrary} indices in the conclusions. Namely, rules $\tyvar$ and $\tylam_{2}$ introduce an arbitrary $k>0$, and rules $\tylamstar$ and $\tynone$ refer to a dry $\tye$, and dryness only requires the indices to be different from $0$ but otherwise arbitrary. Two lemmas in the next section (\reflemma{env-clos-size} and \reflemma{space-bound-on-current-state}) shall however guarantee that, if the global derivation in which the rules occur types a closed term with $\initty$, such arbitrary choices are in fact strongly constrained, to the point that the indices are actually \emph{uniquely} determined. 

We highlight the main differences w.r.t. the multi-type system presented in Section~\ref{sec:pre}.
\begin{itemize}
	\item \emph{Weakening.} As we have already observed, the approach to weakening is slightly more liberal than in multi types. It is implicitly part of rules $\tylamstar$, $\tylam_{2}$, and $\tynone$. In rules $\tylamstar$ and $\tynone$, a context $\tye$ can be injected in the conclusion, but only if it is dry, and if its domain coincides with the free variables of the typed term. The situation at the level of the \SpKAM for the rule $\tylamstar$ corresponds to final states such as $\spkamstate{\la\var\vartwo}{\esub\vartwo\tm}{\stempty}$. A closure is indeed present, hence the index $k>0$, although it will never be used, hence its type being the empty multi-set. In rule $\tylam_{2}$, which is not present in the multi-type system (because transition $\tokambw$ is not a transition of the \KAM), the variable $\var$ that does not occur free in $\tm$ is intuitively typed with $\emmset^0$ in the premise in the rule. In the conclusion, however, a type $\arr{\emmset^0}{\linty}$ would not be correct, as the index cannot be $0$. Then the rule uses an arbitrary index $k>0$. The intuition is that  the \SpKAM closure corresponding to $\emmset$, before being eliminated by rule $\tokambw$, has size strictly greater than zero.
	
	Please notice that all these considerations ensure the invariant for which for any judgment $\tjudg{\tye}{\tm}{\linty}$ or $\tjudg{\tye}{\tm}{\mty^k}$ one has $\dom{\tye}=\fv\tm$.
	\item \emph{Separate multi-set rules.} In the traditional multi-type rules $\tynone$ and $\tymany$ are incorporated inside rule $\tyapp$. This was impossible to do in the closure-type system because the rule $\tynone$ was necessarily different from the rule $\tymany$, and not just the special case when there are zero premises. This is because of the way we deal with weakenings: weakening is present just in the rule $\tynone$, while it is not in rule $\tymany$.
	
	Rules $\tynone$ and $\tymany$ correspond to the creation of closures. This is why the index of the created multi-set is $1+\size\tye$. Morally, this corresponds to the closure $(\tm,\lenv)$, where $\tm$ is the typed term and $\lenv$ corresponds to $\tye$. Hence, we have $\size{(\tm,\lenv)}=1+\size\lenv=1+\size\tye$.
	\item  \emph{Unchaining.} The type rule $\tyapp_{2}$ is the type-theoretic equivalent of the \SpKAM transition rule $\tokamseav$, that implements the unchaining optimization. If one forgets the indices, the rule is just a specialization of the rule $\tyapp_{1}$, from which it can be derived (using also an instance of rule $\tyvar$). The derived rule, however, would not have the right indices (because it corresponds to the KAM way of computing that does not unchain), which is why we need an additional rule.
	\item \emph{Coherence.} All the rules which sum type environments have to enforce the coherence relation $\#\tye_i$ between the type environments $\tye_i$ in the premises.
\end{itemize}

\paragraph*{The Weight system.} The intuition behind the weight system is very simple. The weight of the type derivation $\tyd$ ending in $\tjudg{}\tm\initty$ should correspond to the space consumed by the complete run $\run$ of the \SpKAM starting from $\tm$. We know how to  read the size of a \SpKAM state $\state= \spkamstate{\tm}{\lenv}{\stack}$ belonging to $\run$ out of the corresponding type judgment $\tjudg{\tye}{\tm}{\linty}$ belonging to $\tyd$: simply $\size\state=\size\tye+\size\linty$. Then, the space consumed by $\run$ is the maximum of the sizes of all the type judgments occurring in $\tyd$. This is exactly what the weighting system is tracking.

\begin{example}
	We provide the closure-type derivation for the term $(\la\var(\la\vartwo(\la\varthree\var)(\var\vartwo))\var)\Id$ in Figure~\ref{fig:example}.
	\begin{figure}[t]
	\[
	\infer[\tyapp_1]{\tjudgw{}{4} {(\la\var(\la\vartwo(\la\varthree\var)(\var\vartwo))\var)\Id}{\initty}}{
		\infer[\tylam_1]{\tjudgw{}{4}{\la\var(\la\vartwo(\la\varthree\var)(\var\vartwo))\var} {\arr{\mset\initty^1}\initty}}{
			\infer[\tyapp_2]{\tjudgw{\var:\mset\initty^1}{4}{(\la\vartwo(\la\varthree\var)(\var\vartwo))\var}{\initty}}{\infer[\tylam_1]{\tjudgw{\var:\mset\initty^1}{4}{\la\vartwo(\la\varthree\var)(\var\vartwo)} {\arr{\emmset^1}\initty}}{
					\infer[\tyapp_1]{\tjudgw{\var:\mset\initty^1,\vartwo:\emmset^1}{4}{(\la\varthree\var)(\var\vartwo)}{\initty}}{
						\infer[\tylam_2]{\tjudgw{\var:\mset\initty^1}{4}{\la\varthree\var}{\arr{\emmset^3}\initty}}{
							\infer[\tyvar]{\tjudgw{\var:\mset\initty^1}{1}{\var}{\initty}}{}} & \infer[\tynone]{\tjudgw{\var:\emmset^1,\vartwo:\emmset^1}{0}{\var\vartwo}{\emmset^3}}{}}}}} &\hspace{-15pt} \infer[\tymany]{\tjudgw{}{0}{\Id}{\mset\initty^1}}{
								\infer[\tylamstar]{\tjudgw{}{0}{\Id}{\initty}}{}}}
	\]
	\caption{The closure-type derivation for the term $(\la\var(\la\vartwo(\la\varthree\var)(\var\vartwo))\var)\Id$.}
	\label{fig:example}
	\end{figure}
	Also in this case, we can observe the precise correspondence between this type derivation and the execution of the \SpKAM of Example~\ref{ex:spkam}. Not only rules and transitions are in a one-to-one correspondence, but also stack and environment entries (with their sizes) can be seen, respectively, in types and in type environments. Of course, as a consequence, the final weight is 4, as the space consumption of the \SpKAM execution. 
\end{example}

% !TeX spellcheck = en_US
% !TEX root = main.tex
%%%%%%%%%%%%%%%%%%%%%%%%%%%%%%%%%%%%%%%%%%%%%%%%%%%%%%%%%%%%%%%%%%%%%%%%%%%%%%%%
\section{Weights capture the space of the \SpKAM}\label{sec:correctness}
This section is devoted to the proof of our main result, that is, the fact that the weight $w$ of a type judgment $\wtjudg{}{w}{\tm}{\initty}$ captures the space used by the (complete run on $\tm$ of the) \SpKAM. The proof is carried out in a rather standard way. First, we prove soundness, that is the fact that typable terms terminate, and moreover that their evaluation respects the weight. Second, we prove completeness, that is the fact that all terminating terms are typable. 
\subsection{Preliminary properties}
The first property that we prove here is qualitative, that is, it does not concern weights or indices.\footnote{When a definition or a statement does not rest on weights, we do not report them, for the sake of readability.} It is simply the already mentioned fact that the domain of type contexts is exactly the set of free variables of the typed term. This is the type analog of the environment domain invariant of the \SpKAM.
\begin{lemma}[Type contexts domain invariant]
	\label{l:tye-invariant}
	\hfill
	\begin{enumerate}
		\item If $\tyd\pof \tjudg{\tye}{\tm}{\linty}$ then $\dom\tye = \fv\tm$.
		\item If $\tyd\pof \tjudg{\tye}{\tm}{\mty}$ then $\dom\tye = \fv\tm$.
	\end{enumerate}
\end{lemma}

\begin{proof}
	By induction on $\tyd$.
\end{proof}

Then, we have two quantitative properties.
%The first one is that the weight on a judgment bounds the size of the types in that judgment.
%\begin{lemma}[Weights bound types]
%	\label{l:weights-and-judgments}
%	If $\wtjudg{\tye}{\weight}{\tm}{\linty}$ then $\weight\geq \size\tye+ \size\linty$.
%\end{lemma}
%\begin{proof}
%	We prove the statement by induction on the structure of the type derivation. For the axiom rules ($\tyvar$ and $\tylamstar$), the weight is \emph{defined} as $\size\tye+ \size\linty$. For all rules but $\tyapp_{1}$ (that is, for rules $\tylam_{1}$, $\tylam_{2}$, and $\tyapp_{2}$), the statement follows immediately from the \ih applied to the sub-derivation. Then consider rule $\tyapp_{1}$:
%	\[\infer[\tyapp_{1}]{\wtjudg{\tye\uplus 
%			\tyetwo}{\max\{w,v\}}{\tm\tmtwo}{\linty 
%	}}{\wtjudg{\tye}{w}{\tm}{\arr{\mty^{k}}{\linty}}
%		& 
%		\wtjudg{\tyetwo}{v}{\tmtwo}{\mty^{k}}& \tye\# \tyetwo} \]
%	By \ih, $\weight \geq \size\tye + \size{\arr{\mty^{k}}{\linty}} = \size\tye + k + \size\linty$. Since $\wtjudg{\tyetwo}{v}{\tmtwo}{\mty^{k}}$ has been introduced by $\tymany$ or $\tynone$, we have $k = 1+ \size\tyetwo$. Then $\weight \geq  \size\tye + k + \size\linty > \size\tye + \size\tyetwo + \size\linty = \size{\tye\uplus \tyetwo} + \size\linty$ as required. 
%\end{proof}
Before stating them, we need to introduce the definition of typed \SpKAM states. The idea is very simple, and it is due to \citet{deCarvalho18}. In order to state quantitative properties that involve abstract machines, we type the execution of the abstract machine itself. We have already mentioned that type environments correspond to the \SpKAM environments, and that closure types correspond to \SpKAM closures. In \reffig{kamtypes}, the connection is made formal.

% !TeX spellcheck = en_US
% !TEX root = main.tex
%%%%%%%%%%%%%%%%%%%%%%%%%%%%%%%%%%%%%%%%%%%%%%%%%%%%%%%%%%%%%%%%%%%%%%%%%%%%%%%%
\begin{figure}[t]
	\[
	\begin{array}{c@{\hspace{1cm}}c@{\hspace{1cm}}c}
	\infer[\tyenv]{\wtjudg{}{\max\set{\weight_\var|\var\in\dom\lenv}}{\lenv}{\tye}} 
	{\wtjudg{}{\weight_\var}{\lenv(\var)}{\tye(\var)}&&\forall\var\in\dom\lenv}
	&
	\infer[\tyclos]{\wtjudg{}{\max\{\weight,\weighttwo\}}{(\tm,\lenv)}{\mty^{k}}}{ 
		\wtjudg{}{\weight}{\lenv}{\tye} 
		&& 
		\wtjudg{\tye}{\weighttwo}{\tm}{\mty^{k}} }
	\\\\
	\multicolumn{2}{c}{
		\infer[\tystate]{\wtjudg{}{\max_i\{w,v_i,u\}}{\spkamstate\tm\lenv{\clos_1\cdots\clos_n}}{\initty}}{
			\wtjudg{\tye}{w}{\tm}{\arr{\mty_1^{k_{1}}}{\arr{\cdots}{\arr{\mty_n^{k_{n}}}{\initty} } } } 
			&
			\wtjudg{}{u}{\lenv}{\tye}
			&
			\wtjudg{}{v_i}{\clos_i}{\mty_i^{k_{i}}} 
		}
	}
	\end{array}
	\]
	
	\caption{Typing rules for the machine components of the \SpKAM.}
	\label{fig:kamtypes}
\end{figure}

The next lemma states that the size of a closure is equal to the size of its type, for every closure, and similarly for environments.
\begin{lemma}[Types capture the size of closures and environments]
	\label{l:env-clos-size} % \reflemmap{env-clos-size}{one}
	\hfill
	\begin{enumerate}
		\item \label{p:env-clos-size-one} 
		If $\tyd\pof \tjudg{}{\lenv}{\tye}$ then $\size\lenv = \size\tye$.
		\item \label{p:env-clos-size-two} 
		If $\tyd\pof \tjudg{}{\clos}{\mty^{k}}$ then $\size\clos = k$.
	\end{enumerate}
\end{lemma}

This lemma expresses one of the key properties of the type system, which has an almost \emph{magical} feeling. At first sight, indeed, the indices on multi types are completely arbitrary, as the typing rules do not seem to impose a strong constraint. Surprisingly, instead, when one builds the type derivation of a closure, then the indices are uniquely determined, and they capture \emph{exactly} the size of the closure.

% !TeX spellcheck = en_US
% !TEX root = ../main.tex
%%%%%%%%%%%%%%
\begin{proof}
	By mutual induction on $\lenv$ and $\clos$.
	\begin{enumerate}
		\item $\tyd$ has the following form:
		\[
		\infer[\tyenv]{\tjudg{}{\lenv}{\tye}}{\tjudg{}{\lenv(\var)}{\tye(\var)}}\] Let $\tye(\var) = \mty_{\var}^{k_{\var}}$. By \ih (point 2), $\size{\lenv(\var)} = k_{\var}$. 
		Then $\size\lenv = \sum_{\var\in\dom\lenv}\size{\lenv(\var)} =_{\ih}  \sum_{\var\in\dom\lenv}k_{\var} = \size\tye$.
		
		\item $\tyd$ has the following form:
		\[\infer[\tyclos]{\tjudg{}{(\tm,\lenv)}{\mty^{k}}}{ 
			\tjudg{}{\lenv}{\tye} 
			&& 
			\tjudg{\tye}{\tm}{\mty^{k}} }\]
		with $\clos = (\tm,\lenv)$. By \ih (point 1) applied to $\lenv$, we have $\size\lenv = \size\tye$. Since the typing of $\tm$ comes necessarily from a rule $\tynone$ or $\tymany$, we have $k = 1+ \size\tye$. Then $\size\clos = 1 +\size\lenv =_{\ih} 1+ \size\tye = k$.\qedhere
	\end{enumerate}
\end{proof}

The second easy property is the fact that the size of a state is given by the size 
of the types in the judgment for its code, which give the size of the stack and 
the environment.

\begin{lemma}[Judgments capture the size of the corresponding state]
	\label{l:space-bound-on-current-state}
	If \[
	\infer[\tystate]{\tjudg{}{(\tm,\clos_1\cdots\clos_n,\lenv)}{\initty}}{
		\tjudg{\tye}{\tm}{\arr{\mty_1^{k_{1}}}{\arr{\cdots}{\arr{\mty_n^{k_{n}}}{\initty} } } } 
		&
		\tjudg{}{\clos_i}{\mty_i^{k_{i}}} 
		&
		\tjudg{}{\lenv}{\tye}
	}
	\]
	then $\size\tye + \sum_{i=1}^{n} k_{i} = \size\state$.	
\end{lemma}

% !TeX spellcheck = en_US
% !TEX root = ../main.tex
%%%%%%%%%%%%%%
\begin{proof}
	We have:
	\[
	\infer[\tystate]{\tjudg{}{(\tm,\clos_1\cdots\clos_n,\lenv)}{\initty}}{
		\tjudg{\tye}{\tm}{\arr{\mty_1^{k_{1}}}{\arr{\cdots}{\arr{\mty_n^{k_{n}}}{\initty} } } } 
		&
		\tjudg{}{\clos_i}{\mty_i^{k_{i}}} 
		&
		\tjudg{}{\lenv}{\tye}
	}
	\]
	By definition, $\size\state = \size\lenv + \size{\clos_1\cdots\clos_n} =  \size\lenv +\sum_{i=1}^{n}\size{\clos_i}$ . By \reflemma{env-clos-size}, $\size\lenv = \size\tye$ and $\size{\clos_{i}} = k_{i}$ for $1\leq i\leq n$. Then $\size\state = \size\tye + \sum_{i=1}^{n} k_{i}$.
\end{proof}

\subsection{Soundness}
Soundness is the fact that on typed states the \SpKAM terminates. Here it is refined with our space analysis, showing that the weight of the type judgment is exactly the maximum space used by the run of the \SpKAM.

The proof technique is mostly standard. It is based on a subject reduction property plus a space analysis of final states. What is slightly unusual is that  subject reduction is not stated as an independent property, it is instead incorporated into the proof of soundness. This is needed to prove the space bound.

\paragraph*{Space Analysis of final states.} To prove that the weight system 
correctly measures the size of final states, we need an auxiliary lemma ensuring 
that closures and environments typed with empty types and dry type contexts induce 
null weight on their judgment.

\begin{lemma}[Empty types and dry type environments induce null weight]
	\label{l:dry-env-clos} % \reflemmap{dry-env-clos}{one}
	\hfill
	\begin{enumerate}
		\item \label{p:dry-env-clos-one} 
		If $\tyd\pof \wtjudg{}{\weight}{\lenv}{\tye}$ and $\tye$ is dry, then $\weight = 0$.
		\item \label{p:dry-env-clos-two} 
		If $\tyd\pof \wtjudg{}{\weight}{\clos}{\emmset^{k}}$, then $\weight = 0$.
	\end{enumerate}
\end{lemma}

% !TeX spellcheck = en_US
% !TEX root = ../main.tex
%%%%%%%%%%%%%%
\begin{proof}
	By mutual induction on $\lenv$ and $\clos$.
	\begin{enumerate}
		\item $\tyd$ has the following form:
		\[
		\infer[\tyenv]{\wtjudg{}{\max\set{\weight_\var|\var\in\dom\lenv}}{\lenv}{\tye}}{\wtjudg{}{\weight_\var}{\lenv(\var)}{\tye(\var)}}\]
		with $\weight =\max\set{\weight_\var|\var\in\dom\lenv}$. Two cases:
		\begin{enumerate}
			\item $\lenv$ is empty, \ie $\lenv = \stempty$: then $\weight = 0$ and $\tye$ is the empty type environment, for which $\size\tye = 0$. Therefore, we have $\weight = 0 =\size\tye$, validating the statement.
			
			\item $\lenv$ is not empty, \ie $\lenv \neq \stempty$:
			Since $\tye$ is dry,  then $\wtjudg{}{\weight_\var}{\lenv(\var)}{\emmset^{k_{\var}}}$ for some $k_{\var}$. By \ih (point 2), $\weight_{\var}  = 0$. Then $\weight =\max\set{\weight_\var|\var\in\dom\lenv} =0$.
		\end{enumerate}
		\item $\tyd$ has the following form:
		\[\infer[\tyclos]{\wtjudg{}{\max\set{\weighttwo,\weightthree}}{(\tm,\lenv)}{\emmset^{k}}}{ 
			\wtjudg{}{\weighttwo}{\lenv}{\tye} 
			&& 
			\wtjudg{\tye}{\weightthree}{\tm}{\emmset^{k}} }\]
		with $\clos = (\tm,\lenv)$ and $\weight = \max\set{\weighttwo,\weightthree}$. Since the typing of $\tm$ comes necessarily from a rule $\tynone$, $\tye$ is dry, $\weightthree = 0$. Then we can apply the \ih (point 1) to $\lenv$, obtaining $\weighttwo = 0$. Thus we have $\weight = 0$.
	\end{enumerate}
\end{proof}

\begin{proposition}[The weight of final states is their space]
	\label{prop:weight-of-final-states-is-right}
	Let $\state$ be a final state and $\tyd\pof \tjudgw{}{w}{\state}{\initty}$. Then $w = \size\state$.
\end{proposition}
% !TeX spellcheck = en_US
% !TEX root = ../main.tex
%%%%%%%%%%%%%%
\begin{proof}
	Final states have the shape $\spkamstate{\la\var\tm}\lenv\stempty$. Then $\tyd$ has the following shape:
	\[
	\infer[\tystate]{\wtjudg{}{\max\{\weighttwo,\size\tye\}}{\spkamstate{\la\var\tm}\lenv\stempty}{\initty}}{
		\infer[\tylamstar]{\wtjudg{\tye}{\size{\tye}}{\lambda\var.\tm}{\initty}}{
			\dom\tye = \fv{\la\var\tm} & \tye\mbox{ is dry}}
		&
		\wtjudg{}{\weighttwo}{\lenv}{\tye}
	}
	\]
	with $\weight = \max\{\weighttwo,\size\tye\}$. By \reflemmap{dry-env-clos}{one}, $\weighttwo = 0$. Therefore, $\weight = \size\tye =_\text{L.\ref{l:env-clos-size}.\ref{p:env-clos-size-one} }  \size\lenv = \size\state$.
\end{proof}

\paragraph*{Soundness.} Soundness shall be proved by induction on the size of type derivation, which is defined ignoring some typing rules, as follows.
\begin{definition}[Type derivations size]
	The size $\size\tyd$ of a type derivation $\tyd$ is its number of rules without counting rules $\tymany$, $\tynone$, $\tyclos$, and $\tyenv$.
\end{definition}

The subject reduction argument shall need the following auxiliary lemma.

\begin{lemma}[Multi-set splitting]
	\label{l:multiset-splitting} % \reflemmap{multiset-splitting}{env}
	\hfill
	\begin{enumerate}
		\item \emph{Terms}: let $\tyd \pof \wtjudg{\tye}{\weight}{\tm}{\mty^{k}\uplus\mtytwo^{k}}$ with $k= 1+\size\tye$. Then there exist two type derivations $\tyd_{\mty} \pof \wtjudg{\tye_{\mty}}{\weight_{\mty}}{\tm}{\mty^{k}}$ and $\tyd_{\mtytwo} \pof \wtjudg{\tye_{\mtytwo}}{\weight_{\mtytwo}}{\tm}{\mtytwo^{k}}$ such that $\tye_{\mty} \# \tye_{\mtytwo}$, $\tye = \tye_{\mty} \uplus \tye_{\mtytwo}$, $\size\tyd = \size{\tyd_{\mty}} + \size{\tyd_{\mtytwo}}$ and $\weight = \max\set{\weight_{\mty},\weight_{\mtytwo}}$.
		
		\item \emph{Closures}: let $\tyd \pof \wtjudg{}{\weight}{\clos}{\mty^{k}\uplus\mtytwo^{k}}$. Then there exist two type derivations $\tyd_{\mty} \pof \wtjudg{}{\weight_{\mty}}{\clos}{\mty^{k}}$ and $\tyd_{\mtytwo} \pof \wtjudg{}{\weight_{\mtytwo}}{\clos}{\mtytwo^{k}}$ such that $\size\tyd = \size{\tyd_{\mty}} + \size{\tyd_{\mtytwo}}$ and $\weight = \max\set{\weight_{\mty},\weight_{\mtytwo}}$.
		
		\item \label{p:multiset-splitting-env}
		\emph{Environments}: let $\tye$ and $\tyetwo$ summable and $\tyd \pof \wtjudg{}{\weight}{\lenv}{\tye\uplus\tyetwo}$. Then there exist two type derivations $\tyd_{\tye} \pof \wtjudg{}{\weight_{\tye}}{\lenv|_{\dom\tye}}{\tye}$ and $\tyd_{\tyetwo} \pof \wtjudg{}{\weight_{\tyetwo}}{\lenv|_{\dom\tyetwo}}{\tyetwo}$ such that $\size\tyd = \size{\tyd_{\tye}} + \size{\tyd_{\tyetwo}}$ and $\weight = \max\set{\weight_{\tye},\weight_{\tyetwo}}$.
	\end{enumerate}
\end{lemma}

% !TeX spellcheck = en_US
% !TEX root = ../main.tex
%%%%%%%%%%%%%%
\begin{proof}
	\hfill
	\begin{enumerate}
		\item Suppose that $\mty^{k} = \emmset^{k}$. Then $\mty^{k}\uplus\mtytwo^{k} = \mtytwo^{k}$. Now, let $\tye_{\mty}$ be defined as the unique dry type context such that $\dom{\tye_{\mty}} = \dom\tye$ and it is summable with $\tye$---note that necessarily $\size\tye = \size{\tye_{\mty}}$. By \reflemma{tye-invariant}, $\dom\tye = \fv\tm$, and so we obtain the following derivation:
		\[\infer[\tynone]{\wtjudg{\tye_{\mty}}{0}{\tm}{\emmset^{1+\size{\tye_{\mty}}}}}{}\]
		which is the $\tyd_{\mty}$ of the statement. We also take $\tyd_{\mtytwo} \defeq \tyd$, and the statement holds. 
		
		If $\mtytwo^{k} = \emmset^{k}$ the proof is as in the previous case.
		
		Assume now that both $\mty^{k}$ and $\mtytwo^{k}$ are non-empty, say $\mty^{k}= \mset{\linty_{1},\ldots,\linty_{m}}^{k}$ and $\mtytwo^{k}= \mset{\linty_{m+1},\ldots,\linty_{n}}^{k}$. Then $\tyd$ has the following shape:
		\[\infer[\tymany]{\wtjudg{\uplus_{i=1}^n\tye_i}{\max_{i}\set{v_{i}}}{\tm}{\mset{\linty_1,\mydots,\linty_n}^{1+\size{\uplus_{i=1}^n\tye_i}}}}
		{\wtjudg{\tye_i}{v_{i}}{\tm}{\linty_i} & 1\leq i\leq n & \#_i\tye_{i}}\]
		and the two derivations $\tyd_{\mty}$ and $\tyd_{\mtytwo}$ are obtained as follows
		\[\tyd_{\mty}\defeq\ \ \ \infer[\tymany]{\wtjudg{\uplus_{i=1}^m\tye_i}{\max_{i}\set{v_{i}}}{\tm}{\mset{\linty_1,\mydots,\linty_m}^{1+\size{\uplus_{i=1}^m\tye_i}}}}
		{\wtjudg{\tye_i}{v_{i}}{\tm}{\linty_i} & 1\leq i\leq m & \#_i\tye_{i}}\]
		and
		\[\tyd_{\mtytwo}\defeq\ \ \ \infer[\tymany]{\wtjudg{\uplus_{i=m+1}^n\tye_i}{\max_{i}\set{v_{i}}}{\tm}{\mset{\linty_{m+1},\mydots,\linty_n}^{1+\size{\uplus_{i=m+1}^n\tye_i}}}}
		{\wtjudg{\tye_i}{v_{i}}{\tm}{\linty_i} & m+1\leq i\leq n & \#_i\tye_{i}}\]
		which clearly satisfy the statement.
		\item If $\clos = (\tm,\lenv)$ then $\tyd$ has the following form 
		\[\infer[\tyclos]{\wtjudg{}{\max\{\weight,\weighttwo\}}{(\tm,\lenv)}{\mty^{k}}}{ 
			\wtjudg{}{\weight}{\lenv}{\tye} 
			&& 
			\wtjudg{\tye}{\weighttwo}{\tm}{\mty^{k}} }\]
		Then the \ih (point 1) applied to the right premise gives two derivations for $\tm$ with respect to $\mty^{k}$ and $\mtytwo^{k}$. In particular it gives two summable type contexts $\tye_{\mty}$ and $\tye_{\mtytwo}$ with which one applies the \ih (point 3) to the left premise, obtaining two derivations for $\lenv$ with respect to $\tye_{\mty}$ and $\tye_{\mtytwo}$. Pairing the respective derivations for $\tm$ and $\lenv$ one obtains the statement.
		
		\item If $\lenv$ is empty then both $\tye$ and $\tyetwo$ are empty and the statement trivially holds. Otherwise, it follows from applying the \ih (point 2) for each variable in $\dom\tye \cap \dom\tyetwo$.\qedhere
	\end{enumerate}
\end{proof}
\begin{restatable}[Soundness]{theorem}{soundness}\label{thm:soundness}
	Let $\state$ be a \SpKAM reachable state such that there is a derivation $\tyd\pof\tjudgw{}{\weight}{\state}{\initty}$. Then: 
	\begin{enumerate}
		\item \emph{Termination}: there is a run $\run:\state\rightarrow_{\mathrm{Sp\KAM}}^*\state_{f}$ to a final state. Moreover,
		\item \emph{Space bound}: $\weight=\lm\run$.	
	\end{enumerate}
\end{restatable}

% !TeX spellcheck = en_US
% !TEX root = ../main.tex
%%%%%%%%%%%%%%
\begin{proof}
	The proof is by induction on $\size\tyd$ and case analysis on whether $\state$ is final. If $\state$ is final then the run $\run$ in the statement is given by the empty run. Therefore, we have $\lm\run = \size\state$ and the required space bound becomes $\weight= \size\state$, which is given by \refprop{weight-of-final-states-is-right}. If $\state$ is not final then $\state\tospkam\statetwo$. By examining all the transition rules. We set $\linty\defeq\arr{\mty_1^{k_{1}}}{\arr{\cdots}{\arr{\mty_n^{k_{n}}}{\initty}}}$ and $\stack\defeq\stack_1\cdots\stack_n$. We report just one case that illustrates the methodology. The other ones can be found in \begin{LONG} Appendix~\ref{app:correctness}\end{LONG}\begin{SHORT}the associated technical report~\cite{ICFP2022TR}\end{SHORT}.
	
	Case $\tokamseav$, \ie
		$\state=\spkamstate{\tm\var}{\lenv}\stack$ and 
		$\statetwo=\spkamstate\tm{ 
			\lenv|_\tm}{\lenv(\var)\cdot\stack}$. The type derivation $\tyd$ typing  $\state$ has the following shape:
		\[
		\infer[\tystate]{\wtjudg{}{\max_i\{w',u,v_i\}}{\spkamstate{\tm\var}\lenv\stack}{\initty}}{
			\infer[\tyapp_{2}]{\wtjudg{\tye\uplus\var:\mty}{w'}{\tm\var}{\linty} }{\wtjudg{\tye}{w'}{\tm}{\arr{\mty}{\linty}}}
			&
			\tyd_{\lenv}\pof\wtjudg{}{u}{\lenv}{\tye\uplus\var:\mty}
			&
			\wtjudg{}{v_i}{\stack_i}{\mty_i^{k_{i}}} 
		}
		\]
		with $\weight = \max_i\{w',v_i,u\}$. By \reflemma{multiset-splitting}, there are two derivations $\tyd_{\tye}\pof\wtjudg{}{\weightthree_{\tye}}{\lenv|_{\dom\tye}}{\tye}$ and $\tyd_{\var:\mty}\pof\wtjudg{}{\weightthree_{\var:\mty}}{\lenv(\var)}{\var:\mty}$ such that $\size{\tyd_{\lenv}} = \size{\tyd_{\tye}} + \size{\tyd_{\var:\mty}}$ and $\weightthree = \max\set{\weightthree_{\tye}, \weightthree_{\var:\mty}}$. By the environment domain invariant (\reflemma{spkam-invariants}), $\dom\tye = \fv\tm$. Moreover, $\tyd_{\var:\mty}$ is necessarily the conclusion of a unary $\tyenv$ rule of premise $\tyd_{\mty}\pof\wtjudg{}{\weightthree_{\mty}}{\lenv(\var)}{\mty}$ for which $\size{\tyd_{\mty}} = \size{\tyd_{\var:\mty}}$ and $\weightthree_{\mty} = \weightthree_{\var:\mty}$.
		Then, $\statetwo$ can be typed by the following derivation $\tydtwo$:
		\[\tydtwo \defeq 
		\infer[\tystate]{\wtjudg{}{\max_i\{w',u_{\tye},u_{\mty},v_i\}}{\spkamstate\tm{ 
					\lenv|_\tm}{\lenv(\var)\cdot\stack}}{\initty}}{
			\wtjudg{\tye'}{w'}{\tm}{\arr{\mty}{\linty}}
			&
			\tyd_{\tye}\pof \wtjudg{}{u_{\tye}}{\lenv|_\tm}{\tye}
			&
			\tyd_{\mty}\pof \wtjudg{}{u_{\mty}}{\lenv(\var)}{\mty}
			&
			\wtjudg{}{v_i}{\stack_i}{\mty_i^{k_{i}}} 
		}
		\]
		Since $\size{\tyd_{\lenv}} = \size{\tyd_{\tye}} + \size{\tyd_{\var:\mty}} = \size{\tyd_{\tye}} + \size{\tyd_{\mty}}$, we have $\size\tyd > \size\tydtwo$ (because the $\tyapp_{2}$ rule is removed), and so we can apply the \ih, obtaining a run $\runtwo: \statetwo \rightarrow_{\mathrm{Sp\KAM}}^* \state_{f}$ to a final state such that $\max_i\{w',u_{\tye},u_{\mty},v_i\} = \lm\runtwo$. Then there is a run $\run:\state \rightarrow_{\mathrm{Sp\KAM}}^* \state_{f}$, proving the first part of the statement (termination).
		
		For the space bound, note that, since $\weightthree = \max\set{\weightthree_{\tye}, \weightthree_{\var:\mty}} = \max\set{\weightthree_{\tye}, \weightthree_{\mty}}$, we have $\weight = \max_i\{w',u_{\tye},u_{\mty},v_i\}$. Additionally, $\size\state \leq \size\statetwo$, because by the environment domain invariant $\lenv|_\tm$ removes at most $\lenv(\var)$ from $\lenv$, which is however added to the stack. Then $\lm\run = \lm\runtwo = \max_i\{w',u_{\tye},u_{\mty},v_i\} = \weight$, proving the second part of the statement.
		%%%%%%%%%%%%%%%%%%%
		%%%%%%%%%%%%%%%%%%%
\end{proof}

As a corollary, we can transfer back the soundness result from typed \emph{states} to typed \emph{terms}.

\begin{corollary}[Soundness, on terms]\label{cor:soundness}
	Let $\tm$ be a closed $\lambda$-term. If there exists $\tyd\pof\tjudgw{}{\weight}{\tm}{\initty}$, then there exists a complete \SpKAM run $\run$ from $\tm$ such that $\lm\run=w$.
\end{corollary}

\subsection{Completeness}
Completeness is the fact that all states on which the \SpKAM terminates are typable. Here the proof technique is standard: we show that final states are typable, that a subject expansion property holds, and then we infer completeness. We do not perform any space analysis, since it  already follows from the soundness part, once we know that a state is typable.

We need a preliminary notion of \emph{well-formed} state, environment, closure, and stack, defined inductively as follows:
\[
\begin{array}{c@{\hspace{1cm}}c@{\hspace{1cm}}cc}
	\infer{\wf{\stempty}}{} & \infer{\wf{\esub{\var}{\clos}\cdot \lenv}}{\wf{\clos} & \wf{\lenv}} & \infer{(\tm,\lenv)}{\fv{\tm}=\dom{\lenv} & \wf{\lenv}}
\end{array}
\]
\[\begin{array}{c@{\hspace{2cm}}c}
	\infer{\wf{\clos\cdot \stack}}{\wf{\clos} & \wf{\stack}}& \infer{\wf{(\tm,\lenv,\stack)}}{\wf{(\tm,\lenv)} & \wf{\stack}}
\end{array}
\]
This is just a predicate that certifies that a state satisfies the invariant corresponding to Lemma~\ref{l:spkam-invariants}. Clearly, we have the following proposition.
\begin{proposition}\label{prop:reachable-implies-wf} 
	Let $\state$ be a reachable \SpKAM state. Then $\wf{\state}$.
\end{proposition}

\paragraph*{Final states are typable.} To prove that every final state is typable, we need an auxiliary lemma ensuring that closures and environments can always be typed with empty types and dry type contexts.
\begin{lemma}[Closures and environments are typable]
	\label{l:exists-dry-env-clos} % \reflemmap{exists-dry-env-clos}{one}
	\hfill
	\begin{enumerate}
		\item \label{p:exists-dry-env-clos-one} 
		There exists a dry type context $\tye$ and a derivation $\tyd\pof \tjudg{}{\lenv}{\tye}$ for every environment $\lenv$.
		\item \label{p:exists-dry-env-clos-two} 
		There exist $k$ and a derivation $\tyd\pof \tjudg{}{\clos}{\emmset^{k}}$ for every closure $\clos$.
	\end{enumerate}
\end{lemma}

% !TeX spellcheck = en_US
% !TEX root = ../main.tex
%%%%%%%%%%%%%%
\begin{proof}
	By mutual induction on $\lenv$ and $\clos$.
	\begin{enumerate}
		\item If $\lenv$ is empty then $\tye$ is the empty type context. Otherwise, by \ih (point 2), for every closure $\lenv(\var)$ with $\var\in\dom\lenv$ there exists $k_{\var}$ and a derivation $\tyd_\var\pof \tjudg{}{\lenv(\var)}{\emmset^{k_{\var}}}$. Then $\tye$ is defined (only) on $\dom\lenv$ as $\tye(\var) \defeq \emmset^{k_{\var}}$, and $\tyd$ is defined as follows:
		\[
		\infer[\tyenv]{\tjudg{}{\lenv}{\tye}}{\tyd_\var\pof\tjudg{}{\lenv(\var)}{\tye(\var)}}\]
		
		\item Let $\clos = (\tm,\lenv)$. By the environments domain invariant (\reflemma{spkam-invariants}), $\dom\lenv = \fv\tm$. By \ih (point 1), there exists a dry type context $\tye$ and a derivation $\tyd_\lenv\pof \tjudg{}{\lenv}{\tye}$. Since $\dom\lenv = \fv\tm$, we can  apply rule $\tynone$ deriving a judgment $\tjudg{\tye}{\tm}{\emmset}$. Then $\tyd$ is defined as follows:
		\[\infer[\tyclos]{\tjudg{}{(\tm,\lenv)}{\emmset^{1+\size\tye}}}{ 
			\infer[\tynone]{\tjudg{\tye}{\tm}{\emmset^{1+\size\tye}}}{}
			&& 
			\tyd_\lenv\pof\tjudg{}{\lenv}{\tye} 
			 }\qedhere\]
	\end{enumerate}
\end{proof}

\begin{proposition}[Final states are typable]
	\label{prop:final-states-are-typable}
	Let $\state$ be a final state. Then there exists a type derivation $\tyd\pof \tjudg{}{\state}{\initty}$.
\end{proposition}

% !TeX spellcheck = en_US
% !TEX root = ../main.tex
%%%%%%%%%%%%%%
\begin{proof}
	Final states have the shape $\spkamstate{\la\var\tm}\lenv\stempty$. By \reflemmap{exists-dry-env-clos}{one}, there is a type derivation $\tjudg{}{\lenv}{\tye}$ with $\tye$ dry. By the environments domain invariant (\reflemma{spkam-invariants}), $\dom\lenv = \fv{\la\var\tm}$. Then $\tyd$ is defined as follows:
	\[
	\infer[\tystate]{\tjudg{}{\spkamstate{\la\var\tm}\lenv\stempty}{\initty}}{
		\infer[\tylamstar]{\tjudg{\tye}{\lambda\var.\tm}{\initty}}{
			\dom\tye = \fv{\la\var\tm} & \tye \mbox{ is dry}}
		&
		\tjudg{}{\lenv}{\tye}
	}\qedhere
	\]
\end{proof}

\paragraph*{Subject expansion.} The proof of subject expansion is standard. There is only one delicate point, in expanding the application transitions $\tokamseanv$ and $\tokamseav$, where one needs to ensure that the two type contexts for the premises of the application are summable.
\begin{proposition}[Subject expansion]
	\label{prop:sub-exp}
	If $\state\tospkam\statetwo$, where $\wf{\state}$, and there exists 
	$\tydtwo\pof\tjudg{}\statetwo\initty$, then there exists 
	$\tyd\pof\tjudg{}\state\initty$.
\end{proposition}

% !TeX spellcheck = en_US
% !TEX root = ../main.tex
%%%%%%%%%%%%%%
\begin{proof}
	There are three cases that require something more than simply reading backwards the subject reduction argument in the proof of the correctness theorem. One is $\tokambw$, where in addition one needs to type the garbage collected closure, but this is ensured by \reflemmap{exists-dry-env-clos}{one}. The other two are the application transitions $\tokamseanv$ and $\tokamseav$. We treat $\tokamseanv$, the case of $\tokamseav$ is analogous. We have $\state=\spkamstate{\tm\tmtwo}{\lenv}{\clos_1\cdots\clos_n} \tokamseanv \spkamstate\tm{ 
		\lenv|_\tm}{(\tmtwo,\lenv|_\tmtwo)\cdot\clos_1\cdots\clos_n} = \statetwo$.
	The type derivation $\tydtwo$ typing  $\statetwo$ has the following shape:
	\[\tydtwo = 
	\infer[\tystate]{\tjudg{}{\spkamstate\tm{ 
				\lenv|_\tm}{(\tmtwo,\lenv|_\tmtwo)\cdot\clos_1\cdots\clos_n}}{\initty}}{
		\tjudg{\tye}{\tm}{\arr{\mty^{k}}{\linty}}
		&
		\tjudg{}{\lenv|_\tm}{\tye}
		&
		\infer[\tyclos]{\tjudg{}{(\tmtwo,\lenv|_\tmtwo)}{\mty^k}}{\tjudg{\tyetwo}{\tmtwo}{\mty^{k}}&&
			\tjudg{}{\lenv|_\tmtwo}{\tyetwo}}
		&
		\tjudg{}{\clos_i}{\mty_i^{k_{i}}} 
	}
	\]
	Assuming that $\tye \#\tyetwo$, that we shall prove below, the type derivation $\tyd$ for $\state$ is given by:	
	\[
	\tyd \defeq 
	\infer[\tystate]{\tjudg{}{\spkamstate{\tm\tmtwo}\lenv{\clos_1\cdots\clos_n}}{\initty}}{
		\infer[\tyapp_{1}]{\tjudg{\tye\uplus\tyetwo}{\tm\tmtwo}{\linty
		} }{\tjudg{\tye}{\tm}{\arr{\mty^{k}}{\linty}}
			& 
			\tjudg{\tyetwo}{\tmtwo}{\mty^{k}}& \tye\# \tyetwo}
		&
		\tjudg{}{\lenv}{\tye\uplus\tyetwo}
		&
		\tjudg{}{\clos_i}{\mty_i^{k_{i}}} 
	}
	\]
	where the derivation for $\tjudg{}{\lenv}{\tye\uplus\tyetwo}$ is obtained by an omitted and straightforward \emph{multi-sets merging} lemma dual to the multi-sets splitting lemma (\reflemma{multiset-splitting}) used for correctness.
	
	We now prove $\tye \#\tyetwo$, which is the additional bit not present in the proof of subject reduction. Let $\var \in \dom\tye \cap \dom\tyetwo$. Then $\lenv_{\tm}(\var) = \lenv_{\tmtwo}(\var) = \lenv(\var) = \clos$ for some closure $\clos$. By \reflemmap{env-clos-size}{two}, we have both $\tye(\var) = \mtytwo_{1}^{\size\clos}$ and $\tyetwo(\var) = \mtytwo_{2}^{\size\clos}$ for some multi types $\mtytwo_{1}$ and $\mtytwo_{2}$, that is, $\tye \#\tyetwo$.
\end{proof}

\begin{theorem}[Completeness]
	Let $\state$ a reachable state and $\run:\state\rightarrow_{\mathrm{Sp\KAM}}^*\state_{f}$ be a \SpKAM run to a final state $\state_{f}$. Then there exist $\tyd\pof\tjudg{}{\state}{\initty}$.	
\end{theorem}

% !TeX spellcheck = en_US
% !TEX root = ../main.tex
%%%%%%%%%%%%%%
\begin{proof}
	By induction on the length $\size\run$ of $\run$, and notice that all states in $\run$ are reachable and thus well-formed (Prop.~\ref{prop:reachable-implies-wf}). If $\size\run = 0$ then $\state = \state_{f}$, and the existence of a typing derivation for $\state$ is given by \refprop{final-states-are-typable}. If $\size\run > 0$ then $\run$ is given by a transition $\state\tospkam\statetwo$ followed by an execution $\runtwo: \statetwo \rightarrow_{\mathrm{Sp\KAM}}^* \state_{f}$. By \ih, we obtain $\tydtwo\pof\tjudg{}{\statetwo}{\initty}$, and by subject expansion (\refprop{sub-exp}) we obtain $\tyd\pof\tjudg{}{\state}{\initty}$.
\end{proof}

Again, as a corollary, we can transfer the completeness property from \emph{states} to \emph{terms}.
\begin{corollary}[Completeness, on terms]
	If $\tm$ is a closed $\lambda$-term such that there exists a complete \SpKAM run from $\tm$, then there exists $\tyd\pof\tjudg{}{\tm}{\initty}$.
\end{corollary}

Putting together quantitative soundness and completeness gives the quantitative correctness of the closure-type system.
\begin{theorem}[Correctness]\label{thm:correctness}
	Let $\tm$ be a closed $\lambda$-term. Then there exists a complete \SpKAM run $\run$ from $\tm$ such that $\lm\run=w$ if and only if there exists $\tyd\pof\tjudgw{}{\weight}{\tm}{\initty}$.
\end{theorem}

% !TeX spellcheck = en_US
% !TEX root = main.tex
%%%%%%%%%%%%%%%%%%%%%%%%%%%%%%%%%%%%%%%%%%%%%%%%%%%%%%%%%%%%%%%%%%%%%%%%%%%%%%%%
\section{Capturing the \SpKAM low-level time}\label{sec:time}
\paragraph*{Low-level time.} \citet{LICS2022} show that the \SpKAM run on a closed 
term $\tm$ cannot be implemented on Turing machines with an overhead in time that 
is polynomial in the number $n$ of weak head reduction steps from $\tm$ (which is 
the natural cost model for time in \ccbn), nor in the number of \SpKAM steps 
(which is linear in $n$). This is because of the lack of sharing of the \SpKAM 
data structures. Indeed, in transitions $\tokamseanv$ and $\tokamseanv$ the 
environment, of a priori unbounded size, can be duplicated. Of course, this copy 
operation cannot take a constant amount of time, and---as \citet{LICS2022} 
shows---it can actually lead to an exponential overhead.

It is however possible to adopt another more low-level time measure, the \emph{real} implementation time, that is, the time actually taken by the \SpKAM to run. One transition does not count anymore for one unit of time, but we have to take into account the amount of time needed to manage the involved data structures. Since we do not rest on any sharing mechanism, roughly the time consumed by any transition is proportional to the sizes of the involved states\footnote{Since we are interested in measuring time up to a \emph{polynomial} overhead, we are allowed to make some simplifications. For example, in this case, we do not take into account the size of the input pointers, because it can be absorbed in the polynomial overhead.}. Thus, in analogy to what we have done for space we define the low-level time consumption of the \SpKAM in the following way.

\begin{definition}[Low-level run time]
	Let $\run:\compil{\tm_0}\rightarrow^*_{\mathrm{Sp\KAM}}\state$ be a \SpKAM run. Then the low-level time consumption of the run $\run$ is defined as follows:
	\[\lmt{\run}\defeq\sum_{\statetwo\in\run}\size{\statetwo}\]
\end{definition}
Although this low-level time measure can be exponential in the number of $\beta$-steps a term needs to normalize, we have the following reasonability result.\footnote{Intuitively, reasonability holds because in the simulation of Turing machines (the reasonable reference model), \ie in the execution of the encoding of Turing machines into the $\l$-calculus, the exponential gap does not show up.}
\begin{theorem}[\cite{LICS2022}, The \SpKAM is reasonable for low-level time]
	Closed CbN evaluation $\towh$ and the low-level time of the Space KAM $\lmt{\cdot}$ provide a reasonable time cost model for the $\lambda$-calculus.
\end{theorem}

 \paragraph*{The new weighting system.} We now turn to the definition of a weighted type system that is able to capture the \SpKAM low-level time. Actually, the type system is the same of section~\ref{sec:type-system}, what changes is just the weight assignment, that can be found in \reffig{typing-rules-time}. Essentially, all the $\max$'s have been turned into sums. More precisely, one can observe that in every rule (except for $\tynone$ and $\tymany$), the weight $\weight$ of the conclusion with judgment $\tjudgw{\tye}{\weight}{\tm}{\linty}$ is exactly the sum of all the weights of the premises \emph{plus} $\size\tye$ and $\size\linty$, \ie the size of the \SpKAM state corresponding to the judgment in the conclusion. 

\paragraph*{Soundness.} The proof of soundness proceeds as the one in Section~\ref{sec:correctness}. Also in this case we have to extend the type system to the machine components (in \reffig{typing-rules-time}), so that subject reduction/soundness is proved on states, rather than terms.
We need a preliminary lemma stating that the weight of judgments typing empty closure and dry environment is zero. 

\begin{lemma}[Empty types and dry type environments induce null weight]\label{l:dry-env-clos-time}
	\hfill
	\begin{enumerate}
		\item\label{p:dry-env-clos-time-one} If $\tyd\pof \wtjudg{}{\weight}{\lenv}{\tye}$ and $\tye$ is dry, then $\weight = 0$.
		\item\label{p:dry-env-clos-time-two} If $\tyd\pof \wtjudg{}{\weight}{\clos}{\emmset^k}$, then $\weight = 0$.
	\end{enumerate} % \reflemmap{dry-env-clos}{one}
	
\end{lemma}

\begin{proof}
	By mutual induction on $\lenv$ and $\clos$.
	\begin{enumerate}
		\item $\tyd$ has the following form:
		\[
		\infer[\tyenv]{\wtjudg{}{\sum\set{\weight_\var|\var\in\dom\lenv}}{\lenv}{\tye}}{\wtjudg{}{\weight_\var}{\lenv(\var)}{\tye(\var)}}\]
		with $\weight =\sum\set{\weight_\var|\var\in\dom\lenv}$. Two cases:
		\begin{enumerate}
			\item $\lenv$ is empty, \ie $\lenv = \stempty$: then $\weight = 0$ and $\tye$ is the empty type environment, for which $\size\tye = 0$. Therefore, we have $\weight = 0$, validating the statement.
			
			\item $\lenv$ is not empty, \ie $\lenv \neq \stempty$:
			Since $\tye$ is dry,  then $\wtjudg{}{\weight_\var}{\lenv(\var)}{\emmset^{k_{\var}}}$ for some $k_{\var}$. By \ih (point 2), $\weight_\var=0$ and thus $\weight =\sum\set{\weight_\var|\var\in\dom\lenv}=0$.
		\end{enumerate}
		\item   Let $\lenv(\var)\defeq(\tm,\lenv)$. Then $\tyd$ has the following form: 
		
		\[
		\infer[\tyclos]{\wtjudg{}{0+v}{(\tm,\lenv)}{\emmset^{k_{\var}}}}{
			\infer[\tynone]{\wtjudg{\tyetwo}{0}{\tm}{\emmset^{k_{\var}}}}{\tyetwo \text{ is dry}} && \wtjudg{}{v}{\lenv}{\tyetwo}}
		\]
		
		By \ih (point 1) $v=0$. Thus $\weight=v=0$.\qedhere
	\end{enumerate}
\end{proof}

% !TeX spellcheck = en_US
% !TEX root = main.tex
%%%%%%%%%%%%%%%%%%%%%%%%%%%%%%%%%%%%%%%%%%%%%%%%%%%%%%%%%%%%%%%%%%%%%%%%%%%%%%%%
\begin{figure}[t]
	{\small\[
	\begin{array}{c@{\hspace{0.6cm}}c}
	\infer[\tyvar]{\wtjudg{\var:\mset{\linty}^k}{k+\size{\linty}}{\var}{\linty}}{}
	&
	
	\infer[\tylamstar]{\wtjudg{\tye}{\size\tye}{\lambda\var.\tm}{\initty}}{
		\dom\tye = \fv{\la\var\tm} & \tye \mbox{ is dry}}
	\\[10pt]
	\infer[\tylam_{1}]{\wtjudg{\tye}{w+\size\tye+\size\linty+k}{\lambda\var.\tm}{\arr{\mty^{k}}{\linty}}} 
	{\wtjudg{\tye,\var:\mty^{k}}{w}{\tm}{\linty}}
	&
	\infer[\tylam_{2}]{\wtjudg{\tye}{w+\size\tye+\size\linty+k}{\lambda\var.\tm}{\arr{\emmtype^{k}}{\linty}}}
	{\wtjudg{\tye}{w}{\tm}{\linty} & \var\notin\dom\tye}
	\\[10pt]
	\infer[\tymany]{\wtjudg{\uplus_{i=1}^n\tye_i}{\sum_{i}\set{v_{i}}}{\tm}{\mset{\linty_1\mydots\linty_n}^{1+\size{\uplus_{i=1}^n\tye_i}}}}
	{\wtjudg{\tye_i}{v_{i}}{\tm}{\linty_i} & 1\leq i\leq n & \#_i\tye_{i}}
	& \infer[\tynone]{\wtjudg{\tye}{0}{\tm}{\emmset^{1+\size\tye}}}{\dom\tye = \fv{\tm} & \tye  \mbox{ is dry}}
	\\[10pt]
	
	\infer[\tyapp_{1}]{\wtjudg{\tye\uplus 
			\tyetwo}{w+v+\size{\tye\uplus\tyetwo}+\size\linty}{\tm\tmtwo}{\linty 
	}}{\wtjudg{\tye}{w}{\tm}{\arr{\mty^{k}}{\linty}}
		& 
		\wtjudg{\tyetwo}{v}{\tmtwo}{\mty^{k}}& \tye\# \tyetwo & \tmtwo\not\in\mathcal{V}} 
	&
	
	%	\multicolumn{3}{c}{\infer[\tyapp]{\wtjudg{\tye\uplus 
	%	\sum_{i=1}^n \tyetwo_i  }{\max\{w,v_i\}}{\tm\tmtwo}{\linty 
	%		}}{\wtjudg{\tye}{w}{\tm}{\arr{\mset{\lintytwo_1,\ldots,\lintytwo_n}^{\size{
	%		  \tyetwo}+1}}{\linty}}
	%			& 
	%			\mset{\wtjudg{\tyetwo_i}{v_i}{\tmtwo}{\lintytwo_i}}_{i\in\mset{1,\ldots,n}}}}
	%		\\[10pt]
	%
	%		\multicolumn{3}{c}{\infer[\tyapp]{\wtjudg{\tye 
	%		}{w}{\tm\tmtwo}{\linty 
	%				}}{\wtjudg{\tye_{|\tm}}{w}{\tm}{\arr{\mset{\cdot}^{\size{
	%							\tye_{|\tmtwo}}+1}}{\linty}}}}
	%	\\[10pt]
	\infer[\tyapp_{2}]{\wtjudg{\tye\uplus
			\var:\mty^k  
		}{w+\size\tye+k+\size\linty}{\tm\var}{\linty 
	}}{\wtjudg{\tye}{w}{\tm}{\arr{\mty^k}{\linty}}
		&&\tye\#\var:\mty^k}
	\\\\
	\hline
	\\
	\infer[\tyenv]{\wtjudg{}{\sum\set{\weight_\var|\var\in\dom\lenv}}{\lenv}{\tye}} 
	{\wtjudg{}{\weight_\var}{\lenv(\var)}{\tye(\var)}&&\forall\var\in\dom\lenv}
	&
	\infer[\tyclos]{\wtjudg{}{\weight+\weighttwo}{(\tm,\lenv)}{\mty^{k}}}{ 
		\wtjudg{}{\weight}{\lenv}{\tye} 
		&& 
		\wtjudg{\tye}{\weighttwo}{\tm}{\mty^{k}} }
	\\\\
	\multicolumn{2}{c}{
		\infer[\tystate]{\wtjudg{}{w+u+\sum_i v_i}{\spkamstate\tm\lenv\stack}{\initty}}{
			\wtjudg{\tye}{w}{\tm}{\arr{\mty_1^{k_{1}}}{\arr{\cdots}{\arr{\mty_n^{k_{n}}}{\initty} } } } 
			&
			\wtjudg{}{u}{\lenv}{\tye}
			&
			\wtjudg{}{v_i}{\stack_i}{\mty_i^{k_{i}}} 
		}
	}
	\end{array}
	\]}
	
	\caption{Typing rules with the low-level time weights for terms (above) and machine components (below).}
	\label{fig:typing-rules-time}
\end{figure}

\paragraph*{Time analysis of final states.} We need an auxiliary lemma that allows us to prove that the weight system correctly measures the size of final states.

\begin{lemma}[The weight of final states is their low-level time]
	\label{l:weight-of-final-states-is-right-time}
	Let $\state$ be a final state and $\tyd\pof \tjudgw{}{w}{\state}{\initty}$. Then $w = \size\state$.
\end{lemma}
\begin{proof}
	Final states have the shape $\spkamstate{\la\var\tm}\lenv\stempty$. Then $\tyd$ has the following shape:
	\[
	\infer[\tystate]{\wtjudg{}{\weighttwo+\size\tye}{\spkamstate{\la\var\tm}\lenv\stempty}{\initty}}{
		\infer[\tylamstar]{\wtjudg{\tye}{\size\tye}{\lambda\var.\tm}{\initty}}{
			\dom\tye = \fv{\la\var\tm} & \tye\mbox{ is dry}}
		&
		\wtjudg{}{\weighttwo}{\lenv}{\tye}
	}
	\]
	with $\weight = \weighttwo+\size\tye$. By \reflemmap{dry-env-clos-time}{one}, $\weighttwo=0$. Then $\weight =_{\reflemmap{env-clos-size}{one}} \size\lenv =\size\state$.
\end{proof}
Then we are able to state the soundness theorem, this time for the low-level 
time weighting system.\footnote{We have already proved the soundness of the 
type system w.r.t. termination, which does not depend on the weighting 
system. This is why termination is not part of the statement.}
\begin{restatable}[Low-level time soundness]{theorem}{timesoundness}
	Let $\state$ be a \SpKAM reachable state such that there is a derivation $\tyd\pof\tjudgw{}{\weight}{\state}{\initty}$. Then the run $\run:\state\rightarrow_{\mathrm{Sp\KAM}}^*\state_{f}$ to the final state is such that
	$\weight=\lmt\run$.
\end{restatable}

% !TeX spellcheck = en_US
% !TEX root = ../main.tex
%%%%%%%%%%%%%%
\begin{proof}
	The proof is by induction on $\size\tyd$ and case analysis on whether $\state$ is final. If $\state$ is final then the run $\run$ in the statement is given by the empty run, containing only $\state$. Therefore, we have $\lmt\run = \size\state$ and the required time bound becomes $\weight= \size\state$, which is given by \reflemma{weight-of-final-states-is-right-time}. If $\state$ is not final then $\state\tospkam\statetwo$. By examining all the transition rules. We set $\linty\defeq\arr{\mty_1^{k_{1}}}{\arr{\cdots}{\arr{\mty_n^{k_{n}}}{\initty}}}$ and $\stack\defeq\stack_1\cdots\stack_n$.
	We report just one case to illustrate the methodology, the other ones can be found in \begin{LONG}Appendix~\ref{app:time}\end{LONG}\begin{SHORT}the associated technical report~\cite{ICFP2022TR}\end{SHORT}.
	
	Case $\tokamsub$, \ie $\state=\spkamstate\var
		{\esub\var{(\tmtwo,\lenv)}}\stack$ and
		$\statetwo=\spkamstate\tmtwo\lenv\stack$. The type derivation $\tyd$ typing  $\state$ has the following shape:
		
		\[
		\infer[\tystate]{\wtjudg{}{k+\size\linty+u+w'+\sum_i v_i}{\spkamstate\var
				{\esub\var{(\tmtwo,\lenv)}}\stack}{\initty}}{
			\infer[\tyvar]{\wtjudg{\var:\mset{\linty}^k}{k+\size{\linty}}{\var}{\linty}}{} 
			&
			\infer[\tyenv]{\wtjudg{}{u+w'}{\esub\var{(\tmtwo,\lenv)}}{\var:\mset\linty^k}} {
				\infer[\tyclos]{
					\wtjudg{}{u+w'}{(\tmtwo,\lenv)}{\mset\linty^k}
				}{
					\wtjudg{\tye}{w'}{\tmtwo}{\mset\linty^k }
					&
					\wtjudg{}{u}{\lenv}{\tye} 
				} 
			}
			&
			\wtjudg{}{v_i}{\stack_i}{\mty_i^{k_{i}}} 
		}
		\]
		with $\weight = k+\size\linty+u+w'+\sum_i v_i$. The target state $\statetwo$ can be typed by the following derivation $\tydtwo$:
		\[\tydtwo\defeq
		\infer[\tystate]{\wtjudg{}{w'+u+\sum_i v_i}{\spkamstate\tmtwo\lenv\stack}{\initty}}{
			\wtjudg{\tye}{w'}{\tmtwo}{\linty } 
			&
			\wtjudg{}{u}{\lenv}{\tye}
			&
			\wtjudg{}{v_i}{\stack_i}{\mty_i^{k_{i}}} 
		}
		\]
		Since the $\tyvar$ rule is removed, we have $\size\tyd>\size\tydtwo$, and so we can apply the \ih, obtaining a run $\runtwo: \statetwo \rightarrow_{\mathrm{Sp\KAM}}^* \state_{f}$ to a final state such that $w'+u+\sum_i v_i = \lmt\runtwo$.
		
		For the time bound, note that $\lmt\run = \size\state+\lmt\runtwo=w$.
\end{proof}

As a corollary, we can transfer back the soundness result from typed \emph{states} to typed \emph{terms}.

\begin{corollary}[Low-level time soundness, on terms]
	Let $\tm$ be a closed $\lambda$-term. If there exists $\tyd\pof\tjudgw{}{\weight}{\tm}{\initty}$, then there exists a complete \SpKAM run $\run$ from $\tm$ such that $\lmt\run=w$.
\end{corollary}

Putting together low-level time soundness and the completeness result of the previous section (which does not rests on the weighting system) one has the low-level time  correctness of the time variant of the closure-type system.
\begin{theorem}[Low-level time correctness]
	Let $\tm$ be a closed $\lambda$-term. Then there exists a complete \SpKAM run $\run$ from $\tm$ such that $\lmt\run=w$ if and only if there exists $\tyd\pof\tjudgw{}{\weight}{\tm}{\initty}$.
\end{theorem}

\begin{example}
	For the sake of completeness, we provide the closure-type derivation of our example term $(\la\var(\la\vartwo(\la\varthree\var)(\var\vartwo))\var)\Id$ decorated with our new weighting system, characterizing low-level time.
	\[
	\infer[\tyapp_1]{\tjudgw{}{11} {(\la\var(\la\vartwo(\la\varthree\var)(\var\vartwo))\var)\Id}{\initty}}{
		\infer[\tylam_1]{\tjudgw{}{11}{\la\var(\la\vartwo(\la\varthree\var)(\var\vartwo))\var} {\arr{\mset\initty^1}\initty}}{
			\infer[\tyapp_2]{\tjudgw{\var:\mset\initty^1}{10}{(\la\vartwo(\la\varthree\var)(\var\vartwo))\var}{\initty}}{\infer[\tylam_1]{\tjudgw{\var:\mset\initty^1}{9}{\la\vartwo(\la\varthree\var)(\var\vartwo)} {\arr{\emmset^1}\initty}}{
					\infer[\tyapp_1]{\tjudgw{\var:\mset\initty^1,\vartwo:\emmset^1}{7}{(\la\varthree\var)(\var\vartwo)}{\initty}}{
						\infer[\tylam_2]{\tjudgw{\var:\mset\initty^1}{5}{\la\varthree\var}{\arr{\emmset^3}\initty}}{
							\infer[\tyvar]{\tjudgw{\var:\mset\initty^1}{1}{\var}{\initty}}{}} & \infer[\tynone]{\tjudgw{\var:\emmset^1,\vartwo:\emmset^1}{0}{\var\vartwo}{\emmset^3}}{}}}}} & \infer[\tymany]{\tjudgw{}{0}{\Id}{\mset\initty^1}}{
			\infer[\tylamstar]{\tjudgw{}{0}{\Id}{\initty}}{}}}
	\]
	The final weight, \ie the low-level time of the evaluation on the \SpKAM is 11. One may easily check that the same result holds if one looks directly at the  \SpKAM execution.
\end{example}

% !TeX spellcheck = en_US
% !TEX root = main.tex
%%%%%%%%%%%%%%%%%%%%%%%%%%%%%%%%%%%%%%%%%%%%%%%%%%%%%%%%%%%%%%%%%%%%%%%%%%%%%%%%
\section{Does the type system induce a model?}\label{sec:semantics}
Intersection type systems are well-known to provide a way to build models of $\lambda$-calculi, \eg filter models~\citep{DBLP:books/daglib/0032840} or relational models~\citep{DBLP:books/cp/BarendregtM22}.
In this section we show that the closure-type system we have presented in Section~\ref{sec:type-system} does \emph{not} give rise to a model, explaining the deep reason why this happens. Several definitions of models appear in the literature, more or less equivalent (categorical~\citep{barendregt_lambda_1984}, syntactical~\citep{Hindley1980-HINLMA-2}, algebraic~\citep{DBLP:journals/iandc/Meyer82}, etc.). The important common point is that from models one can derive $\l$-theories, \ie notions of equality between $\l$-terms. To define them, we first need the general notion of context, \ie $\lambda$-terms containing exactly 
one occurrence of a special symbol, the hole $\ctxhole$, that can appear anywhere and intuitively stands for a removed subterm.
\[
\begin{array}{rrcl}
	\textsc{Contexts}\quad\quad & \Cctx\ni\ctx & \grameq & 
	\ctxhole\midd \lambda x.\ctx \midd \ctx\tmtwo \midd \tm\ctx
\end{array}
\]
The operation of replacing the hole $\ctxhole$ with a term $\tm$ 
in a context $\ctx$, potentially capturing free variables, is noted $\ctxp\tm$ and called \emph{plugging}.
Relations on $\l$-terms are often defined at top-level and subsequently closed by any context, as follows.
\begin{definition}[Contextual closure]\label{def:ctxclosure}
	Let $\relsym\subseteq\Lambda\times\Lambda$ be a relation on $\l$-terms. The \emph{contextual closure of $\relsym$} is the least relation $\Cctxp{\relsym}$ such that $\tm\rel\tmtwo$ entails $\ctxp\tm \ \Cctxp{\relsym}\ \ctxp\tmtwo$, for all $\ctx\in\Cctx$. $\relsym$ is called \emph{compatible} if $\Cctxp{\relsym} \subseteq\relsym$.
\end{definition}
One typical example of context closure is the one at work in $\beta$-\emph{reduction} $\tob$, which is defined as the contextual closure $\tob\defeq \Cctxp{\mapsto_\beta}$ of the $\beta$-rule. Then, $\beta$-\emph{conversion} $=_\beta$ is defined as the reflexive, symmetric, and transitive closure of $\beta$-reduction $\tob$.

In the study of the semantics of programming languages, \emph{program equivalence} is certainly one of the main topics. In particular, we are interested in equating terms that \emph{behave the same}. What does that mean? Many definitions are possible, but the notion of $\l$-theory gives an axiomatic framework that is generally accepted to be the standard one.
\begin{definition}[$\l$-theory~\citep{DBLP:books/cp/BarendregtM22}]
	\begin{itemize}
		\item[]
		\item A relation $\relsym\subseteq\Lambda\times\Lambda$ is  a \emph{congruence} if it is a compatible equivalence.
		\item A relation $\relsym\subseteq\Lambda\times\Lambda$ is an \emph{equational theory}, or \emph{$\l$-theory}, if it is a congruence and contains $\beta$-conversion $=_\beta$, \ie $=_\beta\,\subseteq\, \relsym$.
	\end{itemize}
\end{definition}
Then, we simply define a model as an object which generates a $\l$-theory. We say that a list of pairwise distinct variables $\vec{x}\defeq\var_1,\ldots,\var_n$ is \emph{suitable} for a term $\tm$ when $\fv{\tm}\subseteq\{ \var_1,\ldots,\var_n \}$.
\begin{definition}[Model]
	Let $D$ be a set and $\sem{\cdot}_{(\cdot)}:\Lambda\to\mathcal{V}^*\to D$ an \emph{interpretation} function. Then $\mathcal{M}\defeq(D,\sem{\cdot}_{(\cdot)})$ is a \emph{model} if the equality $=_\mathcal{M}$ induced by $\mathcal{M}$, \ie $\tm=_\mathcal{M} \tmtwo$ when $\sem{\tm}_{\vec{x}}=\sem{\tmtwo}_{\vec{x}}$ for each $\vec{x}$ suitable for $\tm$ and $\tmtwo$, is a $\l$-theory.
\end{definition}

\paragraph*{The Indexed Relational Semantics Is Not a Model.} In relational models, we interpret terms using their types. In particular, the interpretation of a term is given by the set of types that can be assigned to it. In order for the semantics to be \emph{compositional}, we need to extend the type system to open terms, and to consider not only type derivations ending in $\initty$, but also in any type $\linty\in\lintys$. Formally, given a list of variables $\vec{x}\defeq\var_1,\ldots,\var_n$ which is suitable for $\tm$, the indexed relational semantics of $\tm$ for $\vec{x}$ is defined as:
\[
\sem{\tm}_{\vec{x}} \defeq \{\mty_1^{k_1}\to\ldots\to\mty_n^{k_n}\to\linty \mid \exists\, \tyd \pof \tjudg{\var_1:\mty_1^{k_1},\ldots,\var_n:\mty_n^{k_n}}{\tm}{\linty}\}
\]
We call $\mathcal{I}\defeq (\mathcal{P}(\lintys),\sem{\cdot}_{(\cdot)})$ the \emph{indexed relational model} and $=_{\mathcal{I}}$ the induced denotational equality.
Now, is $=_{\mathcal{I}}$ actually a $\l$-theory? The fact that $=_\mathcal{I}$ is an equivalence relation is straightforward. Does $=_\mathcal{I}$ contain $\beta$-conversion, though? No, as the next counterexample shows.
\begin{example} Let us consider the $\lambda$-term $(\la\var\mathsf{I})\vartwo$. Clearly, we have $(\la\var\mathsf{I})\vartwo=_\beta \mathsf{I}$. Now, by Lemma~\ref{l:tye-invariant}, we know that for each type judgment $\tjudg{\tye}{(\la\var\mathsf{I})\vartwo}{\linty}$, we have that $\dom{\tye}=\{y\}$, while for each type judgment $\tjudg{\tyetwo}{\mathsf{I}}{\lintytwo}$, we have that $\dom{\tyetwo}=\emptyset$. Then, there is no possibility of having $\sem{(\la\var\mathsf{I})\vartwo}_\vartwo=\sem{\mathsf{I}}_\vartwo$, implying that $=_\beta\,\not\subseteq\,=_{\mathcal{I}}$.
\end{example}
The (counter)example above shows that indeed $=_{\mathcal{I}}$ is \emph{not} 
a $\lambda$-theory. Why is it the case? The problem lies in the side 
conditions in rules $\tylamstar$ and $\tynone$, \ie in our use of 
weakenings. In fact, these together imply Lemma~\ref{l:tye-invariant}, which 
states that for each type judgment $\tjudg{\tye}{\tm}{\linty}$, we have that 
$\dom{\tye}=\fv{\tm}$. The culprit is that we would like $\tye$, and thus 
$\dom{\tye}$, to be stable under $\beta$-conversion, but clearly $\fv{\tm}$ is 
not, as free variables can be erased during reduction, as in the previous 
example. Moreover, notice that this problem emerges when considering 
\emph{open} terms, as in our example above.\footnote{Actually, in the case 
of open 
terms, even the definition of the \SpKAM is not correct. In particular, the 
rule $\tokamseav$ is not always well-defined anymore. We invite the reader 
to consider the term $(\la\var\mathsf{I})\vartwo$ to observe this behavior.} 
One possible solution could be to remove those side 
conditions and to allow for weakenings in all the axioms. While this tweak 
would provide a $\l$-theory, it would also break the connection with the 
\SpKAM, in particular the correspondence between the size of type 
environments and the size of the \SpKAM environments. We invite the reader 
to try to build the type derivation ending in 
$\tjudg{}{(\la\vartwo(\la\varthree\vartwo))\mathsf{I}}{\initty}$ and to 
observe that the subterm $\la\varthree\vartwo$ could be either typed as 
$\tjudg{\vartwo:\emmset^1}{\la\varthree\vartwo}{\initty}$ or as 
$\tjudg{}{\la\varthree\vartwo}{\initty}$. The corresponding \SpKAM state 
$(\la\varthree\vartwo,\esub\vartwo{(\mathsf{\Id},\stempty)}),\stempty)$ has 
indeed one entry in the environment, and thus the second typing does not 
respect the wanted correspondence between type judgments and states, and 
would thus invalidate Theorem~\ref{thm:correctness}.

% !TeX spellcheck = en_US
% !TEX root = main.tex
%%%%%%%%%%%%%%%%%%%%%%%%%%%%%%%%%%%%%%%%%%%%%%%%%%%%%%%%%%%%%%%%%%%%%%%%%%%%%%%%
\section{Counting different kind of pointers}\label{sec:split}
As we have already detailed in the introduction, the space measure on the Space KAM that we have used until this point, \ie the number of closures multiplied by the size of subterm pointers, is not enough to prove the space reasonability of the \SpKAM. Why this is the case is explained at length in~\citep{LMCS2024}. The intuition is that while executing on the \SpKAM the encoding $\overline{M}\,\overline{i}$ of a Turing machine $M$ on input $i$, one wants to distinguish pointers to (the subterms of the term $\overline{M}$ encoding) the machine $M$, which have constant size, from pointers to (the subterms of the term $\overline{i}$ encoding) the input $i$, which have size $\log(\size{i})$.

The aim of this section is to sketch how the type system of Section~\ref{sec:type-system} can be refined as to take into account different kinds of pointers, thus getting one step closer to providing the exact space measure used in \citep{LMCS2024}. For the sake of clarity, we consider just two kinds of pointers, that we identify using two colors, \red{red} ($\cred$) and \blue{blue} ($\cblue$). 

The idea is that we shall color $\overline{M}\,\overline{i}$ as $\red{\overline{M}}\,\blue{\overline{i}}$ and that the closures containing subterms from $\red{\overline{M}}$ shall be associated to \red{red} pointers while those containing a subterm of $\blue{\overline{i}}$ shall be associated to \blue{blue} pointers. This construction can possibly be generalized to an arbitrary finite number of colors.

\paragraph*{The Split measure.} First, we need to define the \emph{coloring} of a $\l$-term. This is just a map from every (instance of) term constructor, \ie application, abstraction, and variable, appearing in a term to the set $\cols\defeq\{\cred,\cblue\}$. The color of the topmost constructor of a colored $\lambda$-term can be retrieved via the function $\colf\cdot:\Lambda_{\cols}\to\cols$ where $\Lambda_{\cols}$ is the set of colored $\l$-terms. From now on, we shall only consider colored $\lambda$-terms. Note that, since for the space measure we are interested in the color of closures, one might avoid coloring variables. Indeed, closures of the kind $(\var,\lenv)$ are never built, because of the unchaining optimization. Nonetheless, we color variables anyway, for the sake of uniformity and simplicity.

The next step is splitting size of states into a \emph{pair} of natural numbers, where the first one represents the number of \red{red} pointers, and the second one is the number of \blue{blue} pointers. We shall need a notion of sum over pairs, defined as $(h_1,k_1)\doubleplus (h_2,k_2)\defeq (h_1+h_2,k_1+k_2)$, and we overload the $\max$ function over pairs as $\max_i \{(h_i,k_i)\}\defeq (\max_i \{h_i\},\max_i \{k_i\})$. The size of states  is defined as follows.
\begin{definition}[Split abstract space]
	The split abstract space $\size\state$ taken by a \SpKAM state $\state$ is defined on the structure of \SpKAM states as follows.
	\[\begin{array}{ccccccc}
		\textsc{Environments}
		&
		\textsc{Stacks}\\
		\begin{array}{rcl}
			\size{\stempty} & \defeq &(0,0)
			\\
			\size{\esub\var\clos\cdot\lenv} & \defeq & \size{\clos}\doubleplus\size{\lenv}
		\end{array}
		&
		\begin{array}{rcl}
			\size{\stempty} & \defeq &(0,0)
			\\
			\size{\clos\cdot\stack} & \defeq &	\size{\clos}\doubleplus\size{\stack}
		\end{array}
		\\\\
		
		\textsc{Closures} 
		&		
		\textsc{States}\\
		\begin{array}{rcl}
			\size{(\tm,\lenv)} & \defeq & \size\lenv \doubleplus \begin{cases}(1,0) & \text{if }\colf\tm=\cred,\\
			(0,1) & \text{if }\colf\tm=\cblue.\end{cases}
		\end{array}
		& 
		\begin{array}{rcl}			
			\size{(\tm,\lenv,\stack)} &\defeq & \size{\lenv}\doubleplus\size{\stack}
		\end{array}
	\end{array}\]
\end{definition}
The only important detail to notice is the clause for closures, which checks the color of the subterm and increases the size accordingly. The split size a \SpKAM run is then obtained by considering the maximum size of the states reached along that run. The definition looks identical to the one presented in Section~\ref{sec:spkam}. What changes is just the measure on states. We define the maximum over a set of pairs $A$ component-wise: $\max A\defeq (\max_{ (h,k)\in A}h, \max_{(h,k)\in A}k)$.

\begin{definition}[Run split abstract space]
	Let $\run:\compil{\tm_0}\rightarrow_{\mathrm{Sp\KAM}}^*\state$ be a \SpKAM run. Then the split (abstract) space consumption of the run $\run$ is defined as follows:
	\[\lm{\run}\defeq\max_{\statetwo\in\run}\size{\statetwo}\]
\end{definition}

\begin{example}\label{ex:spkam-split}
	Let us consider our example term $(\la\var(\la\vartwo(\la\varthree\var)(\var\vartwo))\var)\Id$, colored as follows: all the subterms of $\la\var(\la\vartwo(\la\varthree\var)(\var\vartwo))\var$ in \red{red}, and $\Id$ in \blue{blue}, \ie we have $(\red{\la\var(\la\vartwo(\la\varthree\var)(\var\vartwo))\var})\blue{\Id}$ (the topmost constructor---an application in this case---never contributes to the space usage, thus for simplicity we do not color it). The colored execution $\run$ is in Figure~\ref{fig:col-run}.
	\begin{figure}[t]
	\[
	\begin{array}{l|l|ll}
		\mathsf{Term}   & 
		\mathsf{Environment} & \mathsf{Stack}\\
		\cline{1-3}
		(\red{\la\var(\la\vartwo(\la\varthree\var)(\var\vartwo))\var})\blue\Id & \stempty & \stempty & \tokamseanv\\
		\red{\la\var(\la\vartwo(\la\varthree\var)(\var\vartwo))\var} & \stempty & (\blue{\Id},\stempty) & \tokambnw\\
		\red{(\la\vartwo(\la\varthree\var)(\var\vartwo))\var} & \esub\var{(\blue{\Id},\stempty)} & \stempty & \tokamseav\\
		\red{\la\vartwo(\la\varthree\var)(\var\vartwo)} & \esub\var{(\blue{\Id},\stempty)} & (\blue{\Id},\stempty) & \tokambnw \\
		\red{(\la\varthree\var)(\var\vartwo)} & \esub\vartwo{(\blue{\Id},\stempty)} \cdot \esub\var{(\blue{\Id},\stempty)}=:\lenv & \stempty & \tokamseanv\\
		\red{\la\varthree\var} & \esub\var{(\blue{\Id},\stempty)} & (\red{\var\vartwo},\lenv) & \tokambw\\
		\red\var & \esub\var{(\blue{\Id},\stempty)} & \stempty & \tokamsub\\
		\blue\Id & \stempty & \stempty
	\end{array}
	\]
	\caption{The colored run of the term $(\red{\la\var(\la\vartwo(\la\varthree\var)(\var\vartwo))\var})\blue{\Id}$.}
	\label{fig:col-run}
	\end{figure}
	In particular, all closures $(\blue{\Id},\stempty)$ are \blue{blue}, and the closure $(\red{\var\vartwo},\lenv)$ is \red{red}. Then, we see immediately that the maximum consumption of space is $\lm{\run}=(1,3)$, \ie one \red{red} closure and three \blue{blue} closures (both the maxima are reached in third-to-last state).
\end{example}

\begin{remark} We highlight that it is possible that for some run $\run$, we have that $\lm\run=(h,k)$, but there is no state $\state\in\run$ such that $\size{\state}=(h,k)$. For example, if there are $\state$ and $\statetwo$ such that $\size\state=(10,0)$ and $\size\statetwo=(0,10)$, and $\state,\statetwo\in\rho$, then $\lm\run\geq(10,10)$. This is because the two components are considered independently. This lax measure does not hinder the proof of space reasonability in~\citep{LICS2022}.
\end{remark}

\paragraph*{The split type system.} The grammar of types needs to be slightly adjusted. In fact, the index counting the size of the closure associated to a multiset has to be split into two: the first one considering the number of \red{red} pointers, and the second one considering the number of \blue{blue} pointers.
\[
\begin{array}{rrcll}
	\textsc{Linear 
		Types}&\linty,\lintytwo&\grameq&\initty\grammarpipe\arr{\mty^\pair}{\linty} 
	\\[3pt]
	\textsc{Closure 
		Types}&\mty^p,\mtytwo^p&\grameq&\mset{\linty_1,\ldots,\linty_n}^p & n\geq 0\\[3pt]
	\textsc{Pairs} & \pair & \defeq & (h,k) & h+k\geq 1
\end{array}
\]
(Dry, summable) type environments, type judgments, and type derivations are defined as in Section~\ref{sec:type-system}, \emph{mutatis mutandis}. In order to define the typing rules, we need to modify the notion of size of types and type contexts.
\[
\begin{array}{c@{\hspace{1cm}}c@{\hspace{1cm}}c}
	\size{\star}\defeq (0,0) &
	\size{\arr{\mty^p}\linty}\defeq p\doubleplus\size{\linty}&
	\size{\var:\mty^p,\tye}\defeq p\doubleplus\size{\tye}
\end{array}
\]
Typing rules are in Figure~\ref{fig:split-rules}. They are mostly identical to the ones of Section~\ref{sec:type-system}, but for the fact that when we create a new multiset (associated to a machine closure) in rules $\tymany$ and $\tynone$, we check if the subterm is either \red{red} or \blue{blue}, and set the index accordingly.

% !TeX spellcheck = en_US
% !TEX root = main.tex
%%%%%%%%%%%%%%%%%%%%%%%%%%%%%%%%%%%%%%%%%%%%%%%%%%%%%%%%%%%%%%%%%%%%%%%%%%%%%%%%
\begin{figure}[t]
	{\small \[
	\begin{array}{c@{\hspace{0cm}}c}
	\infer[\tyvar]{\wtjudg{\var:\mset{\linty}^p}{p\doubleplus\size{\linty}}{\var}{\linty}}{}
	&
	
	\infer[\tylamstar]{\wtjudg{\tye}{\size{\tye}}{\lambda\var.\tm}{\initty}}{
		\dom\tye = \fv{\la\var\tm} & \tye \mbox{ dry}}
	\\[10pt]
	\infer[\tylam_{1}]{\wtjudg{\tye}{w}{\lambda\var.\tm}{\arr{\mty^{p}}{\linty}}} 
	{\wtjudg{\tye,\var:\mty^{p}}{w}{\tm}{\linty}}
	&
	\infer[\tylam_{2}]{\wtjudg{\tye}{\max\{w,\size\tye\doubleplus\size\linty\doubleplus p\}}{\lambda\var.\tm}{\arr{\emmtype^{p}}{\linty}}}
	{\wtjudg{\tye}{w}{\tm}{\linty} & \var\notin\dom\tye}
	\\[10pt]
	\infer[\tymanysr]{\wtjudg{\uplus_{i=1}^n\tye_i}{\max_{i}\set{v_{i}}}{\tm}{\mset{\linty_1\mydots\linty_n}^{(1,0)\doubleplus\size{\uplus_{i=1}^n\tye_i}}}}
	{\wtjudg{\tye_i}{v_{i}}{\tm}{\linty_i} & 1\leq i\leq n & \#_i\tye_{i} & \colf{\tm}=\cred}
	& \infer[\tynonesr]{\wtjudg{\tye}{(0,0)}{\tm}{\emmset^{(1,0) \doubleplus \size\tye  }}}{\dom\tye = \fv{\tm} & \tye  \mbox{ dry} & \colf{\tm}=\cred}
	\\[10pt]
	\infer[\tymanysb]{\wtjudg{\uplus_{i=1}^n\tye_i}{\max_{i}\set{v_{i}}}{\tm}{\mset{\linty_1\mydots\linty_n}^{(0,1)\doubleplus\size{\uplus_{i=1}^n\tye_i}}}}
	{\wtjudg{\tye_i}{v_{i}}{\tm}{\linty_i} & 1\leq i\leq n & \#_i\tye_{i} & \colf{\tm}=\cblue}
	& \infer[\tynonesb]{\wtjudg{\tye}{(0,0)}{\tm}{\emmset^{(0,1) \doubleplus \size\tye  }}}{\dom\tye = \fv{\tm} & \tye  \mbox{ dry} & \colf{\tm}=\cblue}
	\\[10pt]
	
	\infer[\tyapp_{1}]{\wtjudg{\tye\uplus 
			\tyetwo}{\max\{w,v\}}{\tm\tmtwo}{\linty 
	}}{\wtjudg{\tye}{w}{\tm}{\arr{\mty^{p}}{\linty}}
		& 
		\wtjudg{\tyetwo}{v}{\tmtwo}{\mty^{p}}& \tye\# \tyetwo & \tmtwo\not\in\mathcal{V}} 
	&
	
	%	\multicolumn{3}{c}{\infer[\tyapp]{\wtjudg{\tye\uplus 
	%	\sum_{i=1}^n \tyetwo_i  }{\max\{w,v_i\}}{\tm\tmtwo}{\linty 
	%		}}{\wtjudg{\tye}{w}{\tm}{\arr{\mset{\lintytwo_1,\ldots,\lintytwo_n}^{\size{
	%		  \tyetwo}+1}}{\linty}}
	%			& 
	%			\mset{\wtjudg{\tyetwo_i}{v_i}{\tmtwo}{\lintytwo_i}}_{i\in\mset{1,\ldots,n}}}}
	%		\\[10pt]
	%
	%		\multicolumn{3}{c}{\infer[\tyapp]{\wtjudg{\tye 
	%		}{w}{\tm\tmtwo}{\linty 
	%				}}{\wtjudg{\tye_{|\tm}}{w}{\tm}{\arr{\mset{\cdot}^{\size{
	%							\tye_{|\tmtwo}}+1}}{\linty}}}}
	%	\\[10pt]
	\infer[\tyapp_{2}]{\wtjudg{\tye\uplus
			\var:\mty^p  
		}{w}{\tm\var}{\linty 
	}}{\wtjudg{\tye}{w}{\tm}{\arr{\mty^p}{\linty}}
		&&\tye\#\var:\mty^p}
	\end{array}
	\]}
	
	\caption{The split type system.}
	\label{fig:split-rules}
\end{figure}

\begin{example}
	We provide the split type derivation for the term $(\la\var(\la\vartwo(\la\varthree\var)(\var\vartwo))\var)\Id$, colored as in the previous example, \ie we have $(\red{\la\var(\la\vartwo(\la\varthree\var)(\var\vartwo))\var})\blue{\Id}$. We define $\mathbf{0}\defeq (0,0)$.
	\[
	\infer[\tyapp_1]{\tjudgw{}{(1,3)} {(\red{\la\var(\la\vartwo(\la\varthree\var)(\var\vartwo))\var})\blue\Id}{\initty}}{
		\infer[\tylam_1]{\tjudgw{}{(1,3)}{\red{\la\var(\la\vartwo(\la\varthree\var)(\var\vartwo))\var}} {\arr{\mset\initty^{(0,1)}}\initty}}{
			\infer[\tyapp_2]{\tjudgw{\var:\mset\initty^{(0,1)}}{(1,3)}{\red{(\la\vartwo(\la\varthree\var)(\var\vartwo))\var}}{\initty}}{\infer[\tylam_1]{\tjudgw{\var:\mset\initty^{(0,1)}}{(1,3)}{\red{\la\vartwo(\la\varthree\var)(\var\vartwo)}} {\arr{\emmset^{(0,1)}}\initty}}{
					\infer[\tyapp_1]{\tjudgw{\var:\mset\initty^{(0,1)},\vartwo:\emmset^{(0,1)}}{(1,3)}{\red{(\la\varthree\var)(\var\vartwo)}}{\initty}}{
						\infer[\tylam_2]{\tjudgw{\var:\mset\initty^{(0,1)}}{(1,3)}{\red{\la\varthree\var}}{\arr{\emmset^{(1,2)}}\initty}}{
							\infer[\tyvar]{\tjudgw{\var:\mset\initty^{(0,1)}}{(0,1)}{\red{\var}}{\initty}}{}} & \infer[\tynonesr]{\tjudgw{\var:\emmset^{(0,1)},\vartwo:\emmset^{(0,1)}}{\mathbf{0}}{\red{\var\vartwo}}{\emmset^{(1,2)}}}{}}}}} & \hspace{-2.7cm}\infer[\tymanysb]{\tjudgw{}{\mathbf{0}}{\blue\Id}{\mset\initty^{(0,1)}}}{
			\infer[\tylamstar]{\tjudgw{}{\mathbf{0}}{\blue\Id}{\initty}}{}}}
	\]
	Also in this case, we can observe the precise correspondence between this type derivation and the execution of the \SpKAM of Example~\ref{ex:spkam-split}. Not only rules and transitions are into a one-to-one correspondence, but also stack and environment entries (with their sizes) can be seen, respectively, in types and in type environments. Of course, as a consequence, the final weight is $(1,3)$, as the split space consumption of the \SpKAM execution. 
\end{example}
All the technical development of Section~\ref{sec:correctness} scales smoothly, \emph{mutandis mutandis}. In particular, we have the following characterization:
\begin{theorem}[Split correctness]
	Let $\tm$ be a closed colored $\lambda$-term. Then there exists a complete \SpKAM run $\run$ from $\tm$ such that $\lm\run=w$ if and only if there exists $\tyd\pof\tjudgw{}{\weight}{\tm}{\initty}$.
\end{theorem}

% !TeX spellcheck = en_US
% !TEX root = main.tex
%%%%%%%%%%%%%%%%%%%%%%%%%%%%%%%%%%%%%%%%%%%%%%%%%%%%%%%%%%%%%%%%%%%%%%%%%%%%%%%%
\section{Conclusions}
This paper is set in the recent trend aiming at analyzing complexity properties of $\l$-terms through the use of multi types (a.k.a. non-idempotent intersection types). Taking advantage of a recent result about a reasonable  space measure for the call-by-name $\l$-calculus, we craft a multi-type system able to reflect such a measure on typed $\l$-terms. 

There are at least four interesting directions for future work. First, we 
would 
like to adapt the type system to cover call-by-value evaluation, where a notion of 
reasonable space measure is also known~\citep{LICS2022}. Second, it would be 
interesting to understand what happens in the case of infinite computations. In 
these cases, the space consumption can be either finite, if the machine cycles, or 
infinite. In the former case, a newly devised (coinductive) type system could give 
the space bound. In the latter case, one could be interested in asymptotic 
properties, \eg ``during the (infinite) execution, the space consumption is 
logarithmic in time''. This kind of properties could be useful in applications 
that have to do with infinite data, such as stream processing, and in general with 
lazy infinite (coinductive) data structures. For designing the type system, one 
might possibly build on Vial's study of infinitary evaluation based on a variant 
of multi types \citep{DBLP:conf/lics/Vial17}. Then, we would like to 
investigate if \emph{idempotent} intersection types could be used to 
characterize space consumption. Differently from time, which is obtained by 
summing all the computation steps, indeed, space is obtained by taking a 
maximum over the space consumed by each state, which is an idempotent 
operation. Finally, we would like to formalize our results in a suitable 
proof-assistant. The result seems within reach as the proofs are indeed of 
very syntactical nature. It would seem reasonable to start from Atkey's \texttt{Agda} 
formalization of the correspondence between multi types and the 
KAM~\citep{Atkey}.\footnote{Note added in proof. We have now fully 
formalized 
the results contained in this paper. These will be published/made publicly 
available soon.}

\bibliographystyle{ACM-Reference-Format}
\bibliography{biblio}

\begin{LONG}\newpage
	\appendix
	\section*{Appendix}
	% !TeX spellcheck = en_US
% !TEX root = main.tex
%%%%%%%%%%%%%%%%%%%%%%%%%%%%%%%%%%%%%%%%%%%%%%%%%%%%%%%%%%%%%%%%%%%%%%%%%%%%%%%%
\section{Proofs of Section~\ref{sec:correctness} (Weights capture the space 
of the \SpKAM)}\label{app:correctness}
\soundness*

\begin{proof}
	The proof is by induction on $\size\tyd$ and case analysis on whether $\state$ is final. If $\state$ is final then the run $\run$ in the statement is given by the empty run. Therefore, we have $\lm\run = \size\state$ and the required space bound becomes $\weight= \size\state$, which is given by \refprop{weight-of-final-states-is-right}. If $\state$ is not final then $\state\tospkam\statetwo$. By examining all the transition rules. We set $\linty\defeq\arr{\mty_1^{k_{1}}}{\arr{\cdots}{\arr{\mty_n^{k_{n}}}{\initty}}}$ and $\stack\defeq\stack_1\cdots\stack_n$.
	\begin{itemize}
		\item Case $\tokamseav$, \ie
		$\state=\spkamstate{\tm\var}{\lenv}\stack$ and 
		$\statetwo=\spkamstate\tm{ 
			\lenv|_\tm}{\lenv(\var)\cdot\stack}$. The type derivation $\tyd$ typing  $\state$ has the following shape:
		\[
		\infer[\tystate]{\wtjudg{}{\max_i\{w',u,v_i\}}{\spkamstate{\tm\var}\lenv\stack}{\initty}}{
			\infer[\tyapp_{2}]{\wtjudg{\tye\uplus\var:\mty}{w'}{\tm\var}{\linty} }{\wtjudg{\tye}{w'}{\tm}{\arr{\mty}{\linty}}}
			&
			\tyd_{\lenv}\pof\wtjudg{}{u}{\lenv}{\tye\uplus\var:\mty}
			&
			\wtjudg{}{v_i}{\stack_i}{\mty_i^{k_{i}}} 
		}
		\]
		with $\weight = \max_i\{w',v_i,u\}$. By \reflemma{multiset-splitting}, there are two derivations $\tyd_{\tye}\pof\wtjudg{}{\weightthree_{\tye}}{\lenv|_{\dom\tye}}{\tye}$ and $\tyd_{\var:\mty}\pof\wtjudg{}{\weightthree_{\var:\mty}}{\lenv(\var)}{\var:\mty}$ such that $\size{\tyd_{\lenv}} = \size{\tyd_{\tye}} + \size{\tyd_{\var:\mty}}$ and $\weightthree = \max\set{\weightthree_{\tye}, \weightthree_{\var:\mty}}$. By the environment domain invariant (\reflemma{spkam-invariants}), $\dom\tye = \fv\tm$. Moreover, $\tyd_{\var:\mty}$ is necessarily the conclusion of a unary $\tyenv$ rule of premise $\tyd_{\mty}\pof\wtjudg{}{\weightthree_{\mty}}{\lenv(\var)}{\mty}$ for which $\size{\tyd_{\mty}} = \size{\tyd_{\var:\mty}}$ and $\weightthree_{\mty} = \weightthree_{\var:\mty}$.
		Then, $\statetwo$ can be typed by the following derivation $\tydtwo$:
		\[\tydtwo \defeq 
		\infer[\tystate]{\wtjudg{}{\max_i\{w',u_{\tye},u_{\mty},v_i\}}{\spkamstate\tm{ 
					\lenv|_\tm}{\lenv(\var)\cdot\stack}}{\initty}}{
			\wtjudg{\tye'}{w'}{\tm}{\arr{\mty}{\linty}}
			&
			\tyd_{\tye}\pof \wtjudg{}{u_{\tye}}{\lenv|_\tm}{\tye}
			&
			\tyd_{\mty}\pof \wtjudg{}{u_{\mty}}{\lenv(\var)}{\mty}
			&
			\wtjudg{}{v_i}{\stack_i}{\mty_i^{k_{i}}} 
		}
		\]
		Since $\size{\tyd_{\lenv}} = \size{\tyd_{\tye}} + \size{\tyd_{\var:\mty}} = \size{\tyd_{\tye}} + \size{\tyd_{\mty}}$, we have $\size\tyd > \size\tydtwo$ (because the $\tyapp_{2}$ rule is removed), and so we can apply the \ih, obtaining a run $\runtwo: \statetwo \rightarrow_{\mathrm{Sp\KAM}}^* \state_{f}$ to a final state such that $\max_i\{w',u_{\tye},u_{\mty},v_i\} = \lm\runtwo$. Then there is a run $\run:\state \rightarrow_{\mathrm{Sp\KAM}}^* \state_{f}$, proving the first part of the statement (termination).
		
		For the space bound, note that, since $\weightthree = \max\set{\weightthree_{\tye}, \weightthree_{\var:\mty}} = \max\set{\weightthree_{\tye}, \weightthree_{\mty}}$, we have $\weight = \max_i\{w',u_{\tye},u_{\mty},v_i\}$. Additionally, $\size\state \leq \size\statetwo$, because by the environment domain invariant $\lenv|_\tm$ removes at most $\lenv(\var)$ from $\lenv$, which is however added to the stack. Then $\lm\run = \lm\runtwo = \max_i\{w',u_{\tye},u_{\mty},v_i\} = \weight$, proving the second part of the statement.
		%%%%%%%%%%%%%%%%%%%
		%%%%%%%%%%%%%%%%%%%
		\item Case $\tokamseanv$, \ie $\state=\spkamstate{\tm\tmtwo}{\lenv}\stack$ 
		and 
		$\statetwo=\spkamstate\tm{ 
			\lenv|_\tm}{(\tmtwo,\lenv|_\tmtwo)\cdot\stack}$.

		The type derivation $\tyd$ typing  $\state$ has the following shape:
		\[
		\infer[\tystate]{\wtjudg{}{\max_i\{w_1,w_2,v_i,u\}}{\spkamstate{\tm\tmtwo}\lenv\stack}{\initty}}{
			\infer[\tyapp_{1}]{\wtjudg{\tye\uplus\tyetwo}{\max\{w_1,w_2\}}{\tm\tmtwo}{\linty } }{\wtjudg{\tye}{w_1}{\tm}{\arr{\mty^{k}}{\linty}}
				& 
				\wtjudg{\tyetwo}{w_2}{\tmtwo}{\mty^{k}}& \tye\# \tyetwo}
			&
			\wtjudg{}{u}{\lenv}{\tye\uplus\tyetwo}
			&
			\wtjudg{}{v_i}{\stack_i}{\mty_i^{k_{i}}} 
		}
		\]
		with $\weight = \max_i\{w_1,w_2,v_i,u\}$. By \reflemmap{multiset-splitting}{env}, there are two derivations $\tyd_{\tye}\pof\wtjudg{}{\weightthree_{\tye}}{\lenv|_{\dom\tye}}{\tye}$ and $\tyd_{\tyetwo}\pof\wtjudg{}{\weightthree_{\tyetwo}}{\lenv|_{\dom\tyetwo}}{\tyetwo}$ such that $\size{\tyd_{\lenv}} = \size{\tyd_{\tye}} + \size{\tyd_{\tyetwo}}$ and $\weightthree = \max\set{\weightthree_{\tye}, \weightthree_{\tyetwo}}$. By the environment domain invariant (\reflemma{spkam-invariants}), $\dom\tye = \fv\tm$ and $\dom\tyetwo = \fv\tmtwo$.		Then, $\statetwo$ can be typed by the following derivation $\tydtwo$:
		\[\tydtwo \defeq 
		\infer[\tystate]{\wtjudg{}{
				\max_i\{w_{1},w_{2},v_i,u_{\tye},u_{\tyetwo}\}}{\spkamstate\tm{ 
					\lenv|_\tm}{(\tmtwo,\lenv|_\tmtwo)\cdot\stack}}{\initty}}{
			\wtjudg{\tye}{w_1}{\tm}{\arr{\mty^{k}}{\linty}}
			&
			\wtjudg{}{u_{\tye}}{\lenv|_\tm}{\tye}
			&
			\infer[\tyclos]{\wtjudg{}{\max\{w_{2},u_{\tyetwo}\}
				}{(\tmtwo,\lenv|_\tmtwo)}{\mty^k}}{\wtjudg{\tyetwo}{w_2}{\tmtwo}{\mty^{k}}&&
				\wtjudg{}{u_{\tyetwo}}{\lenv|_\tmtwo}{\tyetwo}}
			&
			\hspace{-3pt}\wtjudg{}{v_i}{\stack_i}{\mty_i^{k_{i}}} 
		}
		\]
		Since $\size{\tyd_{\lenv}} = \size{\tyd_{\tye}} + \size{\tyd_{\tyetwo}}$, we have $\size\tyd > \size\tydtwo$ (because the $\tyapp_{2}$ rule is removed and rule $\tyclos$ does not count for the size), and so we can apply the \ih, obtaining a run $\runtwo: \statetwo \rightarrow_{\mathrm{Sp\KAM}}^* \state_{f}$ to a final state such that $\max_i\{w_{1},w_{2},v_i,u_{\tye},u_{\tyetwo}\} = \lm\runtwo$. Then there is a run $\run:\state \rightarrow_{\mathrm{Sp\KAM}}^* \state_{f}$, proving the first part of the statement (termination).
		
		For the space bound, note that, since $\weightthree = \max\set{\weightthree_{\tye}, \weightthree_{\tyetwo}}$, we have \\$\max_i\{w_{1},w_{2},v_i,u_{\tye},u_{\tyetwo}\} = \max_i\{w_{1},w_{2},v_i,u\} = w$. Additionally, $\size\state \leq \size\statetwo$, because by the environment domain invariant all entries of $\lenv$ are in $\lenv|_\tm$ or in $\lenv|_{\tmtwo}$. Then $\lm\run = \lm\runtwo = \max_i\{w',u_{\tye},u_{\mty},v_i\} = \weight$, proving the second part of the statement.
		%%%%%%%%%%%%%%%%%%%
		%%%%%%%%%%%%%%%%%%%
		\item Case $\tokambw$, \ie $\state=\spkamstate{\la\var\tm}{\lenv}{\clos\cdot\stack}$ 
		and 
		$\statetwo=\spkamstate\tm{ 
			\lenv}{\stack}$. The type derivation $\tyd$ typing  $\state$ has the following shape:
		\[
		\infer[\tystate]{\wtjudg{}{\max_i\{w',\size\tye+\size\linty+k,u,v,v_i\}}{\spkamstate{\la\var\tm}\lenv{\clos\cdot\stack}}{\initty}}{
			\infer[\tylam_{2}]{\wtjudg{\tye}{\max\{w',\size\tye+\size\linty+k\}}{\lambda\var.\tm}{\arr{\emmtype^{k}}{\linty}}}
			{\wtjudg{\tye}{w'}{\tm}{\linty} & \var\notin\dom\tye}
			&
			\wtjudg{}{u}{\lenv}{\tye}
			&
			\wtjudg{}{v}{\clos}{\emmtype^k}
			&
			\wtjudg{}{v_i}{\stack_i}{\mty_i^{k_{i}}} 
		}\]
		with $\weight = \max_i\{w',\size\tye+\size\linty+k,u,v,v_i\}$. The target state $\statetwo$ can be typed by the following derivation $\tydtwo$:
		\[\tydtwo \defeq
		\infer[\tystate]{\wtjudg{}{\max_i\{w',u,v_i\}}{\spkamstate\tm\lenv\stack}{\initty}}{
			\wtjudg{\tye}{w'}{\tm}{\linty} & \var\notin\dom\tye
			&
			\wtjudg{}{u}{\lenv}{\tye}
			&
			\wtjudg{}{v_i}{\stack_i}{\mty_i^{k_{i}}} 
		}\]
		Since the $\tylam_{2}$ rule is removed, we have $\size\tyd>\size\tydtwo$, and so we can apply the \ih, obtaining a run $\runtwo: \statetwo \rightarrow_{\mathrm{Sp\KAM}}^* \state_{f}$ to a final state such that $\max_i\{w',u,v_i\} = \lm\runtwo$. Then there is a run $\run:\state \rightarrow_{\mathrm{Sp\KAM}}^* \state_{f}$, proving the first part of the statement (termination).
		
		For the space bound, note that:
		\begin{itemize}
			\item $\size\state =  \size\tye+\size\linty+k$ by \reflemma{space-bound-on-current-state}, giving $\weight = \max_i\{w',\size\state,u,v,v_i\}$. 
			\item $v=0$ by \reflemmap{dry-env-clos}{two}, giving $\weight = \max_i\{w',\size\state,u,v_i\}$.
		\end{itemize}
		Now, there are two cases:
		\begin{itemize}
			\item $\lm\run = \lm\runtwo$: that is, $\size\state \leq \lm\runtwo = \max_i\{w',u,v_i\}$. Then $\size\state \leq \max_i\{w',u,v_i\}$, and so $\weight = \max_i\{w',\size\state,u,v_i\} = \max_i\{w',u,v_i\} =_{\ih} \lm\runtwo = \lm\run$.
			
			\item $\lm\run > \lm\runtwo$: that is, $\size\state > \lm\runtwo = \max_i\{w',u,v_i\}$. Then $\size\state > \max_i\{w',u,v_i\}$, and so $\weight = \max_i\{w',\size\state,u,v_i\} = \size\state = \lm\run$.
		\end{itemize}
		
		%%%%%%%%%%%%%%%%%%%
		%%%%%%%%%%%%%%%%%%%
		\item Case $\tokambnw$, \ie $\state=\spkamstate{\la\var\tm}{\lenv}{\clos\cdot\stack}$ 
		and 
		$\statetwo=\spkamstate\tm{ 
			\esub{\var}{\clos}\cdot\lenv}{\stack}$. The type derivation $\tyd$ typing  $\state$ has the following shape:
		\[
		\infer[\tystate]{\wtjudg{}{\max_i\{w',v,v_i,u\}}{\spkamstate{\la\var\tm}\lenv{\clos\cdot\stack}}{\initty}}{
			\infer[\tylam_{1}]{\wtjudg{\tye}{w'}{\lambda\var.\tm}{\arr{\mty^{k}}{\linty}}} 
			{\wtjudg{\tye,\var:\mty^{k}}{w'}{\tm}{\linty}}
			&
			\wtjudg{}{u}{\lenv}{\tye}
			&
			\wtjudg{}{v}{\clos}{\mty^k}
			&
			\wtjudg{}{v_i}{\stack_i}{\mty_i^{k_{i}}} 
		}\]
		with $\weight = \max_i\{w',v,v_i,u\}$. The target state $\statetwo$ can be typed by the following derivation $\tydtwo$:
		\[\tydtwo\defeq
		\infer[\tystate]{\wtjudg{}{\max_i\{w',v,v_i,u\}} {\spkamstate\tm{\esub\var\clos\cdot\lenv}\stack}{\initty}}{
			\wtjudg{\tye,\var:\mty^{k}}{w'}{\tm}{\linty}
			&
			\infer[\tyenv]{\wtjudg{}{\max\{u,v\}}{\esub\var\clos\cdot\lenv}{\tye,\var:\mty^{k}}} {\wtjudg{}{u}{\lenv}{\tye}&\wtjudg{}{v}{\clos}{\mty^k}}
			&
			\wtjudg{}{v_i}{\stack_i}{\mty_i^{k_{i}}} 
		}\]
		Since the $\tylam_{1}$ rule is removed and the $\tyenv$ rule does not count for the size of type derivations, we have $\size\tyd>\size\tydtwo$, and so we can apply the \ih, obtaining a run $\runtwo: \statetwo \rightarrow_{\mathrm{Sp\KAM}}^* \state_{f}$ to a final state such that $\max_i\{w',v,v_i,u\} = \lm\runtwo$. Then there is a run $\run:\state \rightarrow_{\mathrm{Sp\KAM}}^* \state_{f}$, proving the first part of the statement (termination).
		
		For the space bound, note that 
		$$\size\state \ \ =\ \   \size\lenv + \size{\clos\cdot\stack} \ \ =\ \   \size\lenv + \size\clos + \size\stack \ \ =\ \   \size{\esub\var\clos\cdot\lenv} + \size\stack \ \ =\ \  \size\statetwo.$$
		and so the space bound follows from the \ih
		%%%%%%%%%%%%%%%
		%%%%%%%%%%%%%%%
		\item Case $\tokamsub$, \ie $\state=\spkamstate\var
		{\esub\var{(\tmtwo,\lenv)}}\stack$ and
		$\statetwo=\spkamstate\tmtwo\lenv\stack$. The type derivation $\tyd$ typing  $\state$ has the following shape:
		\[
		\infer[\tystate]{\wtjudg{}{\max_i\{k+\size\linty,u,w',v_i\}}{\spkamstate\var
				{\esub\var{(\tmtwo,\lenv)}}\stack}{\initty}}{
			\infer[\tyvar]{\wtjudg{\var:\mset{\linty}^k}{k+\size{\linty}}{\var}{\linty}}{} 
			&
			\infer[\tyenv]{\wtjudg{}{\max\{u,w'\}}{\esub\var{(\tmtwo,\lenv)}}{\var:\mset\linty^k}} {
				\infer[\tyclos]{
					\wtjudg{}{\max\{u,w'\}}{(\tmtwo,\lenv)}{\mset\linty^k}
				}{
					\wtjudg{\tye}{w'}{\tmtwo}{\mset\linty^k }
					&
					\wtjudg{}{u}{\lenv}{\tye} 
				} 
			}
			&
			\wtjudg{}{v_i}{\stack_i}{\mty_i^{k_{i}}} 
		}
		\]
		with $\weight = \max_i\{k+\size\linty,u,w',v_i\}$. The target state $\statetwo$ can be typed by the following derivation $\tydtwo$:
		\[\tydtwo\defeq
		\infer[\tystate]{\wtjudg{}{\max_i\{w',u,v_i\}}{\spkamstate\tmtwo\lenv\stack}{\initty}}{
			\wtjudg{\tye}{w'}{\tmtwo}{\linty } 
			&
			\wtjudg{}{u}{\lenv}{\tye}
			&
			\wtjudg{}{v_i}{\stack_i}{\mty_i^{k_{i}}} 
		}
		\]
		Since the $\tyvar$ rule is removed, we have $\size\tyd>\size\tydtwo$, and so we can apply the \ih, obtaining a run $\runtwo: \statetwo \rightarrow_{\mathrm{Sp\KAM}}^* \state_{f}$ to a final state such that $\max_i\{w',u,v_i\} = \lm\runtwo$. Then there is a run $\run:\state \rightarrow_{\mathrm{Sp\KAM}}^* \state_{f}$, proving the first part of the statement (termination).
		
		For the space bound, $\size\state =  k+\size\linty$ by \reflemma{space-bound-on-current-state}, giving $\weight = \max_i\{\size\state,u,w',v_i\}$. 
		Now, there are two cases:
		\begin{itemize}
			\item $\lm\run = \lm\runtwo$: that is, $\size\state \leq \lm\runtwo = \max_i\{w',u,v_i\}$. Then $\size\state \leq \max_i\{w',u,v_i\}$, and so $\weight = \max_i\{\size\state,u,w',v_i\} = \max_i\{w',u,v_i\} =_{\ih} \lm\runtwo = \lm\run$.
			
			\item $\lm\run > \lm\runtwo$: that is, $\size\state > \lm\runtwo = \max_i\{w',u,v_i\}$. Then $\size\state > \max_i\{w',u,v_i\}$, and so $\weight = \max_i\{\size\state,u,w',v_i\} = \size\state = \lm\run$.\qedhere
		\end{itemize}
		
	\end{itemize}
\end{proof}
	% !TeX spellcheck = en_US
% !TEX root = main.tex
%%%%%%%%%%%%%%%%%%%%%%%%%%%%%%%%%%%%%%%%%%%%%%%%%%%%%%%%%%%%%%%%%%%%%%%%%%%%%%%%
\section{Proofs of Section~\ref{sec:time} (Capturing the \SpKAM low-level time)}\label{app:time}

\begin{lemma}[Multi-set splitting]
	\label{l:multiset-splitting-time} % \reflemmap{multiset-splitting}{env}
	\hfill
	\begin{enumerate}
		\item \emph{Terms}: let $\tyd \pof \wtjudg{\tye}{\weight}{\tm}{\mty^{k}\uplus\mtytwo^{k}}$ with $k= 1+\size\tye$. Then there exists two type derivations $\tyd_{\mty} \pof \wtjudg{\tye_{\mty}}{\weight_{\mty}}{\tm}{\mty^{k}}$ and $\tyd_{\mtytwo} \pof \wtjudg{\tye_{\mtytwo}}{\weight_{\mtytwo}}{\tm}{\mtytwo^{k}}$ such that $\tye_{\mty} \# \tye_{\mtytwo}$, $\tye = \tye_{\mty} \uplus \tye_{\mtytwo}$, $\size\tyd = \size{\tyd_{\mty}} + \size{\tyd_{\mtytwo}}$ and $\weight = \weight_{\mty}+\weight_{\mtytwo}$.
		
		\item \emph{Closures}: let $\tyd \pof \wtjudg{}{\weight}{\clos}{\mty^{k}\uplus\mtytwo^{k}}$. Then there exists two type derivations $\tyd_{\mty} \pof \wtjudg{}{\weight_{\mty}}{\clos}{\mty^{k}}$ and $\tyd_{\mtytwo} \pof \wtjudg{}{\weight_{\mtytwo}}{\clos}{\mtytwo^{k}}$ such that $\size\tyd = \size{\tyd_{\mty}} + \size{\tyd_{\mtytwo}}$ and $\weight = \weight_{\mty}+\weight_{\mtytwo}$.
		
		\item \label{p:multiset-splitting-time-env}
		\emph{Environments}: let $\tye$ and $\tyetwo$ summable and $\tyd \pof \wtjudg{}{\weight}{\lenv}{\tye\uplus\tyetwo}$. Then there exists two type derivations $\tyd_{\tye} \pof \wtjudg{}{\weight_{\tye}}{\lenv|_{\dom\tye}}{\tye}$ and $\tyd_{\tyetwo} \pof \wtjudg{}{\weight_{\tyetwo}}{\lenv|_{\dom\tyetwo}}{\tyetwo}$ such that $\size\tyd = \size{\tyd_{\tye}} + \size{\tyd_{\tyetwo}}$ and $\weight = \weight_{\tye}+\weight_{\tyetwo}$.
	\end{enumerate}
\end{lemma}
\begin{proof}
	The proof goes in the same way of the proof of \reflemma{multiset-splitting}.
\end{proof}

\timesoundness*

\begin{proof}
	The proof is by induction on $\size\tyd$ and case analysis on whether $\state$ is final. If $\state$ is final then the run $\run$ in the statement is given by the empty run. Therefore, we have $\lmt\run = \size\state$ and the required time bound becomes $\weight= \size\state$, which is given by \reflemma{weight-of-final-states-is-right-time}. If $\state$ is not final then $\state\tospkam\statetwo$. By examining all the transition rules. We set $\linty\defeq\arr{\mty_1^{k_{1}}}{\arr{\cdots}{\arr{\mty_n^{k_{n}}}{\initty}}}$ and $\stack\defeq\stack_1\cdots\stack_n$.
	\begin{itemize}
		\item Case $\tokamseav$, \ie
		$\state=\spkamstate{\tm\var}{\lenv}\stack$ and 
		$\statetwo=\spkamstate\tm{ 
			\lenv|_\tm}{\lenv(\var)\cdot\stack}$. The type derivation $\tyd$ typing  $\state$ has the following shape:
		\[
		\infer[\tystate]{\wtjudg{}{w'+k+\size\tye+\size\linty+u+\sum_i v_i}{\spkamstate{\tm\var}\lenv\stack}{\initty}}{
			\infer[\tyapp_{2}]{\wtjudg{\tye\uplus\var:\mty^k}{w'+k+\size\tye+\size\linty}{\tm\var}{\linty} }{\wtjudg{\tye}{w'}{\tm}{\arr{\mty^k}{\linty}}}
			&
			\tyd_{\lenv}\pof\wtjudg{}{u}{\lenv}{\tye\uplus\var:\mty^k}
			&
			\wtjudg{}{v_i}{\stack_i}{\mty_i^{k_{i}}} 
		}
		\]
		with $\weight = w'+k+\size\tye+\size\linty+u+\sum_i v_i$. By \reflemmap{multiset-splitting-time}{env}, there are two derivations $\tyd_{\tye}\pof\wtjudg{}{\weightthree_{\tye}}{\lenv|_{\dom\tye}}{\tye}$ and $\tyd_{\var:\mty}\pof\wtjudg{}{\weightthree_{\var:\mty}}{\lenv(\var)}{\var:\mty}$ such that $\size{\tyd_{\lenv}} = \size{\tyd_{\tye}} + \size{\tyd_{\var:\mty}}$ and $\weightthree = \weightthree_{\tye}+ \weightthree_{\var:\mty}$. By the environment domain invariant (\reflemma{spkam-invariants}), $\dom\tye = \fv\tm$. Moreover, $\tyd_{\var:\mty}$ is necessarily the conclusion of a unary $\tyenv$ rule of premise $\tyd_{\mty}\pof\wtjudg{}{\weightthree_{\mty}}{\lenv(\var)}{\mty}$ for which $\size{\tyd_{\mty}} = \size{\tyd_{\var:\mty}}$ and $\weightthree_{\mty} = \weightthree_{\var:\mty}$.
		Then, $\statetwo$ can be typed by the following derivation $\tydtwo$:
		\[\tydtwo \defeq 
		\infer[\tystate]{\wtjudg{}{w'+u_{\tye}+u_{\mty}+\sum_i v_i}{\spkamstate\tm{ 
					\lenv|_\tm}{\lenv(\var)\cdot\stack}}{\initty}}{
			\wtjudg{\tye'}{w'}{\tm}{\arr{\mty^k}{\linty}}
			&
			\tyd_{\tye}\pof \wtjudg{}{u_{\tye}}{\lenv|_\tm}{\tye}
			&
			\tyd_{\mty}\pof \wtjudg{}{u_{\mty}}{\lenv(\var)}{\mty^k}
			&
			\wtjudg{}{v_i}{\stack_i}{\mty_i^{k_{i}}} 
		}
		\]
		Since $\size{\tyd_{\lenv}} = \size{\tyd_{\tye}} + \size{\tyd_{\var:\mty}} = \size{\tyd_{\tye}} + \size{\tyd_{\mty}}$, we have $\size\tyd > \size\tydtwo$ (because the $\tyapp_{2}$ rule is removed), and so we can apply the \ih, obtaining a run $\runtwo: \statetwo \rightarrow_{\mathrm{Sp\KAM}}^* \state_{f}$ to a final state such that $w'+u_{\tye}+u_{\mty}+\sum_i v_i = \lmt\runtwo$.
		
		For the time bound, note that, since $\weightthree = \weightthree_{\tye}+ \weightthree_{\var:\mty} = \weightthree_{\tye}+ \weightthree_{\mty}$, we have $\weight = w'+k+\size\tye+\size\linty+\weightthree_{\tye}+ \weightthree_{\mty}+\sum_i v_i=\size\state+w'+\weightthree_{\tye}+ \weightthree_{\mty}+\sum_i v_i$. Then $\lmt\run = \size\state+\lmt\runtwo = w'+u_{\tye}+u_{\mty}+\sum v_i+k+\size\tye+\size\linty = \weight$.
		%%%%%%%%%%%%%%%%%%%
		%%%%%%%%%%%%%%%%%%%
		\item Case $\tokamseanv$, \ie $\state=\spkamstate{\tm\tmtwo}{\lenv}\stack$ 
		and 
		$\statetwo=\spkamstate\tm{ 
			\lenv|_\tm}{(\tmtwo,\lenv|_\tmtwo)\cdot\stack}$.

		The type derivation $\tyd$ typing  $\state$ has the following shape:
		\[
		\infer[\tystate]{\wtjudg{}{w_1+w_2+\size{\tye\uplus\tyetwo}+\size\linty +\sum_i v_i+u}{\spkamstate{\tm\tmtwo}\lenv\stack}{\initty}}{
			\infer[\tyapp_{1}]{\wtjudg{\tye\uplus\tyetwo}{w_1+w_2+\size{\tye\uplus\tyetwo}+\size\linty}{\tm\tmtwo}{\linty } }{\wtjudg{\tye}{w_1}{\tm}{\arr{\mty^{k}}{\linty}}
				& 
				\wtjudg{\tyetwo}{w_2}{\tmtwo}{\mty^{k}}& \tye\# \tyetwo}
			&
			\wtjudg{}{u}{\lenv}{\tye\uplus\tyetwo}
			&
			\wtjudg{}{v_i}{\stack_i}{\mty_i^{k_{i}}} 
		}
		\]
		with $\weight = w_1+w_2+\size{\tye\uplus\tyetwo}+\size\linty +\sum_i v_i+u$. By \reflemmap{multiset-splitting-time}{env}, there are two derivations $\tyd_{\tye}\pof\wtjudg{}{\weightthree_{\tye}}{\lenv|_{\dom\tye}}{\tye}$ and $\tyd_{\tyetwo}\pof\wtjudg{}{\weightthree_{\tyetwo}}{\lenv|_{\dom\tyetwo}}{\tyetwo}$ such that $\size{\tyd_{\lenv}} = \size{\tyd_{\tye}} + \size{\tyd_{\tyetwo}}$ and $\weightthree = \weightthree_{\tye}+ \weightthree_{\tyetwo}$. By the environment domain invariant (\reflemma{spkam-invariants}), $\dom\tye = \fv\tm$ and $\dom\tyetwo = \fv\tmtwo$.		Then, $\statetwo$ can be typed by the following derivation $\tydtwo$:
		\[\tydtwo \defeq 
		\infer[\tystate]{\wtjudg{}{
				w_{1}+w_{2}+\sum_i v_i+u_{\tye}+u_{\tyetwo}}{\spkamstate\tm{ 
					\lenv|_\tm}{(\tmtwo,\lenv|_\tmtwo)\cdot\stack}}{\initty}}{
			\wtjudg{\tye}{w_1}{\tm}{\arr{\mty^{k}}{\linty}}
			&
			\wtjudg{}{u_{\tye}}{\lenv|_\tm}{\tye}
			&
			\infer[\tyclos]{\wtjudg{}{w_{2}+u_{\tyetwo}
				}{(\tmtwo,\lenv|_\tmtwo)}{\mty^k}}{\wtjudg{\tyetwo}{w_2}{\tmtwo}{\mty^{k}}&&
				\wtjudg{}{u_{\tyetwo}}{\lenv|_\tmtwo}{\tyetwo}}
			&
			\hspace{-4pt}\wtjudg{}{v_i}{\stack_i}{\mty_i^{k_{i}}} 
		}
		\]
		Since $\size{\tyd_{\lenv}} = \size{\tyd_{\tye}} + \size{\tyd_{\tyetwo}}$, we have $\size\tyd > \size\tydtwo$ (because the $\tyapp_{2}$ rule is removed and rule $\tyclos$ does not count for the size), and so we can apply the \ih, obtaining a run $\runtwo: \statetwo \rightarrow_{\mathrm{Sp\KAM}}^* \state_{f}$ to a final state such that $w_{1}+w_{2}+\sum_i v_i+u_{\tye}+u_{\tyetwo} = \lmt\runtwo$.
		
		For the time bound, note that, since $\weightthree = \weightthree_{\tye}+ \weightthree_{\tyetwo}$, we have $w_{1}+w_{2}+\size{\tye\uplus\tyetwo}+\size\linty+\sum_i v_i+u_{\tye}+u_{\tyetwo} = w_{1}+w_{2}+\size{\tye\uplus\tyetwo}+\size\linty+\sum_i v_i+u = w$. Then $\lmt\run = \lmt\runtwo + \size\state = \weight$.
		%%%%%%%%%%%%%%%%%%%
		%%%%%%%%%%%%%%%%%%%
		\item Case $\tokambw$, \ie $\state=\spkamstate{\la\var\tm}{\lenv}{\clos\cdot\stack}$ 
		and 
		$\statetwo=\spkamstate\tm{ 
			\lenv}{\stack}$. The type derivation $\tyd$ typing  $\state$ has the following shape:
		\[
		\infer[\tystate]{\wtjudg{}{w'+\size\tye+\size\linty+k+u+v +\sum_i v_i}{\spkamstate{\la\var\tm}\lenv{\clos\cdot\stack}}{\initty}}{
			\infer[\tylam_{2}]{\wtjudg{\tye}{w'+\size\tye+\size\linty+k}{\lambda\var.\tm}{\arr{\emmtype^{k}}{\linty}}}
			{\wtjudg{\tye}{w'}{\tm}{\linty} & \var\notin\dom\tye}
			&
			\wtjudg{}{u}{\lenv}{\tye}
			&
			\wtjudg{}{v}{\clos}{\emmtype^k}
			&
			\wtjudg{}{v_i}{\stack_i}{\mty_i^{k_{i}}} 
		}\]
		with $\weight = w'+\size\tye+\size\linty+k+u+v +\sum_i v_i$. The target state $\statetwo$ can be typed by the following derivation $\tydtwo$:
		\[\tydtwo \defeq
		\infer[\tystate]{\wtjudg{}{w'+u+\sum_i v_i}{\spkamstate\tm\lenv\stack}{\initty}}{
			\wtjudg{\tye}{w'}{\tm}{\linty} & \var\notin\dom\tye
			&
			\wtjudg{}{u}{\lenv}{\tye}
			&
			\wtjudg{}{v_i}{\stack_i}{\mty_i^{k_{i}}} 
		}\]
		Since the $\tylam_{2}$ rule is removed, we have $\size\tyd>\size\tydtwo$, and so we can apply the \ih, obtaining a run $\runtwo: \statetwo \rightarrow_{\mathrm{Sp\KAM}}^* \state_{f}$ to a final state such that $w'+u+\sum_i v_i = \lmt\runtwo$.
		
		For the time bound, since $\weighttwo=0$ by \reflemmap{dry-env-clos-time}{two}, we have that $\lmt\run = \size\state+\lmt\runtwo=w$.
		
		%%%%%%%%%%%%%%%%%%%
		%%%%%%%%%%%%%%%%%%%
		\item Case $\tokambnw$, \ie $\state=\spkamstate{\la\var\tm}{\lenv}{\clos\cdot\stack}$ 
		and 
		$\statetwo=\spkamstate\tm{ 
			\esub{\var}{\clos}\cdot\lenv}{\stack}$. The type derivation $\tyd$ typing  $\state$ has the following shape:
		\[
		\infer[\tystate]{\wtjudg{}{\size\tye+\size\linty+k+w'+v +\sum_i v_i +u}{\spkamstate{\la\var\tm}\lenv{\clos\cdot\stack}}{\initty}}{
			\infer[\tylam_{1}]{\wtjudg{\tye}{w'+\size\tye+\size\linty+k}{\lambda\var.\tm}{\arr{\mty^{k}}{\linty}}} 
			{\wtjudg{\tye,\var:\mty^{k}}{w'}{\tm}{\linty}}
			&
			\wtjudg{}{u}{\lenv}{\tye}
			&
			\wtjudg{}{v}{\clos}{\mty^k}
			&
			\wtjudg{}{v_i}{\stack_i}{\mty_i^{k_{i}}} 
		}\]
		with $\weight = \size\tye+\size\linty+k+w'+v +\sum_i v_i +u$. The target state $\statetwo$ can be typed by the following derivation $\tydtwo$:
		\[\tydtwo\defeq
		\infer[\tystate]{\wtjudg{}{w'+v+\sum_i v_i+u} {\spkamstate\tm{\esub\var\clos\cdot\lenv}\stack}{\initty}}{
			\wtjudg{\tye,\var:\mty^{k}}{w'}{\tm}{\linty}
			&
			\infer[\tyenv]{\wtjudg{}{u+v}{\esub\var\clos\cdot\lenv}{\tye,\var:\mty^{k}}} {\wtjudg{}{u}{\lenv}{\tye}&\wtjudg{}{v}{\clos}{\mty^k}}
			&
			\wtjudg{}{v_i}{\stack_i}{\mty_i^{k_{i}}} 
		}\]
		Since the $\tylam_{1}$ rule is removed and the $\tyenv$ rule does not count for the size of type derivations, we have $\size\tyd>\size\tydtwo$, and so we can apply the \ih, obtaining a run $\runtwo: \statetwo \rightarrow_{\mathrm{Sp\KAM}}^* \state_{f}$ to a final state such that $w'+v+\sum_i v_i+u = \lmt\runtwo$.
		
		For the time bound, note that 
		$\lmt\run = \size\state+\lmt\runtwo=w$.
		%%%%%%%%%%%%%%%
		%%%%%%%%%%%%%%%
		\item Case $\tokamsub$, \ie $\state=\spkamstate\var
		{\esub\var{(\tmtwo,\lenv)}}\stack$ and
		$\statetwo=\spkamstate\tmtwo\lenv\stack$. The type derivation $\tyd$ typing  $\state$ has the following shape:
		\[
		\infer[\tystate]{\wtjudg{}{k+\size\linty+u+w'+\sum_i v_i}{\spkamstate\var
				{\esub\var{(\tmtwo,\lenv)}}\stack}{\initty}}{
			\infer[\tyvar]{\wtjudg{\var:\mset{\linty}^k}{k+\size{\linty}}{\var}{\linty}}{} 
			&
			\infer[\tyenv]{\wtjudg{}{u+w'}{\esub\var{(\tmtwo,\lenv)}}{\var:\mset\linty^k}} {
				\infer[\tyclos]{
					\wtjudg{}{u+w'}{(\tmtwo,\lenv)}{\mset\linty^k}
				}{
					\wtjudg{\tye}{w'}{\tmtwo}{\mset\linty^k }
					&
					\wtjudg{}{u}{\lenv}{\tye} 
				} 
			}
			&
			\wtjudg{}{v_i}{\stack_i}{\mty_i^{k_{i}}} 
		}
		\]
		with $\weight = k+\size\linty+u+w'+\sum_i v_i$. The target state $\statetwo$ can be typed by the following derivation $\tydtwo$:
		\[\tydtwo\defeq
		\infer[\tystate]{\wtjudg{}{w'+u+\sum_i v_i}{\spkamstate\tmtwo\lenv\stack}{\initty}}{
			\wtjudg{\tye}{w'}{\tmtwo}{\linty } 
			&
			\wtjudg{}{u}{\lenv}{\tye}
			&
			\wtjudg{}{v_i}{\stack_i}{\mty_i^{k_{i}}} 
		}
		\]
		Since the $\tyvar$ rule is removed, we have $\size\tyd>\size\tydtwo$, and so we can apply the \ih, obtaining a run $\runtwo: \statetwo \rightarrow_{\mathrm{Sp\KAM}}^* \state_{f}$ to a final state such that $w'+u+\sum_i v_i = \lmt\runtwo$.
		
		For the time bound, note that $\lmt\run = \size\state+\lmt\runtwo=w$. \qedhere
		
	\end{itemize}
\end{proof}
\end{LONG}

\end{document}